\documentclass{PoS}
\pdfoutput=1
\usepackage[vcentermath]{youngtab}
\usepackage{subfig}
\usepackage{lscape}

\usepackage{graphicx}
\usepackage{epstopdf}
\usepackage{enumerate}
\usepackage{cite}
\usepackage{tensor}
\usepackage{slashed}
\usepackage{feynmf}
\usepackage{amsmath}
\usepackage{amssymb}
\usepackage{mathrsfs}
\usepackage{lgrind}

\usepackage{bbm}

\numberwithin{equation}{section}

\newcommand {\be} {\begin {equation}}
\newcommand {\ee} {\end {equation}}

\newcommand {\bes} {\begin {equation*}}
\newcommand {\ees} {\end {equation*}}

\newcommand{\eps}{\epsilon}

\newcommand{\tr}{\operatorname{tr}}

\def\CH{{\cal H}}

\newcommand{\beq}{\begin{equation}}
\newcommand{\eeq}{\end{equation}}

\newcommand{\sgn}{\mathrm{sgn}}

\def\be{ \begin{equation} }
\def\ee{ \end{equation} }

\title{TASI Lectures on Large $N$ Tensor Models}

\ShortTitle{Large $N$ Tensor Models}

\author{\speaker{Igor R. Klebanov}\\
       Department of Physics, Princeton University, Princeton, NJ 08544, USA\\
       E-mail: \email{klebanov@princeton.edu}}

\author{Fedor Popov\\
       Department of Physics, Princeton University, Princeton, NJ 08544, USA\\
       E-mail: \email{fpopov@princeton.edu}}

\author{Grigory Tarnopolsky\\
       Department of Physics, Harvard University, Cambridge, MA 02138, USA\\
       E-mail: \email{grtarnop@g.harvard.edu}}

\abstract{The first part of these lecture notes is mostly devoted to a comparative discussion of the three basic large $N$ limits, which apply to fields which are vectors, matrices, or 
tensors of rank three and higher. After a brief review of some physical applications of large $N$ limits, we present 
a few solvable examples in zero space-time dimension. Using models with fields in the fundamental representation of $O(N)$, $O(N)^2$, or $O(N)^3$ symmetry, we
compare their combinatorial properties and highlight a competition between the snail and melon diagrams.  
We exhibit the different methods used for solving the vector, matrix, and tensor large $N$ limits. In the latter example we review how the dominance
of melonic diagrams follows when a special ``tetrahedral" interaction is introduced. 
The second part of the lectures is mostly about the fermionic quantum mechanical 
tensor models, whose large $N$ limits are similar to that in the Sachdev-Ye-Kitaev (SYK) model. The minimal Majorana model with $O(N)^3$ symmetry and
the tetrahedral Hamiltonian is reviewed in some detail; it is the closest tensor counterpart of the SYK model. 
Also reviewed are generalizations to complex fermionic tensors, including a model with $SU(N)^2\times O(N)\times U(1)$ symmetry,
which is a tensor counterpart of the complex SYK model. The bosonic large $N$ tensor models, 
which are formally tractable in continuous spacetime dimension, are reviewed briefly at
the end.
  }

\FullConference{Theoretical Advanced Study Institute Summer School 2017 "Physics at the Fundamental Frontier"\\
         4 June - 1 July 2017\\
         Boulder, Colorado}

\begin{document}

\section{Introduction and Summary}

An important tool in theoretical physics is the study of limits where the number of degrees of freedom, $N_{\rm dof}$, becomes large.
Some models simplify or even become analytically solvable in such a limit. If an (asymptotic) expansion in inverse powers of $N_{\rm dof}$ can be developed, then
models of physical interest are sometimes well approximated by including a small number of terms.  
Three different broad classes of such ``large $N$ limits" have been explored: the vector limit; the matrix (or 't Hooft) limit; and, more recently, the limit which applies to tensors
of rank three and higher. Each of them is addressed with somewhat different techniques. 
   
The most easily tractable, and historically the first,  large $N$ limit applies to
theories where the degrees of freedom transform as $N$-component vectors under a symmetry group. In such theories  
$N_{\rm dof}\sim N$.
One of the first appearances was in the context of statistical mechanics \cite{Stanley:1968gx}.
A well-known example is the $O(N)$ symmetric quantum theory of $N$ scalar fields $\phi^a$ in $d$ dimensions with interaction
$\frac{g}{4} (\phi^a \phi^a)^2$ (for reviews see \cite{Wilson:1973jj,Moshe:2003xn}). 
It is exactly solvable in the large $N$ limit where $gN$ is held fixed, since summation over the necessary class of bubble diagrams is not hard to evaluate.
The $1/N$ expansion may be developed analytically for any space-time dimension $d$.

Another famous large $N$ limit occurs in models of interacting
$N\times N$ matrix fields, so that $N_{\rm dof}\sim N^2$. This limit made its first appearance in the context of generalizing QCD from $SU(3)$ to $SU(N)$ gauge theory, where
the gauge fields are traceless Hermitian $N\times N$ matrices \cite{'tHooft:1973jz}. A 
significant simplification occurs in the 't Hooft large $N$
limit where $g_{\rm YM}^2 N$ is held fixed: the perturbative expansion is dominated by the planar diagrams.\footnote{This is true if the number of flavors $N_f$, 
i.e. fields in the fundamental
representation of $SU(N)$, is held fixed in the 't Hooft limit. However, if $x=N_f/N$ is held fixed in the large $N$ limit, then the leading terms depend on the parameter $x$. 
This is known as the Veneziano limit \cite{Veneziano:1976wm}. This type of limit can be applied to any theory containing both matrix and vector fields.}  
While the `t Hooft large $N$ limit does not make
QCD in dimension above $2$ 
exactly solvable, it has been an important tool in studying its properties. Furthermore, the `t Hooft large $N$ limit was crucial for the discovery and exploration of the
Anti-de Sitter/Conformal Field Theory (AdS/CFT) correspondence \cite{Maldacena:1997re,Gubser:1998bc,Witten:1998qj}, which has been a major research direction for 
over 20 years. 
For introductions to the AdS/CFT correspondence
you may consult the lectures by O. DeWolfe \cite{DeWolfe:2018dkl}, J. Erdmenger \cite{Erdmenger:2018xqz}, and D. Harlow \cite{Harlow:2018fse} in this volume, 
lectures at the earlier TASI schools including \cite{Klebanov:2000me,DHoker:2002nbb,Maldacena:2003nj}, and the comprehensive review 
\cite{Aharony:1999ti}.  

Besides the gauge theories, the `t Hooft large $N$ limit applies to matrix models, such as the integral over of a Hermitian matrix $\Phi$
with single-trace interactions like $g_3 \tr \Phi^3$ (see section \ref{Hermitmat}).  
Such matrix models are exactly solvable in the large $N$ limit keeping $g_3^2 N$ fixed, in the special low-dimensional
cases $d\leq 1$ \cite{Brezin:1977sv}. Here the planar graphs may be thought of as discretized random surfaces. 
Tuning $g_3^2 N$ to a special value where a random surface becomes macroscopic \cite{Gross:1989vs,Brezin:1990rb,Douglas:1989ve} 
(for reviews see \cite{Ginsparg:1993is,Klebanov:1991qa}) has
taught us a lot about the two-dimensional quantum gravity, which can be mapped to the quantum Liouville theory \cite{Polyakov:1981rd}.

\begin{figure}[h!]
                \centering
                \includegraphics[width=16cm]{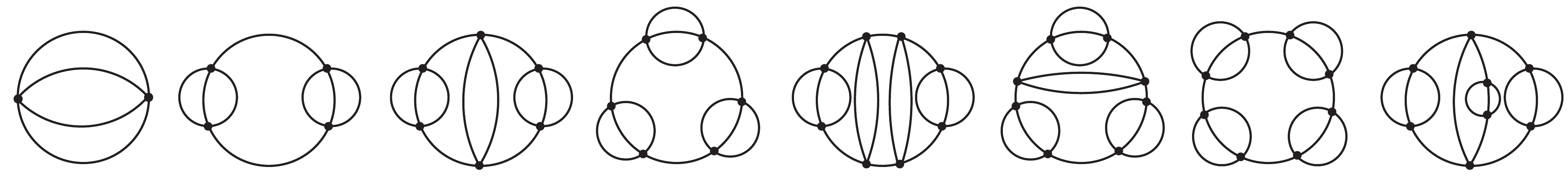}
\caption{All the melonic vacuum diagrams up to order $g^8$.}
                \label{MelonsEx}
\end{figure}

\begin{figure}
	\centering
	\includegraphics[scale=1]{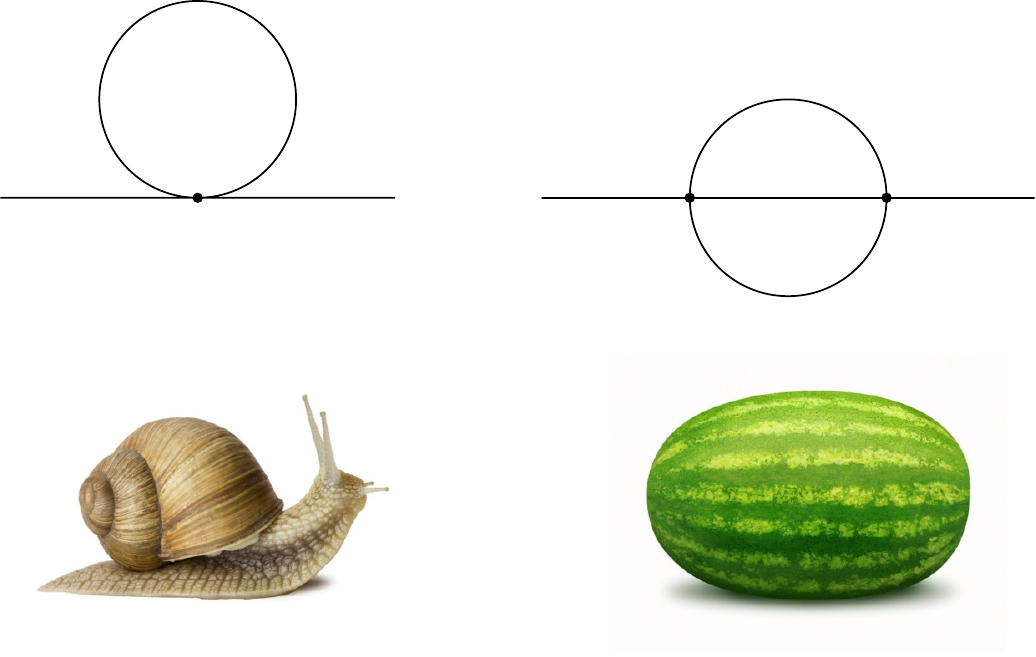}
	\caption{A snail vs. a (water)melon.}
	\label{svm}
\end{figure}

In view of these classic results, it is natural to study theories with rank-$m$ tensor degrees of freedom $\phi^{a_1 \ldots a_m}$, where each index takes $N$ values so that 
$N_{\rm dof}\sim N^m$ \cite{Ambjorn:1990ge,Sasakura:1990fs,Gross:1991hx}. 
Since the complexity of taking the large $N$ limit increases from $m=1$ to $m=2$, one might expect that the tensor models
with $m>2$ are much more difficult than the matrix models.  
However, 
by choosing the interactions appropriately, it is possible to find models with $m>2$ where a large $N$ limit is solvable
\cite{Gurau:2009tw,Gurau:2011aq,Gurau:2011xq,Bonzom:2011zz,Tanasa:2011ur,Bonzom:2012hw,Carrozza:2015adg,Witten:2016iux,Klebanov:2016xxf,Giombi:2018qgp}
(for reviews, see  \cite{Gurau:2011xp,Tanasa:2015uhr,Delporte:2018iyf}). 
The perturbative expansion is then dominated by special classes of Feynman diagrams.
While the original hope for applications of these tensor models lay with quantum gravity above
two dimensions, starting in October 2016 a new physical connection was opened up \cite{Witten:2016iux,Klebanov:2016xxf} with the Sachdev-Ye-Kitaev (SYK) model 
\cite{Sachdev:1992fk,1999PhRvB..59.5341P, 2000PhRvL..85..840G,Kitaev:2015,Kitaev:2017awl} (for recent reviews of the SYK model, 
see \cite{Sarosi:2017ykf,Rosenhaus:2018dtp}).
The $q=4$ version of the SYK model
involves a large number $N_{\rm SYK}$ of  
Majorana fermions $\psi^i$, 
interacting via the quartic Hamiltonian
\beq
H_{\rm SYK} = \sum_{i_{1}<i_{2}<i_{3}<i_{4}} J_{i_{1}i_{2}i_{3}i_{4}} \psi^{i_{1}} \psi^{i_{2}} \psi^{i_{3}} \psi^{i_{4}}\ , 
\label{SYKHamilt}
\eeq
where each $J_{i_{1}i_{2}i_{3}i_{4}}$ is a Gaussian random variable with standard deviation $\sim N^{-3/2}_{\rm SYK}$.
The closest tensor counterpart \cite{Klebanov:2016xxf} of this model contains $N^3$ Majorana fermions $\psi^{abc}$, 
$a,b,c=1, \ldots, N$, 
\begin{align}
\{\psi^{a b c}, \psi^{a' b' c'}\} =\delta^{a a'} \delta^{b b'}   \delta^{c c'}\, , \label{comrel}     
\end{align}
whose interactions are
governed by the $O(N)^3$ symmetric ``tetrahedral" Hamiltonian
\begin{align}
H =  \frac{g} {4} \bigg (\psi^{abc}\psi^{ab'c'} \psi^{a'bc'}\psi^{a'b'c} -  \frac{1 } {4} N^4 \bigg ) \, .
\label{Hequal}
\end{align} 
In this case the large $N$ limit has to be taken keeping $g^2 N^3$ fixed. Then the diagrammatic expansion is dominated by the so-called melonic diagrams
(see figure \ref{MelonsEx}); they are obtained by iterating the melon (or sunset) propagator insertion shown on the right in figure \ref{svm}.

Besides the three ``basic" large $N$ limits mentioned above, there are 
more exotic examples where $N_{\rm dof}$ can scale as a fractional power of $N$ in the large $N$ limit. Such limits
are possible in presence of additional symmetries, such as extended supersymmetry. For example, in the context of the ABJM model \cite{Aharony:2008ug}, 
which is the $U(N)_k\times U(N)_{-k}$
Chern-Simons gauge theory in $d=3$ coupled to massless matter, it is possible to take the large $N$ limit while keeping $k$ fixed (this is is different from
the 't Hooft limit where $N/k$ is held fixed). In this so-called "M-theory limit", there is strong
evidence that the ABJM theory is dual to M-theory
on the   $AdS_4\times S^7/Z_k$ background, and one finds $N_{\rm dof}\sim k^{1/2} N^{3/2}$ 
\cite{Klebanov:1996un,Drukker:2010nc,Herzog:2010hf,Marino:2011eh} (for reviews see \cite{Marino:2016new,Pufu:2016zxm}).
An even more exotic situation is when the Chern-Simons theory is $U(N)_{k_1}\times U(N)_{k_2}$ with $k_1+ k_2\neq 0$; then 
$N_{\rm dof}\sim |k_1+k_2|^{1/3} N^{5/3}$ \cite{Jafferis:2011zi}. Such ``exotic" large $N$
limits are fascinating, but we will not discuss them further in these lectures. 

In section \ref{apps} we review some classic applications of the vector and matrix large $N$ limits, including the $O(N)$ magnets and $SU(N)$ gauge theories. 
We also discuss briefly the planar ${\cal N}=4$ supersymmetric Yang-Mills theory in $d=4$ and the AdS/CFT correspondence.  
Section \ref{snailsmelons} contains 
a few solvable large $N$ examples in $d=0$. Using integrals over bosonic variables in the fundamental representation of $O(N)$, $O(N)^2$, or $O(N)^3$ symmetry, we
compare the different combinatorial properties of the three basic large $N$ limits. We highlight a competition between the snail and melon diagrams illustrated in figure \ref{svm}.  
In section \ref{Theproof} we explain, following \cite{Klebanov:2016xxf,Carrozza:2015adg}, 
why the melonic diagrams are dominant in large $N$ theories with $O(N)^3$ symmetry and
the tetrahedral quartic interaction. In section \ref{oncube} we review in some detail that $O(N)^3$ symmetric Majorana quantum mechanics (\ref{Hequal}) and its comparison with the SYK
model (\ref{SYKHamilt}). We explain why the $O(N)^3$ tensor model Hamiltonian is much more sparse than in the SYK model with $N^3$ fermions.
Nevertheless, some quantities in the two models have the same large $N$ behavior, which can be shown using the Schwinger-Dyson equations.
A novel feature of the tensor model is that the number of $SO(N)^3$ invariant $2k$-particle operators grows as $k! 2^k$ \cite{Bulycheva:2017ilt}; 
as a result, the Hagedorn temperature vanishes
in the large $N$ limit. In section \ref{compferm} we review the complex fermionic tensor model with $SU(N)^2\times O(N)\times U(1)$ symmetry \cite{Klebanov:2016xxf}, 
which is a tensor counterpart of the
complex SYK model \cite{Sachdev:2015efa,Davison:2016ngz}. We also present new results on a complex bipartite model with $O(N)^3$ symmetry; its spectrum  contains an operator with a complex scaling dimension.
Such complex dimensions of the form $\frac{d}{2}+ i \alpha(d)$ also appear in some large $N$ bosonic tensor models, which are briefly reviewed in section \ref{Bonten}. 
However, in the $O(N)^3$ model with the sixth-order ``prismatic" interaction \cite{Giombi:2018qgp}, 
there are ranges of $d$ where the large $N$ theory appears to be free of the complex scaling
dimensions.

These notes are far from a comprehensive review, but we hope that they will give the reader a sense of the variety of large $N$ models that
have been studied over many years. It is clear that their exploration is far from over. 
There may well be even more sophisticated large $N$ limits that are yet to be discovered, and interesting quantum theories which realize them.

\section{Some Applications of the Vector and Matrix Large $N$ Limits}

\label{apps}

\begin{figure}
		\centering
		\includegraphics[scale=0.5]{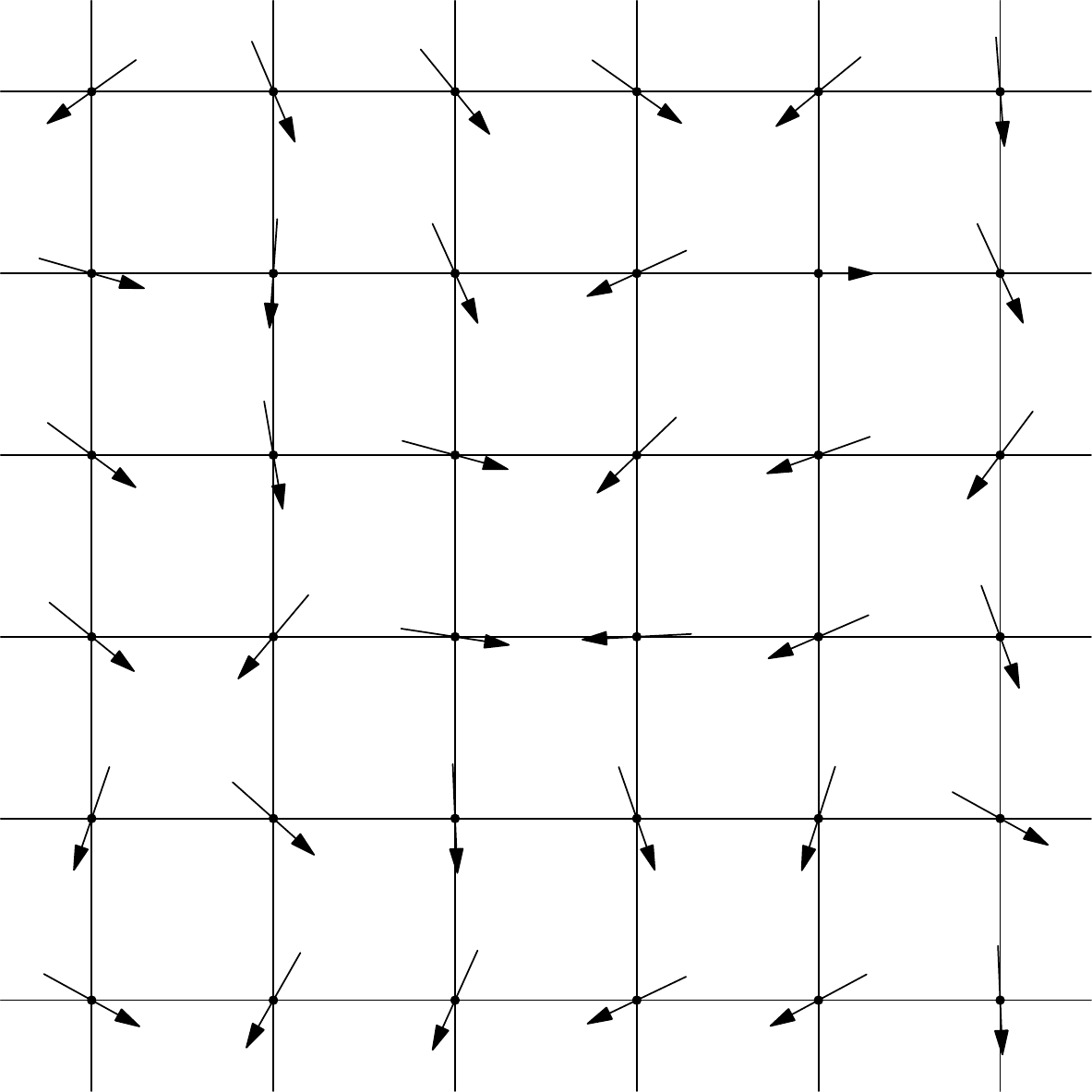}
		\caption{A configuration of spins in an $O(2)$ model on a two-dimensional lattice.}
		\label{fig:magnet}
	\end{figure}

A classic application of the vector large $N$ limit is
to the $O(N)$ ferromagnet, which is described by the following energy with $J>0$ (see figure \ref{fig:magnet})
	\be
	\mathcal{E} = - J \sum_{\langle ij\rangle } \vec{n}_i \cdot \vec{n}_j,\quad \vec{n} = \left(n^1,\ldots, n^N\right), \quad \vec{n}^2 = 1\, ,
	\ee 
where $\langle ij\rangle$ denotes the nearest neighbor lattice sites.
	The partition function is 
	\be
	Z = \sum_{\{\vec n_i\}} e^{-\beta \mathcal{E}}\, ,
	\ee
where the sum is over all possible choices of $\vec n$ vectors at the lattice sites.
	This model has a second order phase transition at a critical value of $\beta J$, 
near which it is described by the Euclidean field theory, as reviewed in \cite{Wilson:1973jj,Moshe:2003xn}:
	\be
	S = \int d^d x \left[\frac12 \left(\nabla \phi_i\right)^2 + \frac{m^2}{2} \phi_i^2 + \frac{g}{4} (\phi_i^2)^2 \right]\,.
\label{quartic}	
\ee
	The phase transition separates the phases with $m^2>0$ and $m^2 < 0$. For $N>2$ the second-order transition is present for $d>2$, while for $N=1,2$
it is present also for $d=2$.
 In general the field theory (\ref{quartic}) is super-renormalizable in $d<4$, and for $2< d< 4$ it flows to the
interacting infrared fixed point. In $d=4-\eps$ the IR stable zero of the beta function is at weak coupling: 
\be
	\beta_g = -\epsilon g + \frac{N+8}{8\pi^2} g^2 +O(g^3) \Rightarrow g_* = \frac{8\pi^2}{N+8}\epsilon +O(\epsilon^2)\ . 
	\ee
Substituing $g_*$ into scaling dimension of various operators produces their $\epsilon$ expansion.
The field $\phi_i$ is the spin operator, and $\phi_i^2$ is  the "energy" operator -- it is the simplest $O(N)$ invariant ``single-trace" operator. Their scaling
dimensions have $\epsilon$ expansions 
\be
	\Delta_{\phi_i}=1- \frac12 \epsilon + \frac{N+2}{4(N+8)^2}\epsilon^2+ O(\epsilon^3) , 
\qquad \Delta_{\phi^2} = 2 -\frac{6}{N+8}\epsilon +O(\epsilon^2) \,.
	\ee
Extending them to higher orders in $\epsilon$ and extrapolating to $\epsilon=1$ gives accurate estimates of the scaling dimensions in the $d=3$
critical $O(N)$ model for all values of $N$, including $N=1$ which corresponds to the Ising model.

Another tool for studying the critical $O(N)$ model is the $1/N$ expansion. It may be developed for the IR fixed point in $2<d<4$ by introducing 
the auxiliary field $\sigma$ (for a review, see the TASI 2015 lectures \cite{Giombi:2016ejx}):
\be
	S_\sigma = \int d^d x \left[\frac12 \left(\nabla \phi_i\right)^2 - \frac {i}{2} \sigma \phi_i^2 + \frac{1}{4 g} \sigma^2 \right]\,.
	\ee
After integrating out the $N$ fields $\phi^i$, the field $\sigma$ acquires an induced effective action of order $N$, 
which dominates in the IR over the $\sigma^2$ term.
Using the induced $\sigma$ propagator, which is of order $1/N$, leads 
to the $1/N$ expansions for the scaling dimensions.
While they are available as functions of $d$ \cite{Moshe:2003xn,Fei:2014yja}, we will just state them for the physically interesting case $d=3$:
	\be
	\Delta_{\phi_i}=\frac{1} {2}+ \frac{4}{3 \pi^2 N}- \frac{256}{27 \pi^4 N^2}+ \ldots\ , \qquad
 \Delta_{\phi^2} = 2 - \frac{32 }{3 \pi^2 N} + \frac{32 (16- 27 \pi^2)}{27 \pi^4 N^2}+ \ldots\, .
	\ee
Precise results \cite{Kos:2013tga} from applications of the numerical conformal bootstrap \cite{Rattazzi:2008pe} (for a recent review, see \cite{Poland:2018epd}) 
to the $d=3$ critical $O(N)$ model
 show an excellent match with these $1/N$ expansions for $N> 10$.
 
The $d=3$ critical $O(N)$ model has another interesting application -- to higher spin quantum gravity. It has been conjectured \cite{Klebanov:2002ja}
that its $O(N)$ singlet sector is dual to the
minimal Vasiliev higher-spin theory in $AdS_4$ \cite{Vasiliev:1990en}. This conjecture and its generalizations have passed a number of non-trivial tests; 
for reviews see \cite{Giombi:2012ms,Giombi:2016ejx}.

Since its introduction in 1974 \cite{'tHooft:1973jz}, the 't Hooft large $N$ limit has had a multitude of applications. The recent ones include lattice calculations of bound state masses in $SU(N)$
gauge theory for moderate values of $N$, and their large $N$ extrapolations. For example, in the 3-dimensional pure glue $SU(N)$ theory, where the numerical results are particularly
accurate, it was found \cite{Athenodorou:2016ebg} that the masses of low-lying glueballs exhibit a very good fit with the expansions
\be
\frac{m} {g_{\rm YM}^2 N} = a_0 + \frac{a_1}{N^2} + \frac{a_2}{N^4}+ \ldots
\ee
for values of $N$ in ranging from $2$  to $16$. This constitutes a nice non-perturbative check of the large $N$ expansion which appears in the 't Hooft limit.

\begin{figure}
	\centering
	\includegraphics[scale=1.5]{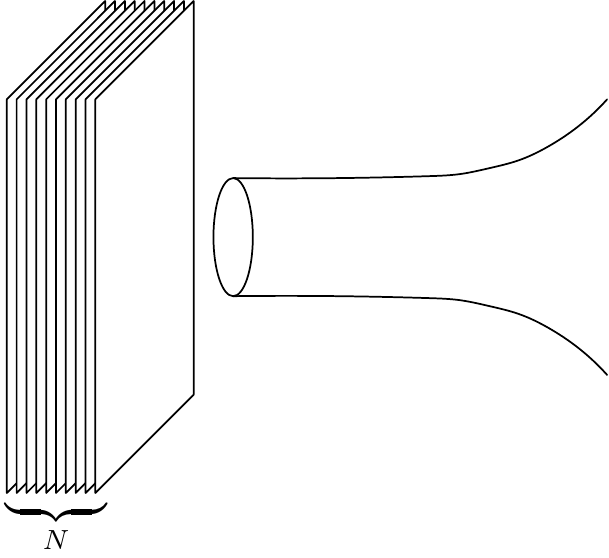}
	\caption{A stack of $N$ D3-branes and the curved background it creates.}
	\label{fig:d3brane}
\end{figure}	

The  't Hooft large $N$ limit also plays an important role in the correspondence \cite{Maldacena:1997re,Gubser:1998bc,Witten:1998qj} between 
the ${\cal N}=4$ supersymmetric $SU(N)$ gauge in $d=4$ and the type
IIB string theory on the $AdS_5\times S^5$ background. To arrive at this correspondence, it is convenient 
\cite{Klebanov:1997kc,Maldacena:1997re}
to begin with a stack of a large number $N$ of Dirichlet 3-branes
(for an original review of D-branes, see the TASI 1996 lectures \cite{Polchinski:1996na}). At low energies, this stack has two equivalent descriptions illustrated in
figure \ref{fig:d3brane}.
The first involves the ${\cal N}=4$ supersymmetric
$SU(N)$ gauge theory weakly coupled to the type IIB closed superstrings. Since D-branes carry Ramond-Ramond charges \cite{Polchinski:1995mt}, 
the second description involves type IIB 
closed superstring theory in the background of the extreme Ramond-Ramond charged 3-brane, which has
the metric
\be\label{d3metric}
ds^2 = h^{-\frac12}(r)\left(-dt^2 + dx_i^2\right)+h^\frac12(r)\left(dr^2 + r^2 d\Omega_5^2\right),\qquad h(r)=1 + \frac{L^4}{r^4}\ ,
\ee
where $d\Omega_5^2$ is the metric of a round unit 5-sphere.
Equating the ADM tension of this extreme gravitational background to $N$ times the D3-brane tension gives 
\be
L^4 = \lambda \alpha'^2\ , \qquad \lambda\equiv  g_{\rm YM}^2 N\ ,
\ee
which implies that the curvature is small everywhere in string units when the 't Hooft coupling $\lambda$ is large. The low-energy limit may be taken 
directly in the geometry by sending
$r\rightarrow 0$ \cite{Maldacena:1997re}, which corresponds to omitting the first term $1$ in $h(r)$. The resulting metric
\be
ds^2 \approx \frac{r^2}{L^2}\left(-dt^2+dx_i^2\right) + L^2 \frac{dr^2}{r^2} + L^2 d\Omega_5^2 
\ee
is that of a product of (the Poincar\' e patch of) the 5-dimensional negatively curved 
anti-de Sitter space, $AdS_5$, and the 5-dimensional positively curved sphere, $S^5$. The two 5-dimensional spaces have equal curvature radii $L$.

The dimensionless parameter which measures the effects of quantum gravity, i.e. the string loop corrections in $AdS_5\times S^5$, is $G_{10}/L^8$, where
$G_{10}$ is the ten-dimensional Newton constant. This is found to be of order
$1/N^2$. Therefore, the 't Hooft large $N$
limit corresponds to the classical limit of string theory. While the AdS/CFT correspondence has been conjectured to be valid at any value of $\lambda$, a crucial additional 
simplification occurs when it is taken to be very large. In this regime, the string theory may be well-approximated by the effective supergravity, and the effect of higher-derivative
stringy currections to the effective action is suppressed.
This provides a remarkable application of the methods of Einstein's theory of gravity -- this time to strongly coupled large $N$ QFT! 

The AdS/CFT dictionary \cite{Gubser:1998bc,Witten:1998qj} includes the one-to-one correspondence between gauge invariant scalar operators of scaling dimension $\Delta$
and scalar fields of mass-squared
\be 
m^2= \frac{\Delta (\Delta-4)}{L^2}\ .
\ee 
For example, we can consider the chiral primary operators 
\be
O_k= 	{\rm tr}\, \Phi^{(i_1} \Phi^{i_2} \ldots \Phi^{i_k)}\, ,
\ee
which are symmetric traceless polynomials made of the
six scalar fields $\Phi_i$ contained in the ${\cal N}=4$ SYM theory; the scalars are in the adjoint representation of the gauge group and in the $6$ of the 
R-symmetry group $SU(4)\sim SO(6)$. 
	These operators are protected by supersymmetry and have exact dimension $\Delta = k$ (this can be checked
perturbatively at small $\lambda$). On the AdS side they are dual to the Kaluza-Klein modes on the $S^5$ which indeed have $m^2 L^2= k(k-4)$, $k=2, 3, \ldots$
\cite{Kim:1985ez}. This provided one of the first tests of the AdS/CFT correspondence.

The planar  ${\cal N}=4$ supersymmetric Yang-Mills theory has another remarkable property, the exact integrability 
\cite{Minahan:2002ve}
(for a comprehensive review see \cite{Beisert:2010jr}). 
The integrability has allowed matching of the perturbative expansions of some quantities, evaluated using the planar diagrams for $\lambda \ll 1$, to the predictions of
string theory in weakly curved $AdS_5\times S^5$ valid for $\lambda\gg 1$.  
Consider, for example, another operator made out of the adjoint scalars: an $SO(6)$ singlet known
	as the Konishi operator, 
	\be	
	O_{Konishi} = \sum^6_{i=1} {\rm tr}\, \Phi_i^2 \ .
	\ee
Its dimension
is not protected by supersymmetry, as can be seen from the perturbative planar expansion
\cite{Fiamberti:2007rj,Bajnok:2008bm}
	\be
	\Delta_{Konishi} = 2 + \frac{12}{(4\pi)^2} \lambda -  \frac{48}{(4\pi)^4}\lambda^2 +  \frac{336}{(4\pi)^6} \lambda^3 + 
\frac{-2496+576 \zeta(3)-1440 \zeta(5)} {(4\pi)^8} \lambda^4   + O(\lambda^5)\ .\label{konishicft}
	\ee  
The non-perturbative methods to describe the dimension of this operator as a function of $\lambda$ using the exact integrability were developed in 
\cite{Gromov:2009tv,Gromov:2013pga}.
They have been used to develop the strong coupling expansion \cite{Gromov:2009zb,Gromov:2011bz,Gromov:2014bva,Hegedus:2016eop}  
	\be
	\Delta_{Konishi} = 2 \lambda^{1/4}-2 + 2\lambda^{-1/4} + \left (\frac12 - 3\zeta(3)\right ) \lambda^{-3/4} + 
\left (\frac {15}{2} \zeta(5)  +6 \zeta(3)+ \frac{1}{2} \right ) \lambda^{-5/4}+
O\left (\lambda^{-7/4}\right) \ .
\label{deltaK}
	\ee
In fact, the growth of dimensions of unprotected operators as $\lambda^{1/4}$ was one of the first predictions  of the AdS/CFT correspondence
 \cite{Gubser:1998bc}.
The massive closed superstring states have $m^2 = \frac{4n}{\alpha'}, n =1,\ldots$.
The dimension of operator dual to such a massive string in $AdS_5\times S^5$ is 
	\be
	\Delta = 2 + \sqrt{4+\left(m L\right)^2} = 2 \sqrt{n} \lambda^{1/4}+ \ldots, \qquad  \lambda \gg 1\,.
	\ee
The Konishi operator is dual to the first massive state, $n=1$, and this explains why the coefficient of the first term in (\ref{deltaK}) is $2$. 
The next three terms in the strong coupling expansion (\ref{deltaK}) 
also agree with calculations \cite{Roiban:2011fe,Beccaria:2012xm} using superstring theory in $AdS_5\times S^5$. Thus, studies of the scaling
dimension of the Konishi operator in the large $N$ limit of the ${\cal N}=4$ supersymmetric $SU(N)$ gauge theory have led to highly non-trivial tests of
the AdS/CFT correspondence.

\section{Vector, Matrix and Tensor Models: Snails vs. Melons}
\label{snailsmelons}

While in the previous sections we gave a broad survey of the history and relevance of various large $N$ limits, in this section we will focus on the comparison of three
"basic" large $N$ limits. Two of them, the vector and matrix large $N$ limits are widely known and have been studied for many years.
The third applies only to theories with tensor degrees of freedom of rank $3$ and higher. For such tensor theories with specially chosen interactions, it can be shown that
the diagrammatic expansion of the path integral is dominated by the so-called "melonic" graphs. 
The two-loop melon propagator correction (better known as the sunset graph) in $\phi^4$ theory is shown in the right part of figure \ref{svm}. 
The different large $N$ limits are characterized by the competition of this diagram with the one-loop snail diagram, shown in the left part of that figure.
  
 We call these three limits basic because their existence can be shown through combinatorial analysis alone and does not hinge on specific dimensionality
or additional symmetries of the theory. 
Therefore, instead of $d$-dimensional QFT we will first consider the $d=0$ examples, which are simply integrals. 
These examples also provide good practice for deriving the symmetry factors of various Feynman diagrams.

\subsection{$\phi^4$ models in $d=0$}

As a warm-up let us consider a one-field example, which is the $d=0$ $\phi^4$ theory. Here the partition function is simply an integral over one real variable:
\begin{equation}
Z(g)= \int_{-\infty}^\infty {d\phi\over \sqrt{2\pi}} e^{-\frac{\phi^2}{2} - g \frac{\phi^4}{24}}\,.
\label{partfun}
\end{equation}
This integral may be expanded in powers of $g$ using the integral
 \begin{equation}
I_n= \int_{-\infty}^\infty {d\phi\over \sqrt{2\pi}} (\phi^2/2)^n e^{-\alpha \phi^2/2}= (-1)^n \partial_\alpha^n \alpha^{-1/2}\ ,
\end{equation}
giving
\begin{equation}
Z(g)= 1- {g\over 8} + {35 g^2 \over 384}- {385 g^3 \over 3072}+ \mathcal{O}(g^4)\,.
\end{equation}
In fact, in this simple example the integral may be evaluated exactly:
\begin{equation}
Z(g)= \sqrt{3\over 2\pi g} e^{3\over 4g} K_{1/4}\left ({3\over 4g} \right ) \ ,
\label{exactBessel}
\end{equation}
where $K_{\alpha}(x)$ is the modified Bessel function.

The Feynman rules corresponding to (\ref{partfun}) assign the factor $1$ to the propagator and $-g$ to the quartic vertex. 
The ``free energy" $\log Z$ should be given by expansion in connected vacuum amplitudes. 
The only diagram appearing at order $g$ is the ``figure eight" graph which has symmetry factor $1/8$. 
This graph may also be thought of as a snail diagram with two legs connected.

\begin{figure}[h!]
                \centering
                \includegraphics[width=8cm]{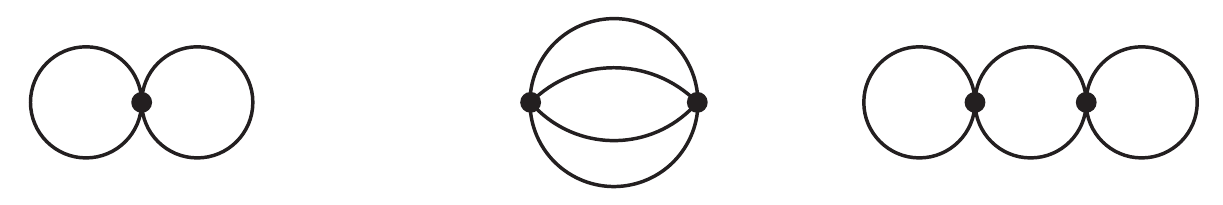}
\caption{The three vacuum diagrams up to order $g^2$: ``figure eight," ``melon," and ``triple bubble."}
                \label{Diags}
\end{figure}

At order $g^2$ there are two contributing connected graphs:
the melon and the triple bubble graph. In the melon the two vertices are connected by 4 lines, and we find the symmetry factor 
${1\over 2 \cdot 4!}=\frac{1}{48}$.  
The bubble graph may be thought of as two snail diagrams with their legs connected, and it has symmetry factor ${1\over 2^4}=\frac{1}{16}$. 
Adding up these connected vacuum graphs, we therefore find
 \begin{equation}
\log Z(g)= - {g\over 8} + {g^2 \over 48} + {g^2 \over 16} +\mathcal{O}(g^3)= - {g\over 8} + {g^2 \over 12} +\mathcal{O}(g^3)\ .
\end{equation}
This agrees with the expansion of the logarithm of the exact result (\ref{exactBessel}):
\begin{equation}
\log Z(g)= - {g\over 8} + {g^2 \over 12} -{11 g^3 \over 96} +{17 g^4 \over 72} -{619 g^5 \over 960} + {709 g^6 \over 324} -  
{858437 g^7 \over 96768} +
\mathcal{O}(g^8)\ .
\end{equation}
We note that the coefficients alternate in sign but their magnitude grows rapidly.
 From the properties of Bessel functions we know that
at high orders it grows as $n!$. Therefore, the small $g$ expansion is only asymptotic. Nevertheless, at small $g$ inclusion of the first few orders reproduces the exact
result with good precision.  

Now let us extend this discussion to the case of multiple real variables $\phi^i$ where $i=1,\ldots, n$. 
The partition function may in general be written as
\begin{equation}
Z=  \prod_{i=1}^n \int_{-\infty}^\infty {d\phi_i\over \sqrt{2\pi}} \exp \left (-\frac{1} {2} \phi^i \phi^i  - \frac{1} {24} C_{ijkl}\phi^i \phi^j \phi^k \phi^l \right )\ ,
\label{partfungen}
\end{equation}
where $C_{ijkl}$ is a fully symmetric tensor. Various particular models may be obtained by imposing special symmetries on this tensor.

Using the connected vacuum Feynman graphs with propagator 
\begin{equation}
\langle \phi^i \phi^j \rangle = \delta^{ij}\ ,
\end{equation}
 we obtain the following general expansion:
\begin{equation}
\log Z= - {C_{iijj}\over 8} + {C_{ijkl} C_{ijkl} \over 48} + {C_{iikl} C_{jjkl} \over 16} +\mathcal{O}(C^3) \ .
\label{genexp}
\end{equation}
The first term comes from the figure eight diagram; the second from the melon; and the third from the tripple bubble.

\subsection{Vector of $O(N)$} 

Let us set $n=N$ and impose the $O(N)$ symmetry on the model.  Then $\phi^i$, $i=1, 2, \ldots N$ transforms in the fundamental representation, and 
we take
\begin{equation}
C_{ijkl} = {g\over 3} \left ( \delta_{ij} \delta_{kl} + \delta_{ik} \delta_{jl} + \delta_{il} \delta_{jk} \right) \ ,
\end{equation}
which turns the model into
\begin{equation}
Z^{\rm vector} (g) = \prod_{i=1}^N \int_{-\infty}^\infty {d\phi_i\over \sqrt{2\pi}} e^{-\phi^i \phi^i /2 - g (\phi^i \phi^i)^2/24}\,.
\label{partfunvector}
\end{equation}

From (\ref{genexp}) we obtain the expansion
\begin{equation}
{\log Z^{\rm vector} (g)\over N}  = - {N+2\over 24} g +  {N+2\over 144}g^2 +  {(N+2)^2\over 144}g^2 + \mathcal{O}(g^3)\ ,
\label{partfunvecexp}
\end{equation}
where the second term comes from the melon graph, and the third from the bubble graph. 
To insure that ${\log Z^{\rm vector} (g)\over N}$ is finite in the large $N$ limit, we must keep $\lambda= gN$ fixed. Then the melon graph is suppressed by $1/N$ while the bubble
graphs, which originate from snail diagrams, survive. In fact, by drawing the index structure of the graphs it is not hard to see that the only surviving ones involve chains of bubbles. 
Thus, in vector models, the snails beat the melons. 

In the large $N$ limit the free energy has the structure
\begin{equation}
{\log Z^{\rm vector} (g)\over N}  = f_0 (\lambda) + N^{-1} f_1 (\lambda)+ \ldots\,, 
\label{partfungenstruct}
\end{equation}
where $f_0(\lambda)= -\lambda/24 + \lambda^2/144 + \mathcal{O}(\lambda^3)$. 
The function $f_0(\lambda)$ may be determined non-perturbatively using the standard method of introducing an auxiliary variable $\sigma$, so that  
\begin{equation}
Z^{\rm vector} (g) = \int_{-\infty}^\infty \prod_{j=1}^N {d\phi_j\over \sqrt{2\pi}} \int d\sigma \sqrt{\frac {6}{\pi g} }
\exp \left (- {6 N \sigma^2 \over \lambda} -  {\phi^k \phi^k (1+ 2i \sigma)\over 2}\right )\ .
\label{partfunvectoraux}
\end{equation}
After performing the Gaussian integral over $\phi^j$ we find
\begin{equation}
Z^{\rm vector} (g) = \sqrt{\frac {6}{\pi g} } \int d\sigma
\exp \left (- {6 N \sigma^2 \over \lambda} - {N\over 2} \log (1+ 2 i \sigma) \right ) \ .
\end{equation}
For large $N$ the integral is dominated by the saddle point located at $\sigma=- i \tilde \sigma$ where
\begin{equation}
{12\tilde \sigma\over \lambda}= {1\over 1+ 2 \tilde\sigma}\,.
\end{equation}
The solution of this quadratic equation which matches onto the perturbation theory is
 \begin{equation}
\tilde \sigma (\lambda) = {\sqrt{1+ {2\lambda \over 3}}-1 \over 4}\ ,
\end{equation}
and we find
\begin{equation}
f_0^{\rm vector} (\lambda) = {6 \tilde \sigma^2 \over \lambda} - {1\over 2} \log (1+ 2 \tilde \sigma)=  -{\lambda\over 24} + {\lambda^2\over 144} -{5\lambda^3\over 2592}
+{7\lambda^4\over   10368}-{7\lambda^5\over  25920}
 + \mathcal{O}(\lambda^6)\ ,
\end{equation}
which agrees with our Feynman graph calculations.
To all orders in $\lambda$,
\begin{equation}
f_0^{\rm vector} (\lambda) = \sum_{k=1}^{\infty} (-\lambda)^k \frac{1} {4k(k+1) 6^k } {2k \choose k} \,.
\end{equation}
In this series the coefficients decrease, so it is convergent for sufficiently small $|\lambda |$. This is one of the advantages of the large $N$ limit -- the functions
that appear order by order in $1/N$ have perturbation series with a finite radius of convergence.

Let us note a remarkable fact: $f_0 (\lambda)$ makes sense even for negative $\lambda$, so long as it is greater than $\lambda_c=-3/2$ (for $\lambda<\lambda_c$ it is ambiguous due
a branch cut). Thus, a large $N$ limit may be defined even for potentials that are not bounded from below.
The expansion for $\lambda> \lambda_c$ is
\begin{equation}
f_0^{\rm vector} (\lambda) = {2\log 2 -1 \over 4} - {1\over 6} (\lambda- \lambda_c) + \frac{2} {9} \sqrt{\frac{2}{3}} (\lambda- \lambda_c)^{3/2} 
 + \mathcal{O}\big ((\lambda-\lambda_c)^2 \big)\ .
\end{equation}
It is common to parametrize the leading singular term as $(\lambda- \lambda_c)^{2-\gamma}$, and $\gamma$ is called the susceptibility exponent. 
We find $\gamma_{\rm vector}=1/2$, which is characteristic
of the branched polymers \cite{Ambjorn:1985az}.

\subsection{$O(N)\times O(N)$ symmetric real matrix model} 

\begin{figure}[h!]
                \centering
                \includegraphics[width=3cm]{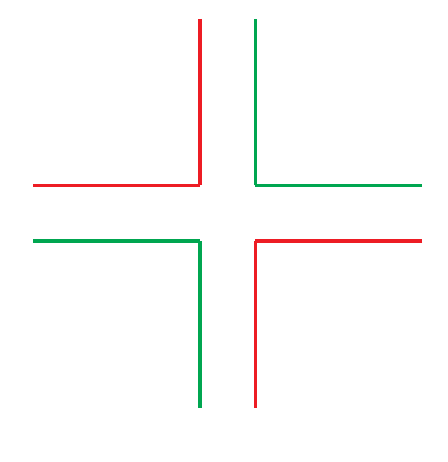}
                \caption{The resolved quartic vertex of the real matrix model (\ref{partfunmatrix}).}
                \label{FatVertex}
\end{figure}

Now let us consider $n=N^2$ real degrees of freedom $\phi^{ab}$, $a,b=1, \ldots, N$, and impose
$O(N)\times O(N)$ symmetry, so that $\phi^{ab}$ are in the bi-fundamental representation.
The two indices of the matrix are distinguishable, and each one is acted on by a different $O(N)$ group:
\begin{align}
&\phi^{ab} = M_{1}^{aa'} M_{2}^{bb'} \phi^{a'b'}, \quad  M_{1}\in O(N)_1,\quad M_{2}\in O(N)_2 \,.
\end{align} 
We will study the matrix integral
\begin{equation}
Z^{\rm matrix} (g) = \prod_{a,b} \int_{-\infty}^\infty {d\phi^{ab}\over \sqrt{2\pi}} 
\exp \left ( 
-\frac{1} {2} \phi^{ab} \phi^{ab}  - \frac {g} {24} \phi^{a_1 b_1} \phi^{a_1 b_2} \phi^{a_2 b_1} \phi^{a_2 b_2} \right )\,.
\label{partfunmatrix}
\end{equation}
The propagator
\begin{equation}
\langle \phi^{a_1 b_1}  \phi^{a_2 b_2} \rangle = \delta^{a_1 a_2} \delta^{b_1 b_2}\ ,
\end{equation}
may be represented by a double line consisting of a red and a green strand, while the interaction vertex is shown in
figure \ref{FatVertex}.

Using the matrix notation, we may write
\begin{equation}
\phi^{ab} \phi^{ab}= \tr (\phi \phi^T)\ , \qquad
\phi^{a_1 b_1} \phi^{a_1 b_2} \phi^{a_2 b_1} \phi^{a_2 b_2}= \tr (\phi \phi^T \phi \phi^T )\ .
\end{equation}
This demonstrates the invariance of the model under $\phi \rightarrow M_1 \phi M_2^T$.
There is one more $O(N)\times O(N)$ invariant quartic term: 
\begin{equation}
V_{dt}= \frac{g_{dt}}{24} \tr (\phi \phi^T) \tr(\phi \phi^T )
\end{equation}
To achieve a smooth large $N$ limit, the double-trace coupling has to be scaled as $g_{dt}\sim N^{-2}$, while the single-trace coupling as $g\sim N^{-1}$. 
In this scaling, models including double-trace couplings are tractable \cite{Das:1989fq,Klebanov:1994kv}, but we will not discuss them further. 

We can write the single-trace term as
\begin{equation}
\tr (\phi \phi^T \phi \phi^T )=
\phi^{a_1 b_1} \phi^{a_2 b_2} \phi^{a_3 b_3} \phi^{a_4 b_4}\delta^{a_1 a_2} \delta^{a_3 a_4} \delta^{b_1 b_3} \delta^{b_2 b_4}\ .
\end{equation}
Appropriately symmetrizing this product of Kronecker symbols we obtain the version of tensor $C^{ijkl}$ appropriate for this model. 

Using the connected vacuum Feynman graphs
we obtain the expansion
\begin{equation}
{\log Z^{\rm matrix} (g)\over N^2}  = - \frac{1}{24}  (2 N+1) g +  \frac{1}{288}  \left(N^2+2 N+3\right) g^2 + \frac{1}{144}  (2 N+1)^2 g^2  + \mathcal{O}(g^3)\ ,
\label{partfunmatrixexp}
\end{equation}
where the second term comes from the melon graph, and the third from the bubble graph. 
In order to keep ${\log Z^{\rm matrix} (g)\over N^2}$ finite, we keep $\lambda= gN$ fixed in the large $N$ limit. 
Now we see that both the melon and bubble graphs contribute at leading order. Thus, in matrix models, the competition between snails and melons results in a draw.

\begin{figure}[h!]
                \centering
                \includegraphics[width=10cm]{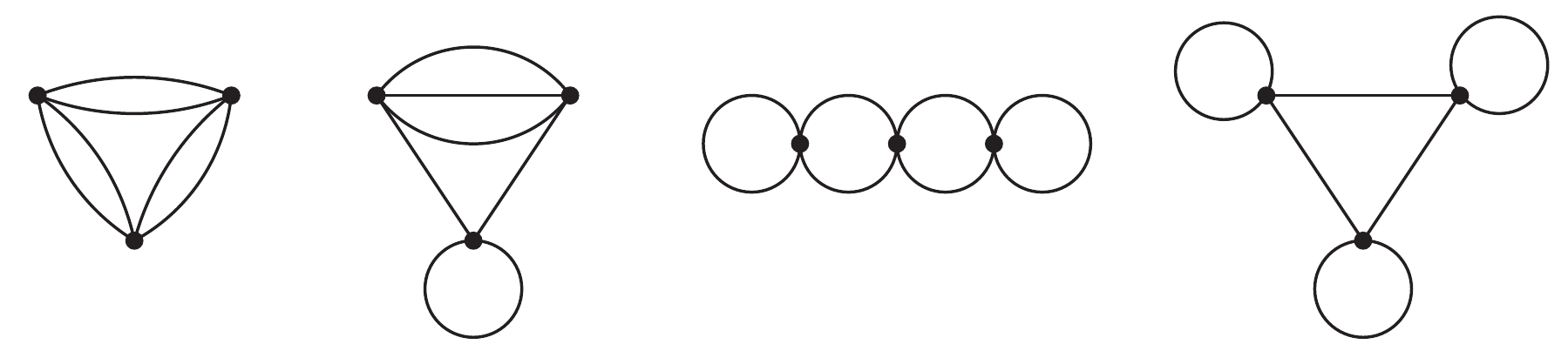}
\caption{Third order vacuum diagrams.}
                \label{Diagsg3}
\end{figure} 

Continuing to the diagrams of order $g^{3}$, we find  
\begin{align}
\left.{\log Z^{\rm matrix} (g)\over N^2}\right|_{g^{3} \textrm{order}}   =&-\frac{1}{48}\cdot \frac{1}{27}  \left(N^3+4 N^2+13 N+9\right)g^{3} -\frac{1}{24}\cdot\frac{1}{18} (2 N+1) \left(N^2+2 N+3\right)g^{3} \notag\\
&-\frac{1}{32}\cdot\frac{1}{27}  (2 N+1)^3 g^{3} - \frac{1}{48}\cdot \frac{1}{27} (2N+1)^3 g^{3}\, ,
\end{align}
where the four terms correspond to the four diagrams shown in Figure \ref{Diagsg3}, respectively. All of them contribute in the large $N$ limit.
In fact, 't Hooft proved \cite{'tHooft:1973jz} that all the planar graphs, i.e. the graphs of spherical topology, are dominant.
To demonstrate this, it is convenient to rescale the matrix $\phi^{ab}\rightarrow \sqrt N \phi^{ab}$. Then
\begin{equation}
Z^{\rm matrix} \sim \prod_{a,b} \int_{-\infty}^\infty {d\phi^{ab}\over \sqrt{2\pi}} 
\exp \left ( 
-\frac{N} {2} \phi^{ab} \phi^{ab}  - \frac {N\lambda} {24} \phi^{a_1 b_1} \phi^{a_1 b_2} \phi^{a_2 b_1} \phi^{a_2 b_2} \right )\,.
\label{partfunmatrixproof}
\end{equation}
Now each propagator carries a factor $1/N$, and each vertex factor $N\lambda$. Also, each face of the graph contains an index loop and 
contributes a factor of $N$. So, the net power of $N$ for a Feynman graph with $V$ vertices, $F$ faces and $E$ edges 
is $N^{V+F-E}= N^{\chi}$, where $\chi$ is the Euler characteristic.
Since $\chi=2-2 g$, where $g$ is the genus of the graph, we see that the graphs contributing at order $N^2$ are the graphs of genus $0$, i.e. of spherical topology.

In the large $N$ limit the free energy (\ref{partfunmatrixexp}) has the structure
\begin{equation}
{\log Z^{\rm matrix} (g)\over N^2}  = f_0^{\rm matrix} (\lambda) + 
N^{-1} f_{1/2}^{\rm matrix} (\lambda)+
N^{-2} f_1^{\rm matrix} (\lambda)+ \ldots \ ,
\label{partfunmatstruct}
\end{equation}
where $f_g(\lambda)$ is the sum over graphs of genus $g$. We see that the leading correction is due to the non-orientable surfaces of genus $1/2$, which is $RP_2$.
For the Hermitian matrix model, discussed in section \ref{Hermitmat}, such non-orientable surfaces of odd Euler characteristic do not appear.
Our perturbative calculation gives 
$f_0^{\rm matrix}(\lambda)= -  \lambda/12+  \lambda^2/32 -\lambda^{3}/48+ \mathcal{O}(\lambda^4)$.

Let us obtain the exact expression for $f_0^{\rm matrix}(\lambda)$, which is analogous to the one we obtained for the vector model. 
To do this we represent the real $N\times N$ matrix using the singular value decomposition:
\be
\phi= L \kappa R^T\ ,
\ee 
where $\kappa$ is a diagonal matrix of real non-negative singular values $\kappa_a$, and $R$ and $L$ are two independent $O(N)$ matrices. Integrating them out, we find
 \begin{equation}
Z^{\rm matrix} (g) \sim  \prod_{a} \int_{0}^\infty d\kappa_a |\Delta (\kappa^2)| 
e^{ -N \sum_{b=1}^N  \left ( 
\frac{1} {2}\kappa_b^2    + \frac {\lambda} {24} \kappa_b^4 \right )}
\ ,
\end{equation}
where $\Delta (\kappa^2)= \prod_{a<b} (\kappa_a^2- \kappa_b^2)$ is the Vandermonde determinant. 
A way to understand this form of the Jacobian is to note that $\kappa_a^2$ are the eigenvalues of the real symmetric matrix $\phi^T \phi$. 
Introducing the singular value density $\rho(\kappa)$, 
\be
\int_{0}^\infty d\kappa \rho(\kappa)=1\ ,
\ee
we see that in the large $N$ limit it is governed by the effective potential 
\be
V_{eff} = \int_0^\infty d\kappa \rho(\kappa) \left  ( \frac{1} {2}\kappa^2   + \frac {\lambda} {24} \kappa^4 \right )-
\frac{1} {2} \int_0^\infty d\kappa d\kappa' \rho(\kappa) \rho(\kappa') \log |\kappa^2- (\kappa')^2|\ .
\ee

\begin{figure}[h!]
                \centering
                \includegraphics[width=10cm]{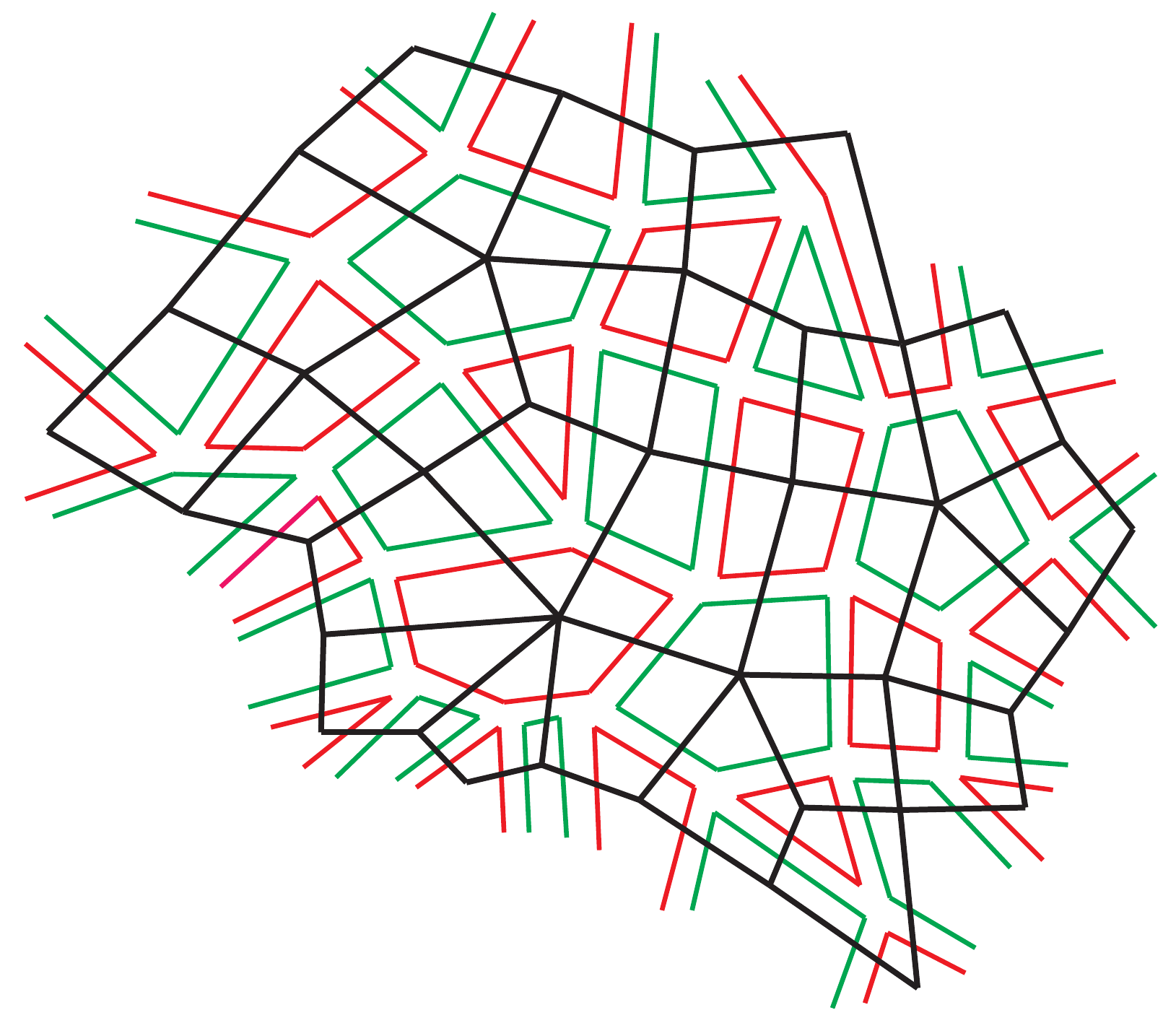}
\caption{A secton of a resolved Feynman graph for the real matrix model, which has alternating red and green loops. Its dual lattice is made of randomly connected squares.}
                \label{RanLat}
\end{figure}

Now it is convenient \cite{Klebanov:2003wg, Dalley:1991qg} 
to introduce the symmetric function $\tilde\rho (\kappa) = (\rho(\kappa)+\rho(-\kappa))/2$, which is defined on the entire real axis, so that
\be
V_{eff} = \int_{-\infty}^\infty d\kappa \tilde \rho(\kappa) \left  ( \frac{1} {2}\kappa^2   + \frac {\lambda} {24} \kappa^4 \right )- 
\int_{-\infty}^\infty d\kappa d\kappa' \tilde \rho(\kappa) \tilde \rho(\kappa') \log |\kappa- \kappa'|\ .
\ee
The singular integral equation which follows from this was solved in \cite{Brezin:1977sv}: 
 \begin{align}
& \tilde \rho (\kappa) = \frac {1} {\pi} \left ( \frac {1} {2} +  \frac {\lambda} {6} a^2  + \frac {\lambda} {12} \kappa^2 \right )\sqrt{4 a^2 - \kappa^2}\ , \notag \\
& a^2(\lambda) = {\sqrt{1+ 2\lambda }-1 \over \lambda}\,.
\label{density}
\end{align}
We see that $\tilde \rho(\kappa)$ is a symmetric function with support between $-2a(\lambda)$ and $2a(\lambda)$.
We finally have
\begin{align}
&\rho(\kappa)=2 \tilde \rho (\kappa)\ , \qquad \kappa>0\ , \\ 
&\rho(\kappa)=0\ , \qquad\qquad \kappa<0\ . \notag 
\end{align}
As $\lambda\rightarrow 0$, $a\rightarrow 1$, and it approaches the classic Wigner semicircle law found for the Gaussian matrix models.
The non-Gaussian effects deform the density to the more general function (\ref{density}). Substituing (\ref{density}) into $V_{eff}$ we find
\begin{equation}
f_0^{\rm matrix} (\lambda) = \frac {1} {24}  (a^2(\lambda)-1) (9- a^2(\lambda))- \frac {1} {2} \log a^2(\lambda)=  
 -{\lambda\over 12} + {\lambda^2\over 32} -{\lambda^3\over 48}
+{7\lambda^4\over   384}
 + \mathcal{O}(\lambda^5)\,.
\end{equation}
Similarly to the free energy in the vector case, $f_0^{\rm matrix} (\lambda)$ is well defined for $\lambda> \lambda_c$ where $\lambda_c=-1/2$. 
Expanding $f_0^{\rm matrix} (\lambda)$ near $\lambda_c$ we find that the leading singular term is now $\sim (\lambda- \lambda_c)^{5/2}$ corresponding to 
$\gamma_{\rm matrix}=-1/2$.
This is the well-known susceptibility exponent of the pure two-dimensional quantum gravity \cite{Gross:1989vs,Brezin:1990rb,Douglas:1989ve}. 
In the limit $\lambda\rightarrow \lambda_c$ the discretized random square lattices, which are the dual lattices to the Feynman graphs for 
the matrix integral (\ref{partfunmatrixproof}), become large (a section of such a lattice is shown in figure \ref{RanLat}).
Therefore, in this limit it is possible to define the  continuum limit of two-dimensional quantum gravity. 

\subsection{$SU(N)$ symmetric Hermitian matrix model} 
\label{Hermitmat}

Now let us consider a somewhat different matrix integral. It involves a Hermitian matrix $\Phi^i_j$, $i,j=1, \ldots, N$, and we impose
$SU(N)$ symmetry. $\Phi$ is in the adjoint representation, i.e. it transforms as
\begin{align}
\Phi = U \Phi' U^\dagger \, ,
\end{align}   
where $U\in SU(N)$. An interesting integral to consider is
\begin{equation}
Z^{\rm Hermitian} (g) = \prod_{i,j} \int_{-\infty}^\infty {d {\rm Re} \Phi^i_j\over \sqrt{2\pi}} {d {\rm Im} \Phi^i_j\over \sqrt{2\pi}} 
\exp \tr \left ( 
-\frac{1} {2} \Phi^2  - \frac {g_3} {6} \Phi^3 \right )\,.
\label{partfunmatrixHermit}
\end{equation}
This may be viewed as a toy model for interactions of gluons, and the large $N$ limit is taken keeping $\lambda= g_3^2 N$ fixed.

\begin{figure}[h!]
                \centering
                \includegraphics[width=2cm]{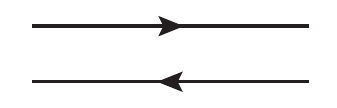}
\caption{Hermitian matrix propagator in the double line representation.}
                \label{matrixprop}
\end{figure} 

\begin{figure}[h!]
                \centering
                \includegraphics[width=6cm]{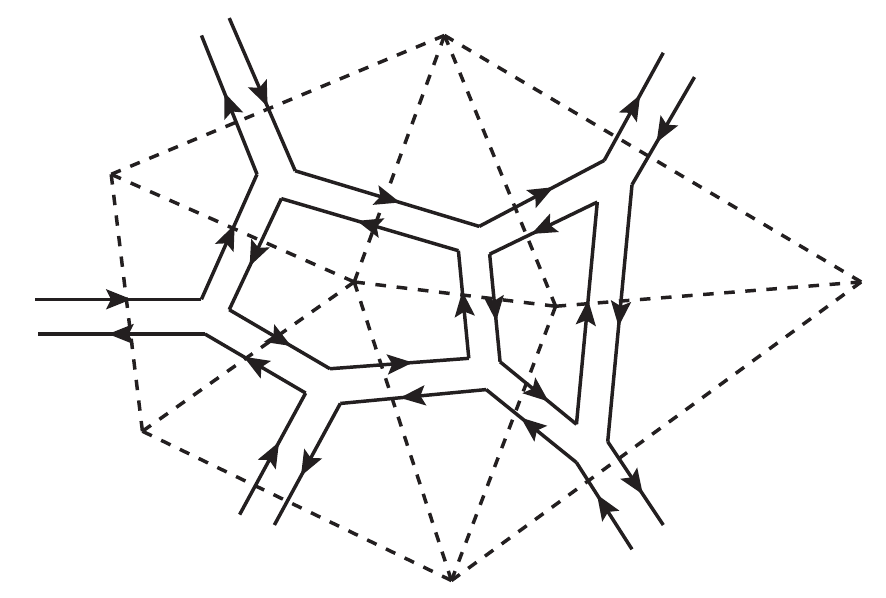}
\caption{A section of a planar diagram in the Hermitian matrix model, which represents an orientable triangulated random surface. }
                \label{matrixdiagram}
\end{figure}

The propagator 
\begin{equation}
\langle \Phi^{i_1}_{j_1}  \Phi^{i_2}_{j_2}\rangle = \delta^{i_1}_{ j_2} \delta^{i_2}_{j_1}\ ,
\end{equation}
may be represented using double lines with opposite directions (see Figure \ref{matrixprop}).
The graphs dual to the Feynman graphs are now made of triangles, so that this integral may be interpreted in terms of orientable triangulated surfaces (see Figure \ref{matrixdiagram}).
If we also impose the condition that $\Phi$ is traceless, so that it is truly in the adjoint representation of $SU(N)$, then the propagator becomes 
\begin{equation}
\langle \Phi^{i_1}_{j_1}  \Phi^{i_2}_{j_2}\rangle = \delta^{i_1}_{ j_2} \delta^{i_2}_{j_1}- \frac {1} {N} \delta^{i_1}_{ j_1} \delta^{i_2}_{j_2}\ .
\end{equation}
The tracelessness condition removes some of the tadpole graphs.

We can decompose a Hermitian matrix as
$\Phi= V \kappa V^\dagger$, where $\kappa$ is a diagonal matrix of real eigenvalues $\kappa_a$, which add up to zero if $\Phi$ is traceless,
and $V$ is an $SU(N)$ matrix. 
Integrating over $V$ gives \cite{Brezin:1977sv}
 \begin{equation}
Z^{\rm Hermitian} (g) \sim  \prod_{a} \int_{-\infty}^\infty d\kappa_a \Delta^2 (\kappa) 
\exp \left ( -\sum_{b=1}^N  \left ( 
\frac{1} {2}\kappa_b^2    + \frac {g_3} {6} \kappa_b^3 \right ) \right )
\,.
\end{equation}
Now, the critical behavior is $f_0(\lambda)\sim (\lambda_c - \lambda)^{5/2}$ \cite{Brezin:1977sv}, which is again characterized by the susceptibility exponent
$\gamma_{\rm matrix}=-1/2$.

\subsection{$O(N)^3$ symmetric real tensor model}

\begin{figure}[h!]
                \centering
                \includegraphics[width=3.5cm]{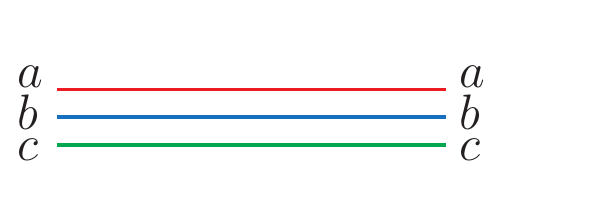}
                \caption{A resolved representation of the propagator (\ref{tensorprop}).}
                \label{Propagator}
\end{figure}

\begin{figure}[h!]
                \centering
                \includegraphics[width=13cm]{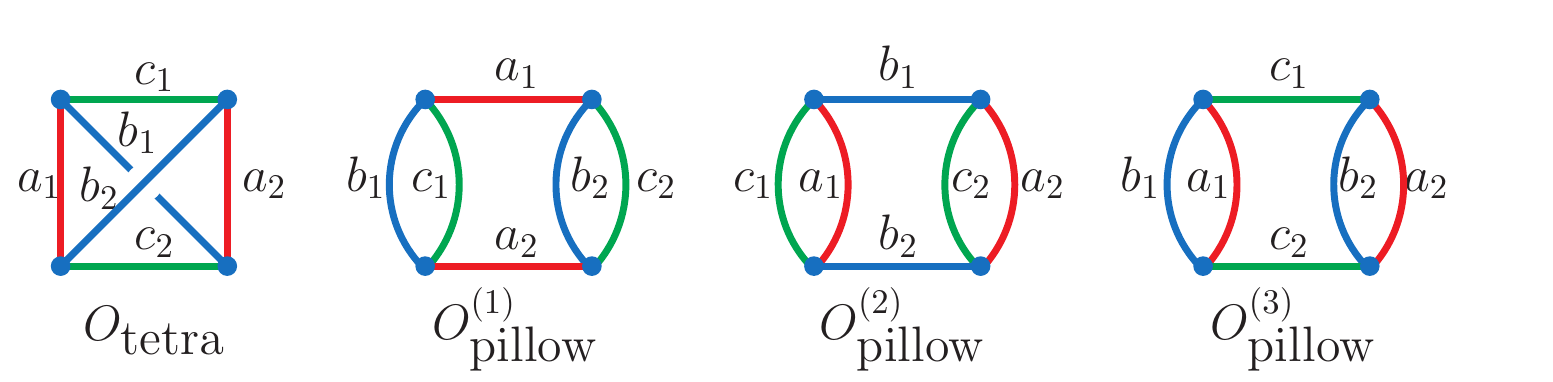}
                \caption{All the single-sum four-particle operators, the tetrahedron and the three pillows, with the index contractions shown explicitly.}
                \label{O4ops}
\end{figure}

\begin{figure}[h!]
                \centering
                \includegraphics[width=10cm]{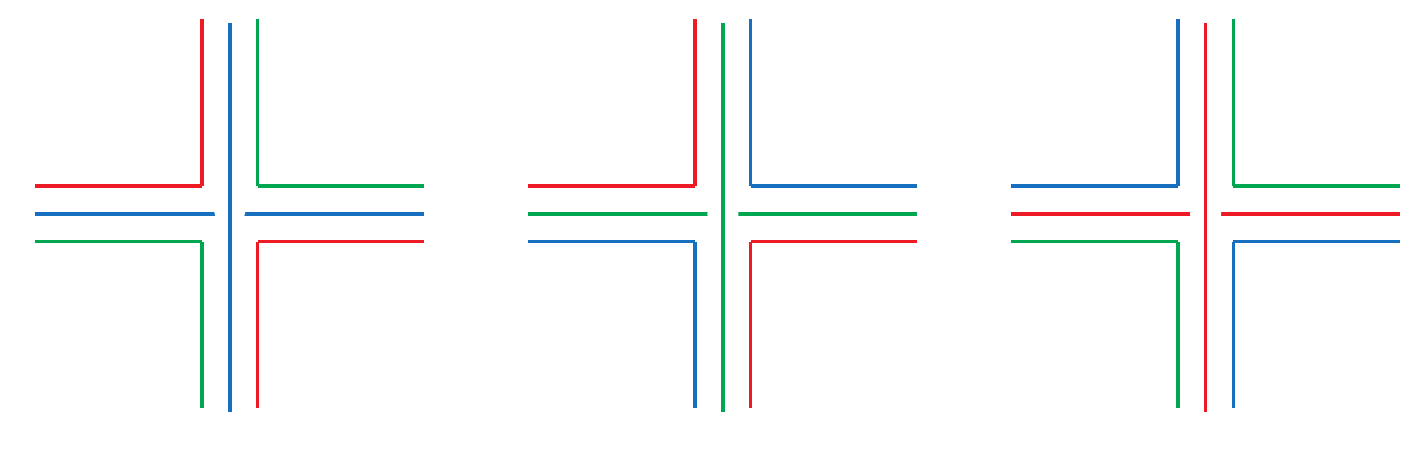}
                \caption{Three equivalent ways to represent the resolved tetrahedral vertex.}
                \label{vertex}
\end{figure}

Now let us consider $N^3$ real degrees of freedom $\phi^{abc}$, $a,b,c=1, \ldots, N$, and impose
$O(N)^3$ symmetry, so that $\phi^{abc}$ are in the tri-fundamental representation.\footnote{A natural generalization is to 
have $a=1, \ldots, N_1$,  $b=1, \ldots, N_2$, $c=1, \ldots, N_3$, leading to a tensor model with $O(N_1)\times O(N_2)\times O(N_3)$ symmetry. 
Then $\phi^{abc}$ may be thought of as a collection of $N_2$ matrices, each one $N_1\times N_3$. 
A limit where $N_2$ is taken to infinity first was studied in \cite{Ferrari:2017ryl, Ferrari:2017jgw}.}
The 3 indices of a tensor are distinguishable, and each one is acted on by a different $O(N)$ group:
\begin{align}
&\phi^{abc} = M_{1}^{aa'} M_{2}^{bb'}M_{3}^{cc'}\phi^{a'b'c'},  \nonumber \\
& M_{1}\in O(N)_1,\quad M_{2}\in O(N)_2, \quad M_{3} \in O(N)_3\,.
\end{align} 
The most general $O(N)^3$ invariant quartic potential has the form
\begin{align}
V_4 &= \frac {g} {24} \phi^{a_1 b_1 c_1} \phi^{a_1 b_2 c_2} \phi^{a_2 b_1 c_2} \phi^{a_2 b_2 c_1}+ \notag \\
 & \frac{g_{p1}}{24} \phi^{a_{1}b_{1}c_{1}}\phi^{a_{2}b_{1}c_{1}}\phi^{a_{2}b_{2}c_{2}}\phi^{a_{1}b_{2}c_{2}} +
\frac{g_{p2}}{24} \phi^{a_{1}b_{1}c_{1}}\phi^{a_{1}b_{2}c_{1}}\phi^{a_{2}b_{1}c_{2}}\phi^{a_{2}b_{2}c_{2}}+
\frac{g_{p3}}{24} \phi^{a_{1}b_{1}c_{1}}\phi^{a_{1}b_{1}c_{2}}\phi^{a_{2}b_{2}c_{2}}\phi^{a_{2}b_{2}c_{1}} \notag \\
&+ \frac{g_{ds}}{24} \left (\phi^{a_{1}b_{1}c_{1}}\phi^{a_{1}b_{1}c_{1}}\right )^2\ .
\end{align}
The first is the ``tetrahedral" quartic term \cite{Carrozza:2015adg,Klebanov:2016xxf}, 
which is the leftmost diagram in figure \ref{O4ops}; the next three are the so-called pillow terms which are the remaining three diagrams in the figure.
The final term is the double-sum term. We will be interested in the large $N$ limit where the tetrahedral coupling is dominant and scales as $g\sim N^{-3/2}$, 
while the remaining couplings scale to zero faster: $g_{p}\sim N^{-2}$,  and $g_{ds}\sim N^{-3}$ \cite{Carrozza:2015adg,Giombi:2017dtl}. 
Therefore, we will include $g$ only and study the integral
\begin{equation}
Z^{\rm tensor} (g) = \prod_{a,b,c} \int_{-\infty}^\infty {d\phi^{abc}\over \sqrt{2\pi}} 
\exp \left ( 
-\frac{1} {2} \phi^{abc} \phi^{abc}  - \frac {g} {24} \phi^{a_1 b_1 c_1} \phi^{a_1 b_2 c_2} \phi^{a_2 b_1 c_2} \phi^{a_2 b_2 c_1} \right )\,.
\label{partfuntensor}
\end{equation} 
Even though this quartic term is not bounded from below for $N>2$, it is possible to develop formal perturbative expansion in $g$ using the propagator
\begin{equation}
\langle \phi^{a_1 b_1 c_1}  \phi^{a_2 b_2 c_2} \rangle = \delta^{a_1 a_2 } \delta^{b_1 b_2} \delta^{c_1 c_2}\,.
\label{tensorprop}
\end{equation}
For the purposes of counting the powers of $N$, we can draw the resolved (or ``stranded") graphs where the strands are of three different colors, corresponding to the indices
transforming under the three different $O(N)$ groups. The propagator is 
shown in figure \ref{Propagator}, and the tetrahedral vertex in figure \ref{vertex}.
The proof of melon dominance following \cite{Carrozza:2015adg,Klebanov:2016xxf}
will be reviewed in the following section, but
first let us study the low orders in perturbation theory, as we did  for the vector and matrix models.

\begin{figure}[h!]%
    \centering
    \subfloat{{\includegraphics[width=5.5cm]{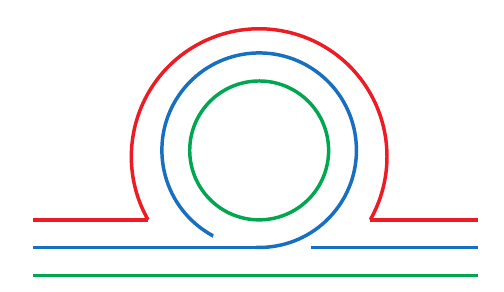} }}%
    \qquad
    \subfloat{{\includegraphics[width=6cm]{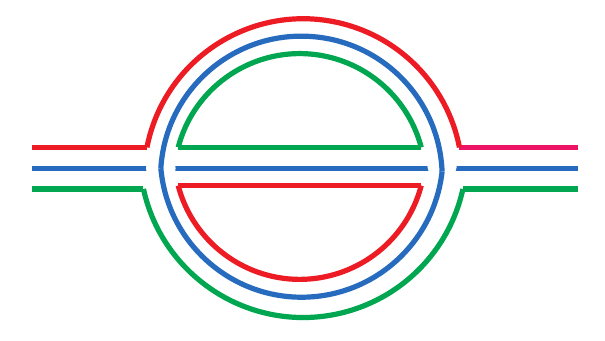} }}%
    \caption{a) The snail propagator correction has one index loops and scales as $gN\sim \frac{\lambda}{\sqrt N}$,\\ b) The melon propagator correction has
three index loops and scales as $g^2 N^3\sim \lambda^2$. }
    \label{snailandmelonprop}%
\end{figure}

Using the connected vacuum Feynman graphs 
we obtain the expansion
\begin{equation}
{\log Z^{\rm tensor} (g)\over N^3}  = -\frac{1}{8}  N g +  \frac{1}{288}  \left(N^3+3 N+2\right) g^2 + \frac{1}{16} N^{2} g^2  + \mathcal{O}(g^3)\ ,
\label{partfuntensorexp}
\end{equation}
where the first term comes from the figure eight, the second from the melon, and the third from the triple bubble graph. 
In order to keep ${\log Z^{\rm tensor} (g)\over N^3}$ finite, we keep $\lambda= gN^{3/2}$ fixed in the large $N$ limit. 
Now we see that the melon contributes while the figure eight and triple bubble graphs are suppressed. So, finally the melons are winning!
Another way to see this is by comparing the melon and snail propagator corrections, whose resolved form is shown in figure \ref{snailandmelonprop}.
The melon diagram has
three index loops and scales as $g^2 N^3\sim \lambda^2$, while the snail diagram has one index loop and scales as $gN\sim \frac{\lambda}{\sqrt N}$. 

\begin{figure}[h!]
                \centering
                \includegraphics[width=10cm]{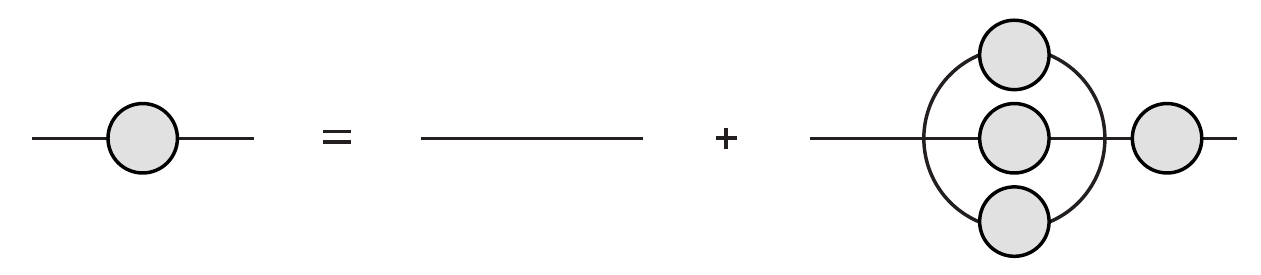}
\caption{Schwinger-Dyson equation for the two-point function.}
                \label{SDeq}
\end{figure}

Because of the melon dominance, in the large $N$ limit the free energy behaves as
\begin{equation}
{\log Z^{\rm tensor} (g)\over N^3}  = f_0^{\rm tensor} (\lambda) + O(N^{-1/2})\,,
\label{partfuntensorstruct}
\end{equation}
where $f_0^{\rm tensor} (\lambda)$ sums the contributions of melonic vacuum diagrams only (see figure \ref{MelonsEx}).
To solve for $f_0^{\rm tensor} (\lambda)$ one can use the Schwinger-Dyson equation for the full two-point function 
$G (\lambda)$ implied by the diagram for self-energy
 \cite{Bonzom:2011zz}
\begin{equation}
G^{-1}
(\lambda) = 1+ \Sigma(\lambda)\ , \qquad \Sigma(\lambda)=- \frac {\lambda^{2}}{36} G_{\textrm{melons}}(\lambda)^{3}\,.
\end{equation}
This may be written as (see figure \ref{SDeq})
\begin{equation}
G(\lambda) = 1+ \frac {\lambda^{2}}{36} G (\lambda)^{4}\,.
\label{SDeqn}
\end{equation}

Then free energy is obtained from the two-point function $G$ through the relation 
\begin{equation}
G(\lambda) = 1+4 \lambda   \partial_{\lambda} f_0^{\rm tensor} (\lambda)\ ,
\label{userel}
\end{equation}
which follows from the equation
\begin{equation}
\frac{1}{Z^{\rm tensor} (g)}\prod_{a,b,c} \int_{-\infty}^\infty {d\phi^{abc}\over \sqrt{2\pi}} \frac {\partial} {\partial \phi^{a' b' c'}} \bigg (\phi^{a' b' c'}  
\exp \left ( 
-\frac{1} {2} \phi^{abc} \phi^{abc}  - \frac {g} {24} \phi^{a_1 b_1 c_1} \phi^{a_1 b_2 c_2} \phi^{a_2 b_1 c_2} \phi^{a_2 b_2 c_1} \right ) \bigg )=0\ .
\end{equation}
Applying the derivative gives
\begin{equation}
N^3 - N^3 G + 
\frac{4g}{Z^{\rm tensor} (g)} \frac {\partial} {\partial g} Z^{\rm tensor} (g)=0\,,
\end{equation}
which is equivalent to (\ref{userel}).

The solution of (\ref{SDeqn}) may be written as
\begin{align}
& G(\lambda) = \frac{\sqrt 3}{(2\lambda)^{1/2} v^{1/4}} \left ( (1+ 4 v)^{1/4} - (2- (1+ 4 v)^{1/2})^{1/2} \right )
   \ , \nonumber \\
& v(\lambda) =\frac {(\lambda/3)^{2/3}}{2}
\bigg [ \left ( 1+ \sqrt {1- \frac {2^6 \lambda^2}{3^5} }\right )^{1/3} + 
\left ( 1- \sqrt {1- \frac {2^6 \lambda^2}{3^5} }\right )^{1/3}\bigg ] \,.
\end{align} 
The explicit series is \cite{Bonzom:2011zz,Klebanov:2017nlk}
 \begin{equation}
f_0^{\rm tensor} (\lambda) = \sum_{n=1}^\infty a_{2n} \left (\frac {\lambda}{6}\right )^{2n} \ ,
\label{vacgraphs3} 
\end{equation}
where 
\begin{equation}
a_2=\frac {1} {8} \ , \quad a_4= \frac {1} {4} \ , \quad a_6= \frac {11} {12} \ , \quad a_8= \frac {35} {8},\quad \ldots ,\quad a_{2n}= \frac{1}{8n(4n+1)}{4n+1 \choose n}\,.
\end{equation}
From the exact large $N$ solution we find that the leading singular behavior of $f_0^{\rm tensor} (\lambda)$ as $\lambda$ 
approaches a critical value is $(\lambda_c^2- \lambda^2)^{3/2}$, where $\lambda^{2}_{c} =3^{5}/2^{6} $, and the susceptibility exponent is
$\gamma_{\rm tensor}=1/2$, just as in the vector model. Therefore, the theory is again in the branched polymer phase \cite{Ambjorn:1985az}. 

\section{Melonic Dominance}
\label{Theproof}

In this section we demonstrate the melonic dominance in theories with
$O(N)^3$ symmetry, both in the fermionic and bosonic cases, and for any $d$.
The presentation 
follows that in \cite{Klebanov:2016xxf}, and the arguments are analogous to those in \cite{Carrozza:2015adg}.
We will ignore the coordinate dependence of fields and just focus on the index structure.

The propagator (\ref{tensorprop}) has the index structure depicted in figure \ref{Propagator}. 
The three colored strands (or wires) represent propagation of the 
three indices of the $\phi^{abc}$ field. 
The tetrahedral vertex has the index structure depicted in the figure \ref{vertex}. 
There are three equivalent ways to draw the vertex; for concreteness we will use the first way. "Forgetting" the middle lines we obtain the standard matrix model vertex as in figure \ref{FatVertex}.

 \begin{figure}[h!]
                \centering
                \includegraphics[width=14cm]{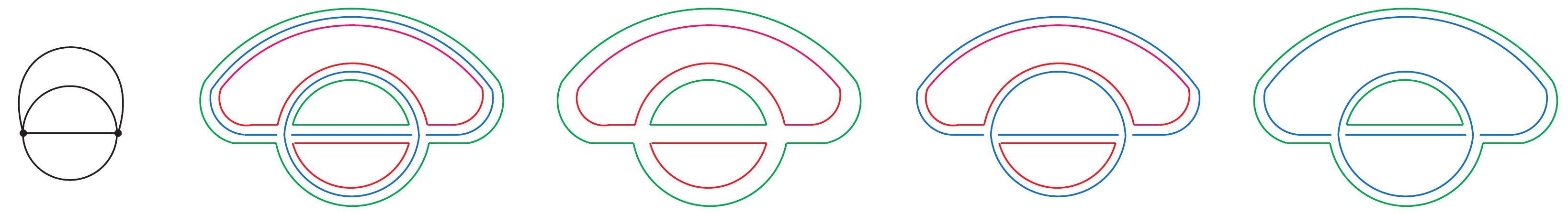}
                \caption{A melonic second-order diagram and all its double-line subgraphs.}
                \label{ExampleDg2p3}
\end{figure} 

 \begin{figure}[h!]
                \centering
                \includegraphics[width=14cm]{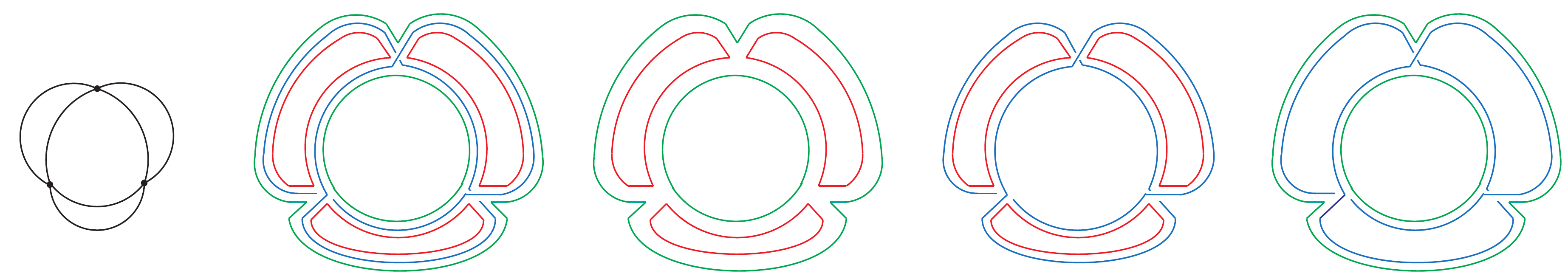}
                \caption{A non-melonic third-order diagram and all its double-line subgraphs.}
                \label{ExampleDg3p3}
\end{figure}

Let us consider the vacuum Feynman diagrams. Examples of melonic and non-melonic diagrams with their resolved representations and double-line subgraphs  
are depicted in figures \ref{ExampleDg2p3} and  \ref{ExampleDg3p3}. 
Each resolved Feynman diagram consists of  loops of three different colors and is proportional to $N^{f_{\rm total}}$, where $f_{\rm total}$ is the total number of index loops. 
Suppose we ``forget'' all the loops of some particular color 
in our diagram \cite{Witten:2016iux}, as in figures \ref{ExampleDg2p3} and  \ref{ExampleDg3p3}. Then what remains is a double-line (or ribbon) graph of the kind one finds in
matrix models. One can count the number of all index loops $f$ in this graph using the Euler characteristic $\chi$
\begin{align}
f= \chi +e-v\,, \label{Euler}
\end{align}
where $e$ is the number of edges and $v$ is the number of vertices. In our theory we obviously have $e= 2v$, therefore $f= \chi+v$. 
We can forget red, blue or green loops, and in each case we get a double-line graph made of the remaining two colors. If we forget, say, all red wires, then using the formula
(\ref{Euler}) we find $f_{bg}=\chi_{bg}+v$, where $f_{bg}=f_{b}+f_{g}$ is the number of blue and green loops and $\chi_{bg}$ is the Euler characteristic of this blue-green graph. Analogously we get $f_{rg}=\chi_{rg}+v$ and $f_{br}=\chi_{br}+v$. Adding up these formulas we find 
\begin{align}
f_{bg}+f_{rg}+f_{br} = 2(f_{b}+f_{g}+f_{r})=\chi_{bg}+\chi_{br}+\chi_{rg}+3v\,.
\end{align}
Thus, the total number of loops is 
\begin{align}
f_{\rm total} = f_{b}+f_{g}+f_{r}= \frac{3v}{2}+3 -g_{bg}-g_{br}-g_{rg}\,, \label{fandgen}
\end{align}
where $g=1-\chi/2$ is the genus of a graph. Because $g\geqslant 0$ we obtain 
\begin{align} 
f_{\rm total} \leqslant 3+\frac{3v}{2}\,.
\end{align}
This provides a simple proof that the maximal scaling of a vacuum graph with $v$ vertices is $\sim N^3 \lambda^v$. 
Now  the goal is to show that the equality $f_{\rm total} = 3+3v/2$ is satisfied  only for the melonic diagrams. We will call the graphs which 
satisfy $f_{\rm total}= 3+3v/2$ the maximal graphs. Thus we should argue that maximal graphs are necessarily melonic. 
We note that,  due to (\ref{fandgen}), each double-line subgraph of a maximal graph has genus zero.

Now let us classify all loops in our graph according to how many vertices they pass through (a loop can pass
the same vertex twice). Let us denote by $\mathcal{F}_{s}\geqslant 0$ the number of loops, which pass through $s$ vertices. 
For a maximal graph
\begin{align}
f_{\rm total}=\mathcal{F}_{2}+\mathcal{F}_{3}+\mathcal{F}_{4}+\mathcal{F}_{5}+\ldots = 3+ \frac{3v}{2}\ ,
\label{sumrule}
\end{align}
where we set $\mathcal{F}_{1}=0$. 
Indeed, a snail insertion into a propagator, which is the only way of obtaining an index loop of length $1$, is suppressed by a factor of $\sqrt{N}$ 
(see figure \ref{snailandmelonprop}).
Now, since each vertex must be passed $6$ times, we also get 
\begin{align} 
2\mathcal{F}_{2} +3\mathcal{F}_{3}+4\mathcal{F}_{4}+5\mathcal{F}_{5}+\dots = 6v\,.
\end{align}
Combining this with (\ref{sumrule}), we find 
\begin{align} 
2\mathcal{F}_{2} +\mathcal{F}_{3} = 12+\mathcal{F}_{5}+2\mathcal{F}_{6}+\dots\,. \label{maineqforF}
\end{align}
Now our goal  is to show that $\mathcal{F}_{2} >0$ using this formula (in fact, $\mathcal{F}_{2} \geqslant 6$, but all we will need is that it is non-vanishing).

\begin{figure}[h!]%
    \centering
    \subfloat{{\includegraphics[width=3.5cm]{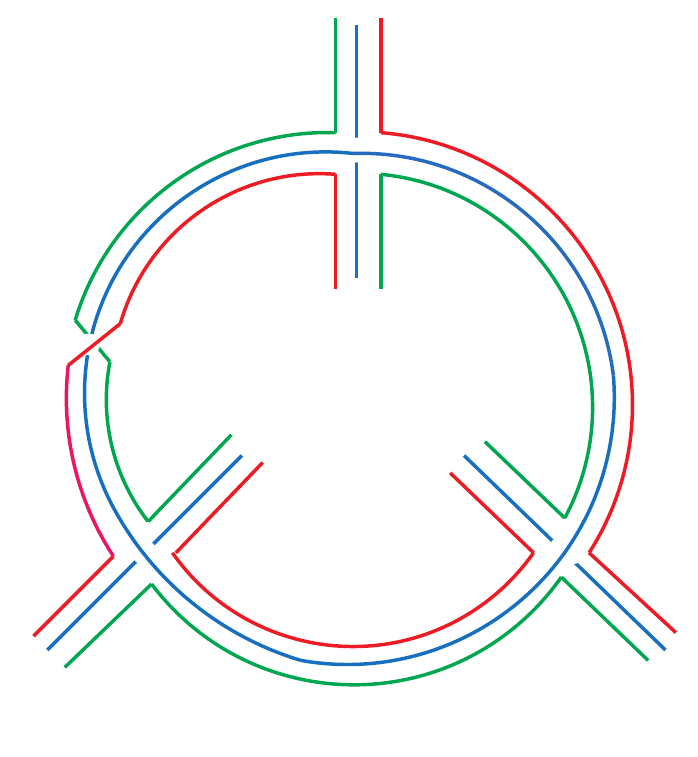} }}%
    \qquad
    \subfloat{{\includegraphics[width=3.5cm]{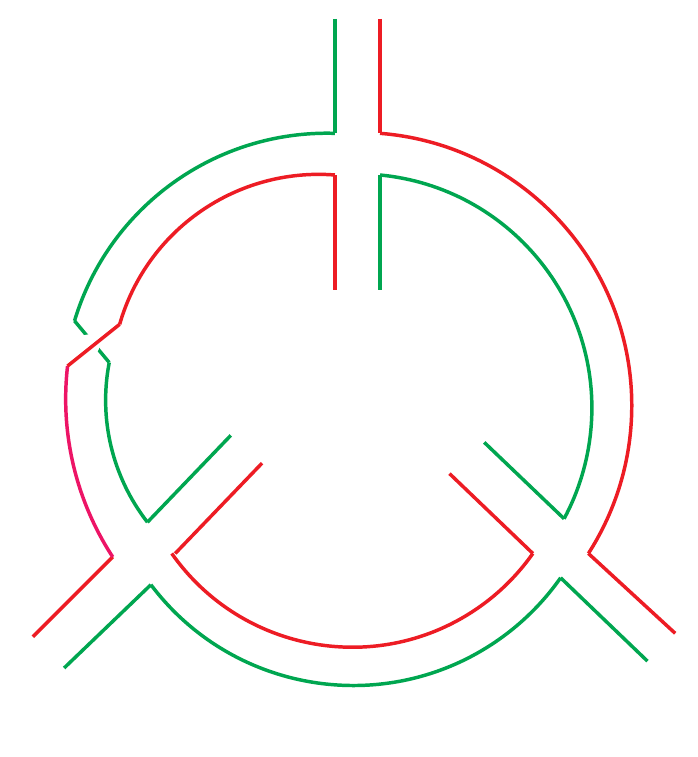} }}%
    \caption{a) Local part of a graph with a middle index loop passing  through 3 vertices. b) The same figure where the middle index has been ``forgotten." }
    \label{Twsited3vertLoop}%
\end{figure}

Let us first argue that a maximal graph must have $\mathcal{F}_{3}=0$.
To have $\mathcal{F}_{3}>0$ we need a closed index loop passing through 3 vertices. Without a loss of generality we can assume that this loop is formed by the middle lines in each vertex (blue lines). The only possibility with a closed loop of an internal (blue) index, which passes through three vertices, is shown in fig. \ref{Twsited3vertLoop} a). After "forgetting" the color of this loop we get the ribbon graph  in fig. \ref{Twsited3vertLoop} b), which is non-planar due a twisted propagator. 
So, a graph with $\mathcal{F}_{3}>0$ cannot be maximal. Thus, setting  $\mathcal{F}_{3}=0$ in (\ref{maineqforF}), 
we deduce that a maximal graph should have $\mathcal{F}_{2}>0$.

\begin{figure}[h!]%
    \centering
    \subfloat{{\includegraphics[width=3.5cm]{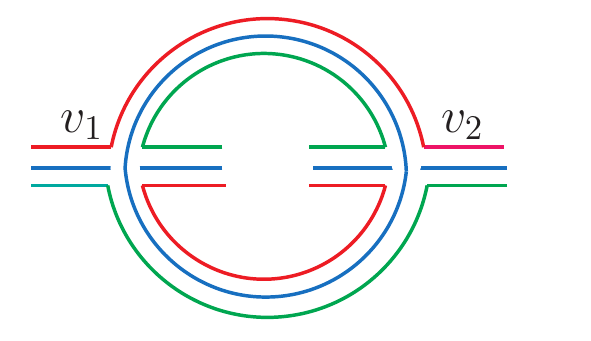} }}%
    \qquad
    \subfloat{{\includegraphics[width=3.5cm]{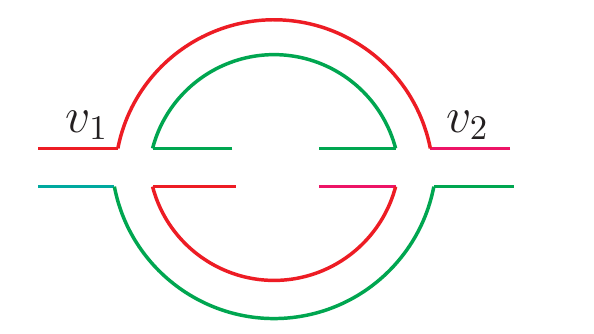} }}%
    \caption{a) Local part of a graph with a middle index loop passing  through two vertices $v_{1}$ and $v_{2}$. b) The same figure where the middle index has been ``forgotten." }
    \label{OnlyPossible2vert}%
\end{figure}

Finally, we need to show that the graphs with $\mathcal{F}_{2}>0$ are melonic. To do this
we will follow Proposition 3 in \cite{Bonzom:2011zz}.
Without a loss of generality we assume that the loop passing through $2$ vertices 
is formed by the middle lines in each vertex (blue lines). The only such possibility 
is shown in fig. \ref{OnlyPossible2vert} a). After "forgetting" the color of this loop we get the ribbon graph  in fig. \ref{OnlyPossible2vert} b).

\begin{figure}[h!]
                \centering
                \includegraphics[width=8cm]{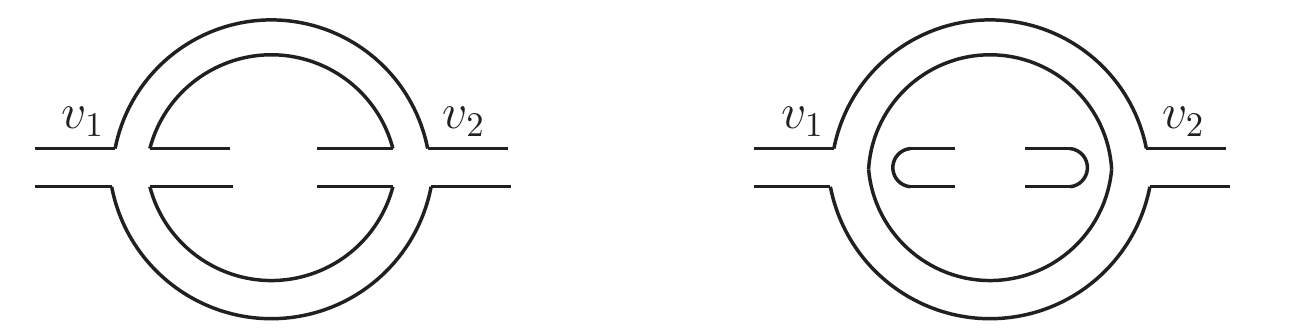}
                \caption{Cutting and sewing lines.}
                \label{2edgesCut}
\end{figure}

\begin{figure}[h!]
                \centering
                \includegraphics[width=3cm]{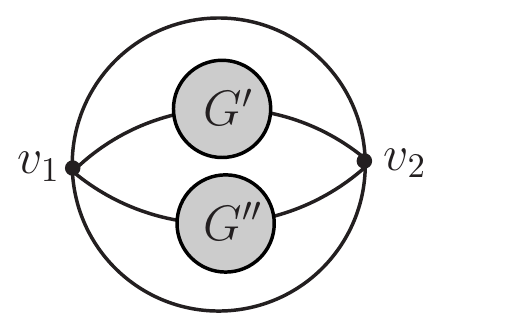}
                \caption{General structure of the maximal graph. }
                \label{GeneralPict1}
\end{figure}

\noindent  Now we uncolor the lines in our ribbon graph, and cut  and sew two edges as in figure \ref{2edgesCut}.  
We cut two edges but did not change the number of loops; therefore, the Euler characteristic of the new graph is $\chi=4$. This is possible only if we separated our original graph into two  genus zero parts.  Therefore, our graph is two-particle reducible for the internal and external couples of lines. Thus, the  whole unresolved  graph  looks like figure \ref{GeneralPict1}.
Then, if graphs $G'$ and $G''$ are empty we get a second-order melon graph as in figure \ref{ExampleDg2p3}.
If they are not empty one can argue (see \cite{Bonzom:2011zz}) that they are also maximal graphs. So, we can recursively apply the same above argument  to them, implying 
that the complete diagram is melonic.

While the proof above applies to $O(N)^3$ theories, we note that the melonic dominance does not require the theory to have multiple $O(N)$ symmetry groups. 
Indeed, it was conjectured in \cite{Klebanov:2017nlk} that the theory of a traceless symmetric
or an antisymmetric
3-index bosonic tensor $\phi^{abc}$ of $O(N)$ with the tetrahedral interaction is dominated by the melonic graphs. This conjecture was substantiated with 
explicit calculations up to a rather high order. 
The conjecture was proved in \cite{Benedetti:2017qxl}, but the combinatorial proof is considerably more complicated than the one presented above for the $O(N)^3$ theory:
indeed, since in the $O(N)$ case the strands have the same color, the trick of ``forgetting" all strands of a given color cannot be applied. 
The proof has been extended to $O(N)$ theories with rank-3 tensors of mixed symmetry \cite{Carrozza:2018ewt}.

\section{The Minimal $O(N)^3$ Tensor Quantum Mechanics}
\label{oncube}

The idea that fermionic tensor models reproduce the SYK-like large $N$ limit, but without disorder, was advanced in \cite{Witten:2016iux}. The quantum mechanical model
constructed there, often called the Gurau-Witten model, contains four species of rank-$3$ tensors and has $O(N)^6$ symmetry. 
Using the $O(N)^3$ symmetric interaction (\ref{partfuntensor}), but replacing the real bosons with Majorana fermions $\psi^{abc}(t)$, 
it is possible to simplify the construction of \cite{Witten:2016iux}. This leads to the minimal SYK-like fermionic quantum mechanical model \cite{Klebanov:2016xxf} with the action
\begin{align}
S = \int d t \Big( \frac i 2 \psi^{abc}\partial_{t}\psi^{abc}- \frac{1}{4}g \psi^{a_{1}b_{1}c_{1}}\psi^{a_{1}b_{2}c_{2}}\psi^{a_{2}b_{1}c_{2}}\psi^{a_{2}b_{2}c_{1}}\Big)\, , \label{FermAct3}
\end{align}
up to an additive constant.
Let us 
emphasize that $\psi^{abc}$ has distinguishable indices, each of which runs from $1$ to $N$. 
At the classical level, i.e. ignoring the right-hand side of (\ref{comrel}), the Fermi statistics implies 
\begin{align}
\psi^{a_{1}b_{1}c_{1}}\psi^{a_{1}b_{2}c_{2}}\psi^{a_{2}b_{1}c_{2}}\psi^{a_{2}b_{2}c_{1}}=
 -\psi^{a_{1}b_{2}c_{2}} \psi^{a_{1}b_{1}c_{1}} \psi^{a_{2}b_{1}c_{2}}\psi^{a_{2}b_{2}c_{1}} \,.
\end{align}
After relabeling 
$b_1 \leftrightarrow c_2$ and $b_2 \leftrightarrow c_1$ we get the relation 
\begin{align}
\psi^{a_{1}b_{1}c_{1}}\psi^{a_{1}b_{2}c_{2}}\psi^{a_{2}b_{1}c_{2}}\psi^{a_{2}b_{2}c_{1}}=
 -\psi^{a_{1}c_{1}b_{1}} \psi^{a_{1}c_{2}b_{2}} \psi^{a_{2}c_{2}b_{1}}\psi^{a_{2}c_{1}b_{2}} \ .
\end{align}
This demonstrates the vanishing of the tetrahedral interaction term in the $O(N)$ symmetric theory with fermions in any irreducible 
$3$-index representation: fully symmetric, fully anti-symmetric or mixed symmetry.  

Thus, the theory (\ref{FermAct3}) with $O(N)^3$ symmetry appears to be the simplest possible tensor counterpart of the SYK model.\footnote{
This tensor-SYK correspondence can be generalized to the versions of SYK model where the Hamiltonian couples $q>4$ fermions.
The corresponding tensor model involves a Majorana tensor with $q-1$ distinguishable indices \cite{Klebanov:2016xxf,Narayan:2017qtw}. The generalized tetrahedral interaction preserving
$O(N)^{q-1}$ symmetry is unique for $q=6$ \cite{Klebanov:2016xxf}, but there is a growing set of possibilities for $q\geq 8$ \cite{Gubser:2018yec}.
The large $N$ limit is taken keeping $g^2 N^{(q-1)(q-2)/2}$ fixed.
}  
The $SO(N)^3$ symmetry may be gauged by the replacement
\begin{equation}
\partial_t \psi^{abc}  \rightarrow (D_t \psi)^{abc} = \partial_t \psi^{abc} + A_1^{a a'} \psi^{a' b c} + A_2^{b b'} \psi^{a b' c} + A_3^{c c'} \psi^{a b c'}
\ ,
\end{equation}
where $A_i$ is the gauge field corresponding to the $i$-th $SO(N)$ group. In $d=1$ the gauge fields are non-dynamical, and their only effect is to restrict the states to be
annihilated by the symmetry charges
\begin{equation}
Q_{1}^{aa'}= \frac {i}{2} [\psi^{abc},\psi^{a'bc}]\ , \qquad
Q_{2}^{b b'}= \frac {i}{2} [\psi^{ab c},\psi^{a b' c}]\ , \qquad
Q_{3}^{c c'}= \frac {i}{2} [\psi^{ab c},\psi^{a b c'}]\ .
\label{SONcharges}
\end{equation}

The spectrum of the Hamiltonian (\ref{Hequal}) has an interesting property: for each eigenstate of energy $E$ there is a corresponding eigenstate of energy $-E$.
To explain the origin of this symmetry, it is useful to introduce unitary operators $P_{ij}$ associated with permutations of 
the $O(N)_i$ and $O(N)_j$ groups \cite{Pakrouski:2018jcc}:
\begin{align}
&P_{23} = P_{23}^\dagger= i^{n(n-1)/2} \prod_a \prod_{b>c} (\psi^{abc}- \psi^{acb} )\ , \notag \\
& P_{12} = P_{12}^\dagger= i^{n(n-1)/2} \prod_c \prod_{a>b} (\psi^{abc}- \psi^{bac} )\  , 
\end{align}
where $n=N^2 (N-1)/2$ is the number of fields in the product.
They satisfy 
\begin{align}
P_{23} \psi^{abc} P_{23}^\dagger =(-1)^{N^2 (N-1)/2} \psi^{acb} \ , \qquad 
P_{12} \psi^{abc} P_{12}^\dagger =(-1)^{N^2 (N-1)/2} \psi^{bac}\ .
\end{align}
These permutations flip the sign of $H$ \cite{Bulycheva:2017ilt,Klebanov:2018nfp,Pakrouski:2018jcc}:
\begin{align}
P_{23} H P_{23}^\dagger =-H\ , \qquad   P_{12} H P_{12}^\dagger =-H \ .
\end{align}
Thus, if $|\Psi\rangle$ is an eigenstate of $H$ with eigenvalue $E$, then 
$P_{12} |\Psi\rangle$ is an eigenstate with eigenvalue $-E$.

We can further define the operator $P$ which implements a cyclic permutation of the three
$O(N)$ groups:
\begin{align}
P=P_{12} P_{23}\ , \qquad  P \psi^{abc} P^\dagger = \psi^{cab}\ .
\end{align}
It has the properties
\begin{align}
P H P^\dagger = H\ , \qquad P^3=I\ ,
\end{align}
thus realizing the $Z_3$ symmetry of the Hamiltonian. For a more complete discussion of the discrete symmetries of the $O(N)^3$ tensor model,
see \cite{Pakrouski:2018jcc}.

A remarkable property of the model (\ref{Hequal}) is that, as $N$ grows, the spectrum of low-lying states becomes dense, and the theory becomes nearly conformal 
(we will demonstrate this in section \ref{SDEquations} using the Schwinger-Dyson equations).
Thus, in the large $N$ limit it is possible to define conformal operators and calculate their scaling dimensions. 
Particularly easy to study is the set of operators  
\begin{equation}
O_2^n= \psi^{abc} (\partial_t^n \psi)^{abc}\ ,
\label{twoparticleops}
\end{equation}
up to a total derivative. These operators are conformal primaries
when $n$ is odd. As discussed in section \ref{multipart}, using the equation of motion repeatedly, we can express them as gauge-invariant 
multi-particle operators without derivatives. 
Operators (\ref{twoparticleops}) are analogous to the ``single Regge trajectory" \cite{Polchinski:2016xgd,Maldacena:2016hyu,Gross:2016kjj} found in the SYK model
\cite{Sachdev:1992fk,1999PhRvB..59.5341P, 2000PhRvL..85..840G,Kitaev:2015,Kitaev:2017awl}. 
In section \ref{SDEquations} we will show that, in the large $N$ limit of the $O(N)^3$ tensor model, the scaling dimensions of these operators are the same as in the SYK model.
But first let us make some comments on the relation between the tensor and SYK models.

\subsection{Comparison of the tensor and SYK Hamiltonians}

Let us compare the Hamiltonian of the $O(N)^3$ tensor model with that of the SYK model with $N_{\rm SYK}=N^3$. 
If we think of $I=(abc)$ as a composite index which takes $N^3$ values,
the Hamiltonian (\ref{Hequal}) may be written in a way similar to that in the Sachdev-Ye-Kitaev model: omitting the overall factor $\frac{g} {4}$,
\begin{align}
&  H = \frac{1} {4!} J_{I_{1}I_{2}I_{3}I_{4}}\psi^{I_{1}}\psi^{I_{2}}\psi^{I_{3}}\psi^{I_{4}}\,,
\end{align}
where 
\begin{align}
&J_{I_{1}I_{2}I_{3}I_{4}} =
\delta_{a_1 a_2} \delta_{a_3 a_4} \delta_{b_1 b_3} \delta_{b_2 b_4} \delta_{c_1 c_4} \delta_{c_2 c_3} 
- \delta_{a_1 a_2} \delta_{a_3 a_4}  \delta_{b_2 b_3} \delta_{b_1 b_4} \delta_{c_2 c_4} \delta_{c_1 c_3}+ 22\ {\rm terms}  \,. 
\end{align}
This definition of $J$ takes values $0, \pm1$, and it is fully antisymmetric under permutations of the indices $I_{k}=(a_{k}b_{k}c_{k})$. 
Let us stress that  this form of $J$ breaks the $O(N^3)$ symmetry of the free theory down to $O(N)^3$.
The tensor $J$ is traceless,
$ J_{I_{1}I_{1}I_{2}I_{3}} = 0$, 
and it satisfies
\begin{align}
\frac{1} {4!}\sum_{\{I_{k}\}}J^{2}_{I_{1}I_{2}I_{3}I_{4}}= \frac{1}{4} N^3 (N-1)^2 (N+2)\ .
\label{tensorscaling}
\end{align}
This is the number of distinct non-vanishing terms in the Hamiltonian.\footnote{If we restore the factor $g^2$, then (\ref{tensorscaling}) becomes 
the combinatorial factor for the simplest melonic
vacuum diagram, which is the leftmost in figure \ref{MelonsEx}. For large $N$ it scales as $\lambda^2 N^3$, in agreement with section \ref{Theproof}.} 
In the SYK model with $N_{\rm SYK} = N^3$ fermions and random quartic couplings $J_{i_{1}i_{2}i_{3}i_{4}} $, the Hamiltonian (\ref{SYKHamilt}) generally involves
\begin{align} 
\frac{1}{4!} N_{\rm SYK} (N_{\rm SYK}-1) (N_{\rm SYK}-2) (N_{\rm SYK}-3) =
\frac{1}{24} N^3 (N^3-1) (N^3-2) (N^3-3)\ , 
\end{align}
distinct terms.
Thus, in the $O(N)^3$ model almost all possible quartic couplings vanish; only a fraction of order $N^{-6}$ is non-vanishing!
For example, for $N=4$, which is the biggest numerical diagonalization so far \cite{Pakrouski:2018jcc}, 
the Hamiltonian contains only $864$ terms out of the $635376$ possible terms, which would be
present in the SYK model with $N_{\rm SYK} = 64$ fermions.
Thus, the detailed structure of quartic couplings in the $O(N)^3$ tensor model is very sparse and highly non-generic from the SYK point of view. 
Nevertheless, the two models have similar large $N$ limits, at least for some quantities.
The sparseness of the tensor model Hamiltonian facilitates applications of the Lanczos method for calculating the spectrum  \cite{Pakrouski:2018jcc}.

In the SYK model with a large number of Majorana fermion species $N_{\rm SYK}$, the energy gaps between low-lying eigenstates are of order $e^{-\alpha N_{\rm SYK}}$,
where $\alpha$ is a positive constant of order $1$.
This was shown numerically in \cite{Maldacena:2016hyu,Garcia-Garcia:2016mno,Cotler:2016fpe,Gur-Ari:2018okm}.
The exponential smallness of the gaps leads to large low-temperature entropy $S_0 = c_0 N_{\rm SYK}$, even though the ground state is non-degenerate.
The normalization constant in the $q=4$ SYK model is \cite{Maldacena:2016hyu}
\be
c_0= \frac 12 \log 2 - \int_0^{1/4} \pi \left (\frac 12 -x \right ) \tan (\pi x) dx \approx 0.23\ .
\ee

In the $O(N)^3$ tensor model, calculation of the low-temperature entropy proceeds by summing the same melonic diagrams as in the SYK model \cite{Benedetti:2018goh}.
Therefore, using $N_{\rm SYK}=N^3$, we find  $S_0 = c_0 N^3$. By analogy with the SYK model, this suggests that the gaps above the ground states are of order 
$e^{-\tilde \alpha N^3}$ for large $N$, where $\tilde\alpha$ is a positive constant of order $1$. 
These tiny gaps should appear in the $SO(N)^3$ invariant part of the spectrum, which includes the ground state.
There is a lower bound on the ground state energy \cite{Klebanov:2018nfp}:
\be
E_{0} > -\frac{g }{16}N^3(N+2) \sqrt{N-1}\ .
\ee 
Nicely, this grows as $\lambda N^3$ in the large $N$ limit, just like in the SYK model the ground state energy grows as $N_{\rm SYK}$ \cite{Maldacena:2016hyu}. 
The splittings between the $SO(N)^3$ non-singlet and singlet states scale to zero as $\lambda/N$ \cite{Choudhury:2017tax,Klebanov:2018nfp}.
So far, these properties of the spectrum have not been possible to
demonstrate via direct numerical diagonalization of the Hamiltonian because the total size of the Hilbert space grows as $2^{N^3/2}$.

Explicit tensor model diagonalizations have been carried out in \cite{Krishnan:2016bvg,Krishnan:2017txw,Klebanov:2018nfp,Krishnan:2018hhu,Pakrouski:2018jcc}.
In the biggest calculation to date \cite{Pakrouski:2018jcc}, the complete spectrum of $SO(4)^3$ invariant states was found in the $O(4)^3$ tensor model.
Since there are only $36$ such states, they do not exhibit the small gaps needed for the study of the nearly conformal behavior.
However, in the $O(6)^3$ tensor model there are over $595 $ million $SO(6)^3$ invariant states \cite{Klebanov:2018nfp}, 
while their energies are bounded as 
\be
- 108 \sqrt 5 g < E < 108 \sqrt 5 g\ .
\ee
Therefore, their spectrum is likely to be very dense, but more work is needed to study their distribution.

\subsection{Schwinger-Dyson equations}
\label{SDEquations}

\begin{figure}[h!]
                \centering
                \includegraphics[width=16cm]{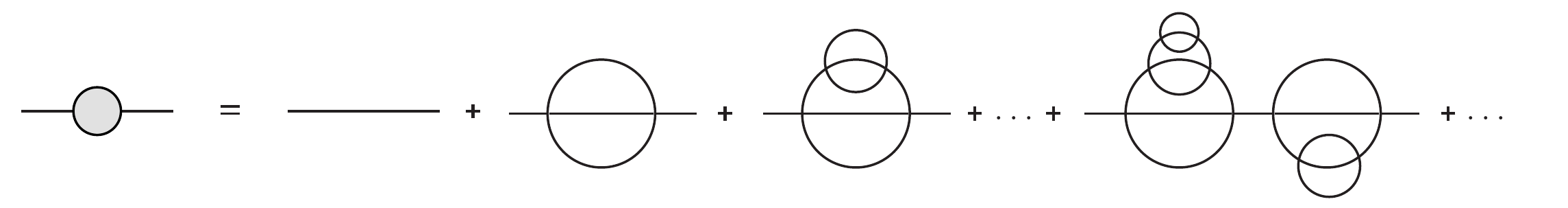}
                \caption{Diagrams contributing to the two point function in the leading  large $N$ order. The line with the gray circle represents the full two point function. Each simple line is the bare propagator. }
                \label{Fullprop}
\end{figure}

Let us study some of the diagrammatics of the $O(N)^3$ symmetric quantum mechanics (\ref{FermAct3}). We will study the ungauged model; the effect of the gauging may be
imposed later by restricting to the gauge invariant operators.
The bare propagator is 
\begin{align}
\langle T(\psi^{abc}(t)\psi^{a'b'c'}(0))\rangle_{0} =  \delta^{aa'}\delta^{bb'}\delta^{cc'}G_{0}(t)  =  \delta^{aa'}\delta^{bb'}\delta^{cc'} \frac{1}{2}\sgn (t)\,.
\end{align}
The full propagator in the large $N$ limit receives corrections from the melonic diagrams represented in figure \ref{Fullprop}. 
Resummation of all melonic diagrams  leads to the  Schwinger-Dyson equation for the two-point function
\begin{align}
G(t_{1}-t_{2}) = G_{0}(t_{1}-t_{2}) + g^{2}N^{3}\int dtdt' G_{0}(t_{1}-t)G(t-t')^{3} G(t'-t_{2})\,, \label{SD2pt}
\end{align}
represented graphically in figure \ref{SDeq}. This equation has the same structure as that
derived in \cite{Polchinski:2016xgd,Maldacena:2016hyu,Gross:2016kjj} for the large $N$ SYK model. 
The solution to (\ref{SD2pt}) in the IR limit is 
\begin{align}
G(t_{1}-t_{2})= -\Big(\frac{1}{4\pi g^{2}N^{3}}\Big)^{1/4}\ \frac{\sgn(t_{1}-t_{2})}{|t_{1}-t_{2}|^{1/2}}\,.
\label{twosolution}
\end{align}

\begin{figure}[h!]
               \centering
              \includegraphics[width=16cm]{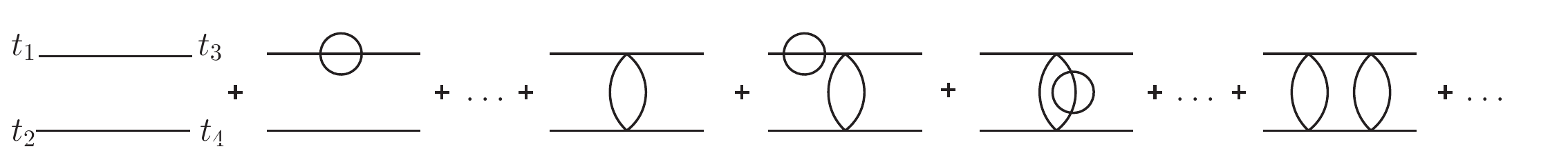}
              \caption{Ladder diagrams contributing to $\Gamma(t_{1},\dots, t_{4})$}
              \label{Ladder2}
\end{figure}

To uncover the spectrum of the bilinear operators in the model, we need to study the 4-point function
$ \langle \psi^{a_1 b_1 c_1 } (t_1)  \psi^{a_1 b_1 c_1 } (t_2) \psi^{a_2 b_2 c_2 } (t_3)  \psi^{a_2 b_2 c_2 } (t_4) \rangle $. 
Its structure is again the same as in the large $N$ SYK model \cite{Maldacena:2016hyu, Polchinski:2016xgd}: 
\begin{align}
\langle \psi^{a_1 b_1 c_1 } (t_1)  \psi^{a_1 b_1 c_1 } (t_2) \psi^{a_2 b_2 c_2 } (t_3)  \psi^{a_2 b_2 c_2 } (t_4) \rangle =N^{6}G(t_{12})G(t_{34}) + \Gamma(t_{1},\dots ,t_{4})\,,
\end{align}
where $\Gamma(t_{1},\dots ,t_{4})$ is given by a series of ladder diagrams depicted in fig \ref{Ladder2}. 

 \begin{figure}[h!]
               \centering
              \includegraphics[width=12cm]{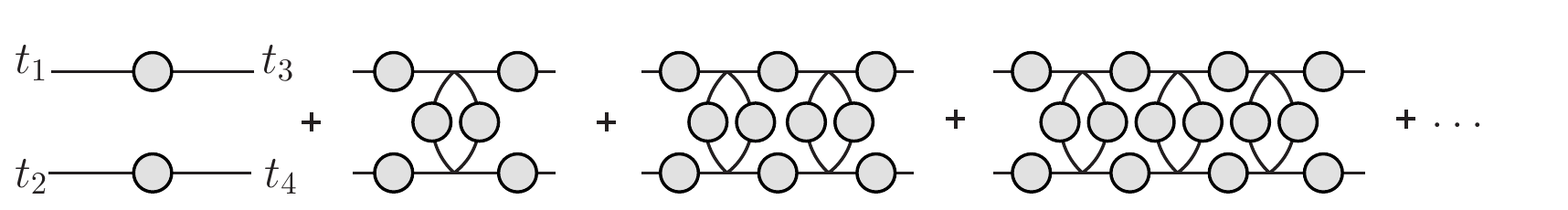}
              \caption{Ladder diagrams contributing to $\Gamma(t_{1},\dots, t_{4})$}
              \label{Ladder3}
\end{figure} 

Resumming the diagrams in fig. \ref{Ladder2} one finds a contribution to $ \Gamma(t_{1},\dots ,t_{4})$ as a series of diagrams  in terms of the full  propagators, see fig. \ref{Ladder3}.
If we denote by $\Gamma_{n}$ the ladder with $n$ rungs, so $\Gamma = \sum_{n} \Gamma_{n}$,  we have
\begin{align}
\Gamma_{0}(t_{1},\dots,t_{4})=N^{3}(-G(t_{13})G(t_{24})+G(t_{14})G(t_{23}))\,.
\end{align}
For the next coefficient one gets 
\begin{align}
\Gamma_{1}(t_{1},\dots,t_{4})=3g^{2}N^{6}\int dt dt'\big(G(t_{1}-t)G(t_{2}-t')G(t-t')^{2}G(t-t_{3})G(t-t_{4})-(t_{3} \leftrightarrow t_{4}) \big)\,,
\end{align}
and one can check further  that 
\begin{align}
\Gamma_{2}(t_{1},\dots,t_{4})=-3g^{2}N^{3}\int dt dt'\big(G(t_{1}-t)G(t_{2}-t')G(t-t')^{2}\Gamma_{1}(t,t',t_{3},t_{4})- (t_{3} \leftrightarrow t_{4})\big)\,.
\end{align}
So, in general, one gets exactly the same recursion relation as in  the SYK model 
\begin{align}
\Gamma_{n+1}(t_{1},\dots, t_{4}) = \int dt dt' K(t_{1},t_{2};t,t') \Gamma_{n}(t,t',t_{3},t_{4})\,,
\end{align}
where  the kernel is 
\begin{align}
K(t_{1},t_{2};t_{3},t_{4})  = -3 g^{2} N^{3}G(t_{13})G(t_{24})G(t_{34})^{2}\,.
\end{align}
In order to find the spectrum of the two-particle operators $O_{2}^n$, following \cite{Maldacena:2016hyu, Gross:2016kjj} one has to solve the integral eigenvalue equation
 \begin{align}
v_{n}(t_{0},t_{1},t_{2}) =g_{\rm antisym}(h)\int dt_{3}dt_{4}K(t_{1},t_{2};t_{3},t_{4})v_{n}(t_{0},t_{3},t_{4})\,, \label{SDopeq}
\end{align}
where  
 \begin{align}
v_{n}(t_{0},t_{1},t_{2})=\langle O_{2}^n (t_{0})\psi^{abc}(t_{1})\psi^{abc}(t_{2})\rangle = \frac{c_{n}\sgn(t_{1}-t_{2})}{|t_{0}-t_{1}|^{h}|t_{0}-t_{2}|^{h}|t_{1}-t_{2}|^{1/2-h}}\,, \label{anteigf}
\end{align}
is the conformal three-point function. 
To find $g(h)$  we compute the integral in (\ref{SDopeq}) using general $d$ dimensional conformal integrals   \cite{Symanzik:1972wj} specified to $d=1$:
 \begin{align}
\int_{-\infty}^{+\infty} du \frac{\sgn(u-t_{1})\sgn (u-t_{2})}{|u-t_{1}|^{\alpha_{1}}|u-t_{2}|^{\alpha_{2}}|u-t_{3}|^{\alpha_{3}}} =l_{\alpha_{1},\alpha_{2}} \frac{\sgn(t_{13})\sgn(t_{23})}{|t_{12}|^{1-\alpha_{3}}|t_{13}|^{1-\alpha_{2}}|t_{23}|^{1-\alpha_{1}}}\,, \label{base1dint}
\end{align}
where $\alpha_{1}+\alpha_{2}+\alpha_{3}=2$ and 
 \begin{align}
l_{\alpha_{1},\alpha_{2}} = \sqrt{\pi}\frac{\Gamma(\frac{1-\alpha_{1}}{2}+\frac{1}{2})\Gamma(\frac{1-\alpha_{2}}{2}+\frac{1}{2})\Gamma(\frac{1-\alpha_{3}}{2})}{\Gamma(\frac{\alpha_{1}}{2}+\frac{1}{2})\Gamma(\frac{\alpha_{2}}{2}+\frac{1}{2})\Gamma(\frac{\alpha_{3}}{2})}\,.
\end{align}
Taking integrals over $t_{3}$ and $t_{4}$ in (\ref{SDopeq}) using (\ref{base1dint}) we find \cite{Maldacena:2016hyu, Gross:2016kjj}
 \begin{align}
g_{\rm antisym}(h)=-\frac{3}{4\pi}l_{\frac{3}{2}-h,\frac{1}{2}}l_{h+\frac{1}{2},\frac{1}{2}}= -\frac{3}{2}\frac{\tan(\frac{\pi}{2}(h-\frac{1}{2}))}{h-1/2}\,.
\end{align}
The scaling dimensions are given by the solutions of $g_{\rm antisym}(h)=1$. The first solution is exact, $h=2$; this is the important mode dual to gravity and
responsible for the quantum chaos in the model 
\cite{Almheiri:2014cka,Polchinski:2016xgd,Maldacena:2016hyu,Jevicki:2016bwu,Maldacena:2016upp,Engelsoy:2016xyb,Jensen:2016pah}.
The further solutions are $h\approx
3.77,\; 5.68,\; 7.63,\;9.60$ 
corresponding to operators $O_2^n$ with $n=3,5,7,9$. In the limit of large $n$, $h_n\rightarrow n+\frac 1 2$. 
This is the expected limit $n+2 \Delta$, where $\Delta=\frac 1 4$ is the scaling dimension of the individual fermion.

\subsection{Multi-particle operators}
\label{multipart}

The model of \cite{Klebanov:2016xxf} contains a rapidly growing number of $SO(N)^3$ invariant $(2k)$-particle operators. 
Since a time derivative may be removed using the equations of motion, we may write the operators in a form where no derivatives are present.
The bilinear singlet operator, $\psi^{abc}\psi^{abc}$, vanishes classically by the Fermi statistics, 
while at the quantum level taking into account (\ref{comrel}), it is a C-number. The first non-trivial operators appear at the quartic level and are shown in figure \ref{O4ops}.
All of them are ``single-sum" operators, i.e. those that correspond to connected diagrams; they cannot be written as products of invariant operators.
On the left is the 
``tetrahedron operator" $O_{\textrm{tetra}}$, which appears in the Hamiltonian (\ref{Hequal}).
The three additional operators in figure \ref{O4ops}, which we denote as $O_{\textrm{pillow}}^{(1)}$,  $O_{\textrm{pillow}}^{(2)}$ and $O_{\textrm{pillow}}^{(3)}$, are the "pillow" operators in the terminology of
\cite{Tanasa:2011ur,Carrozza:2015adg}; they
contain double lines between a pair of vertices.
Up to an additive constant,  
\begin{align}
O_{\textrm{pillow}}^{(1)} = \sum_{a_1 < a_2}
Q_1^{a_{1} a_{2}} Q_1^{a_{1} a_{2}}\ , \qquad  
O_{\textrm{pillow}}^{(2)} =\sum_{b_1 < b_2}
Q_2^{b_{1} b_{2}} Q_2^{b_{1} b_{2}}\ , \qquad  
O_{\textrm{pillow}}^{(3)} = \sum_{c_1 < c_2}
Q_3^{c_{1} c_{2}} Q_3^{c_{1} c_{2}}  \, , 
\end{align}
i.e. they are the quadratic Casimir operators of the three $SO(N)$ groups. 
Since the $SO(N)^3$ charges (\ref{SONcharges}) commute with the Hamiltonian (\ref{Hequal}), so does each of the three pillow operators. 
This means that the scaling dimensions of the pillow operators are unaffected by the interactions, i.e. they vanish. The gauging of the $SO(N)^3$ symmetry sets the charges (\ref{SONcharges}) to zero, so the pillow operators do not appear in the gauged model.

Using the equations of motion we see that the operator 
$O_{\textrm{tetra}}$ is related by the equation of motion to the operator $O_2^1$:
\begin{align}
O_{\textrm{tetra}} = \psi^{abc} (\psi^{3})^{abc}  \propto \psi^{abc}\partial_{t}\psi^{abc}\,.
\end{align}
If we iterate the use of the equation of motion, then all derivatives in an operator may be traded for extra $\psi$-fields. 
Thus, a complete basis of operators may be constructed by
multiplying some number $2k$ of $\psi$-fields and contracting all indices. In this approach, there is a unique operator with $k=2(m+1)$, which is equal to the
Regge trajectory operator $O_2^{2m+1}$. For $m=0$ this operator is $O_{\textrm{tetra}}$, which is proportional to the Hamiltonian.

\begin{figure}[h!]
                \centering
                \includegraphics[width=15cm]{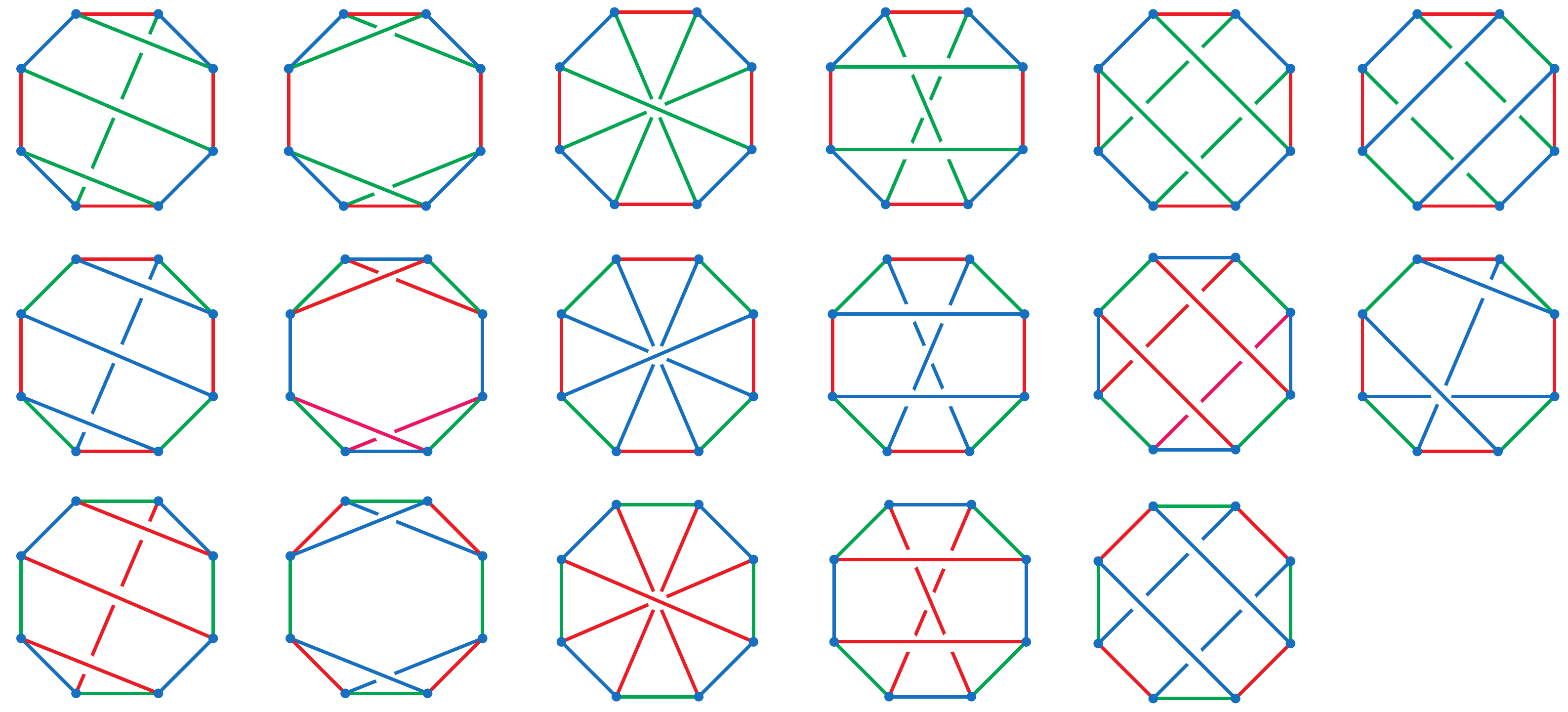}
                \caption{All eight-particle operators in the gauged fermionic model.}
                \label{O8all}
\end{figure}

All the six-particle operators vanish by the Fermi statistics, but there is a number of eight-particle ones.
We will exhibit only the ones not containing bubble insertions.
Having two vertices connected by a double line corresponds to insertion of an
$SO(N)$ charge which vanishes in the gauged model. For this reason we will omit such operators and list only those where there are no double lines.
In \cite{Bulycheva:2017ilt} it was shown that there are 17 inequivalent operators; see figure \ref{O8all}.
For example, the three operators shown in the first column of the figure \ref{O8all} are
\begin{align}
 &O_{1}= \psi^{a_{1}b_{1}c_{1}}\psi^{a_{1}b_{2}c_{2}}\psi^{a_{2}b_{2}c_{1}}\psi^{a_{2}b_{4}c_{4}}\psi^{a_{3}b_{3}
  c_{2}}\psi^{a_{3}b_{1}c_{3}}\psi^{a_{4}b_{4}c_{3}}\psi^{a_{4}b_{3} c_{4}}
= \partial_{t}\psi^{a_{1}b_{1}c_{1}}\partial_{t} \psi^{a_{1}b_{2}c_{2}}\psi^{a_{2}b_{1}c_{2}} \psi^{a_{2}b_{2}c_{1}}
\,,\notag\\
 &O_{2}=  \psi^{a_{1}b_{1} c_{1}}\psi^{a_{1} b_{2} c_{2}}\psi^{a_{2}b_{2} c_{1}}\psi^{a_{2}b_{3} c_{3}}\psi^{a_{3}b_{3}
  c_{2}}\psi^{a_{3} b_{4}c_{4}}\psi^{a_{4} b_{4}c_{3}}\psi^{a_{4} b_{1}c_{4}}
= \partial_{t} \psi^{a_{1}b_{1}c_{1}}\psi^{a_{1}b_{2}c_{2}}\partial_{t} \psi^{a_{2}b_{1}c_{2}} \psi^{a_{2}b_{2}c_{1}}
 \,, \notag \\
  &O_{3}=\psi^{a_{1}b_{1} c_{1}}\psi^{a_{1}b_{2}c_{2}}\psi^{a_{2} b_{2} c_{1}}\psi^{a_{2} b_{3} c_{3}}\psi^{a_{3} b_{1}
  c_{3}}\psi^{a_{3} b_{4} c_{4}}\psi^{a_{4} b_{3} c_{4}}\psi^{a_{4} b_{4}c_{2}}
=\partial_{t} \psi^{a_{1}b_{1}c_{1}}\psi^{a_{1}b_{2}c_{2}} \psi^{a_{2}b_{1}c_{2}} \partial_{t} \psi^{a_{2}b_{2}c_{1}}
 \,.
\label{tetrahedral}
\end{align}
It follows that
\begin{align}
O_{1}+O_{2}+O_{3}\sim \partial_t \psi^{abc} \partial_{t}^{2}\psi^{abc}
\ ,
 \label{melonic_O8}
\end{align}
which up to a total derivative equals the Regge trajectory operator $ O_2^3$.

The higher bubble-free single-sum operators were counted in  \cite{Bulycheva:2017ilt}: there are 
$24$ ten-particle ones, $617$ twelve-particle ones, $4887$ fourteen-particle ones, and $82466$ sixteen-particle ones.
In the bosonic theory, the single-sum operators correspond to connected Feynman diagrams in the theory with three scalar fields and $\phi_1 \phi_2 \phi_3$ interaction.
The number of such diagrams with $2k$ vertices grows as $k!$, and so does the number of invariant single-sum operators in the
bosonic theory \cite{Geloun:2013kta,Beccaria:2017aqc,Itoyama:2017xid,Mironov:2017aqv,Diaz:2017kub,deMelloKoch:2017bvv}.
 In the fermionic theory some operators vanish due to the Fermi statistics, but the factorial growth remains:
in \cite{Bulycheva:2017ilt} it was shown that the number of $(2k)$-particle $SO(N)^3$ invariant single-sum operators grows like $\sim k! 2^k$.
This is much faster than the exponential growth found in string theory. It implies that the Hagedorn temperature vanishes as
$1/\log N$ 
and suggests that the dual description of the tensor models lies ``beyond string theory." Perhaps it is related to M-theory, as suggested by the $N^3$ growth of the 
number of degrees of freedom on the M5-branes \cite{Klebanov:1996un} which may correspond to the M2-branes of trinion topology.

\section{Tensor Models for Complex Fermions}
\label{compferm}

In this section we study two different quantum mechanical models of a complex $3$-tensor $\psi^{abc}$.
One possibility is the model
\begin{align}
S = \int d t \Big( i \bar \psi^{abc}\partial_{t}\psi^{abc}-
\frac{1}{2} g \bar \psi^{a_{1}b_{2}c_{2}}\bar \psi^{a_{2}b_{2}c_{1}} \psi^{a_{1}b_{1}c_{1}}\psi^{a_{2}b_{1}c_{2}} 
\Big)\, . \label{FermAct7}
\end{align}
Its symmetry is
$SU(N)\times O(N) \times SU(N)\times U(1)$, where the $SU(N)$ groups act on the first and third indices.\footnote{
Bosonic models with this symmetry were previously studied in $d=0$ \cite{Tanasa:2011ur,Dartois:2013he,Tanasa:2015uhr}.}
The $U(1)$ acts by a phase rotation, $\psi^{abc}\rightarrow e^{i\alpha} \psi^{abc}$, and the corresponding conserved charge is 
\be
Q= \frac{1}{2} [\bar \psi^{abc}, \psi^{abc}]
\ .
\ee 
The model (\ref{FermAct7}) is the tensor counterpart of
the variant of SYK model where the real fermions are replaced by the complex ones 
\cite{Sachdev:2015efa,Fu:2016vas,Davison:2016ngz,Bulycheva:2017uqj,Yoon:2017nig,Peng:2017spg,Chaturvedi:2018uov}.

Let us study the conformal primary operators of the form  
\begin{align}
{\cal O}_2^{n}= \bar \psi^{abc} (\partial_t^{n} \psi)^{abc}\, \qquad n=0,1, \ldots \ ,
\label{twoparticlecomp}
\end{align}
up to total derivatives (there is a variety of operators made out of the higher powers of the fermionic fields, and some of them are equivalent to
(\ref{twoparticlecomp}) via the equations of motion).
As established in  \cite{Tanasa:2011ur,Dartois:2013he,Tanasa:2015uhr},
the large $N$ limit of the complex model (\ref{FermAct7}) is once again given by the melon diagrams
(the arguments are easier than in section \ref{Theproof} since each index loop passes through an even number of vertices). 
  Let us briefly discuss summing over melonic graphs 
in the model (\ref{FermAct7}) at large $N$. 
The two-point function has the structure
\begin{align}
\langle T(\bar \psi^{abc}(t)\psi^{a'b'c'}(0))\rangle  =  \delta^{aa'}\delta^{bb'}\delta^{cc'}G (t) ,
\end{align}
and $G(t)=- G(-t)$. We find the same Schwinger-Dyson equation as (\ref{SD2pt}); its solution is again (\ref{twosolution}) indicating that the fermion scaling dimension is $\Delta=1/4$.
Now we need to study the 4-point function
$ \langle \bar \psi^{a_1 b_1 c_1 } (t_1)  \psi^{a_1 b_1 c_1 } (t_2) \bar \psi^{a_2 b_2 c_2 } (t_3)  \psi^{a_2 b_2 c_2 } (t_4) \rangle $. It leads to the same
integral eigenvalue equation as (\ref{SDopeq}), but with kernel 
\begin{align}
K(t_{1},t_{2};t_{3},t_{4})  =- g^{2} N^{3} \big ( 2 G(t_{13})G(t_{24})G(t_{34})^{2} -  G(t_{14})G(t_{23})G(t_{34})^{2}\big )\ .
 \end{align}
Now it is possible to have
not only the antisymmetric eigenfunctions as 
in (\ref{anteigf}), but also the symmetric ones
 \begin{align}
v_{2n}(t_{0},t_{1},t_{2})=\langle {\cal O}^{2n}_{2} (t_{0})\psi^{abc}(t_{1})\bar{\psi}^{abc}(t_{2})\rangle = \frac{c_{2n}\sgn(t_{0}-t_{1})\sgn(t_{0}-t_{2})}{|t_{0}-t_{1}|^{h}|t_{0}-t_{2}|^{h}|t_{1}-t_{2}|^{1/2-h}}\,. \label{symeigf}
\end{align} 
This can be justified by noticing that the three point function now is  $\langle {\cal O}^{n}_{2}(t_{0})\psi^{abc}(t_{1})\bar{\psi}^{abc}(t_{2})\rangle$.
We see that for odd $n$ it is 
antisymmetric under $t_{1}\leftrightarrow t_{2}$, while for even $n$ it is symmetric.

\begin{figure}[h!]
                \centering
                \includegraphics[width=12cm]{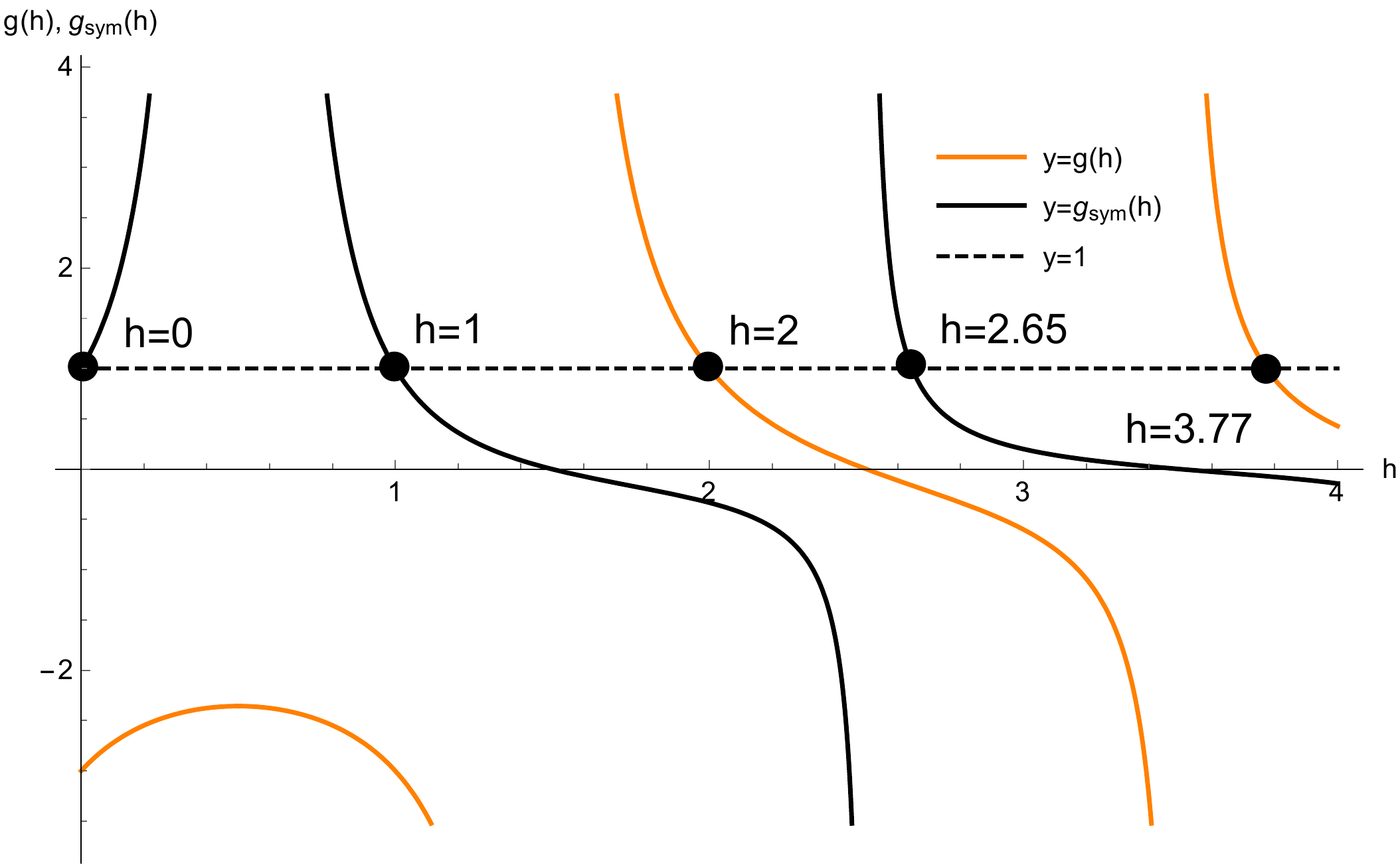}
                \caption{Graphical solution of the equations $g_{\rm antisym} (h) = 1$ (in orange) and  $g_{\rm sym}(h)=1$ (in black). }
                \label{ComplexAnomDim}
\end{figure}

Substituting ansatz (\ref{symeigf}) into the integral equation, and using the integrals (\ref{base1dint}), we find
 \begin{align}
g_{\rm sym}(h)= -\frac{1}{2}\frac{\tan(\frac{\pi}{2}(h+\frac{1}{2}))}{h-1/2}\,.
\label{gsym}
\end{align}
The scaling dimensions of the operators ${\cal O}_2^{n}$ with even $n$ are given by the solutions of
 $g_{\rm sym} (h)=1$. The first eigenvalue is $h=1$, corresponding to the conserved $U(1)$ charge.
The additional values are $h\approx 2.65,\; 4.58,\; 6.55,\; 8.54$ corresponding to the operators with $n=2,4,6,8$ respectively. 
For large $n$ the scaling dimensions approach $n+ \frac 1 2 $ as expected. The numerical results are in good agreement with the asymptotic formula \cite{Maldacena:2016hyu}
\begin{align}
h_n= n+ \frac 1 2 + \frac {1} {\pi n} + {\cal O}(n^{-3})
\end{align}
for $n>2$. 
For  ${\cal O}_2^n$ with odd $n$ the spectrum is the same as for the two-particle operators
(\ref{twoparticleops}) in the Majorana model with
$O(N)^3$ symmetry. The plot of the graphical solution for the scaling dimensions is shown in figure \ref{ComplexAnomDim}, with odd $n$ in orange and even $n$ in black.  

\subsection{Bipartite complex tensor model }

\begin{figure}[h!]
                \centering
                \includegraphics[width=10cm]{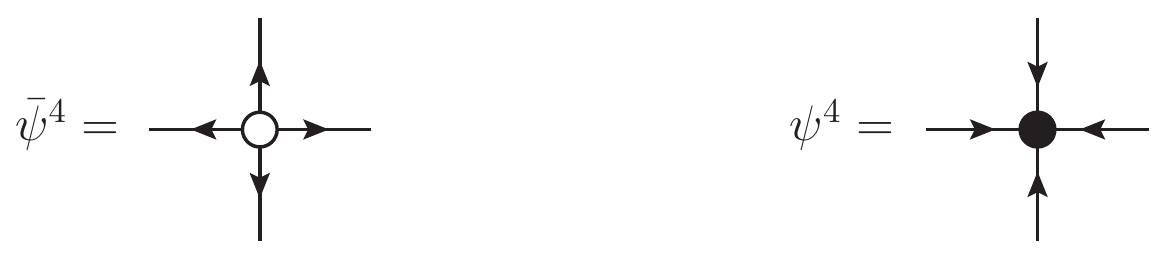}
                \caption{Vertices in the bipartite complex model}
                \label{vertices}
\end{figure}

Another possible  complex model is \cite{Klebanov:2016xxf}:
\begin{align}
S = \int d t \Big( i \bar \psi^{abc}\partial_{t}\psi^{abc}+
\frac{1}{4}g \psi^{a_{1}b_{1}c_{1}}\psi^{a_{1}b_{2}c_{2}}\psi^{a_{2}b_{1}c_{2}}\psi^{a_{2}b_{2}c_{1}}
+ \frac{1}{4} \bar g \bar \psi^{a_{1}b_{1}c_{1}}\bar \psi^{a_{1}b_{2}c_{2}}\bar \psi^{a_{2}b_{1}c_{2}}\bar \psi^{a_{2}b_{2}c_{1}}
\Big)\ . \label{FermAct5}
\end{align}
It is a special case of a more general model with $O(N)^3$ symmetry, which was studied in \cite{Jaewon:2018}.
In the Feynman graph expansion for
the model (\ref{FermAct5})
we denote the $\bar{\psi}^{4}$ and  $\psi^{4}$ vertices by white and black dots, respectively (see figure \ref{vertices}). 
Each Feynman graph necessarily has equal number of black and white dots; therefore, it is called a bipartite graph
(a similar bipartite model for multiple complex fermions was studied in  \cite{Gurau:2016lzk}).

In the model (\ref{FermAct5}) there is no $U(1)$ phase rotation symmetry; the interaction breaks it to $\mathbbm{Z}_{4}$. 
For real $g$ there is an additional symmetry
\begin{align}
\mathbbm{Z}_{2}: \quad \psi \to \bar{\psi}
\, . \label{Z2sym}
\end{align}
If we decompose a complex tensor into two real ones, $\psi^{abc}= \psi_1^{abc}+ i \psi_2^{abc}$, then this $\mathbbm{Z}_{2}$ acts by $ \psi_2^{abc} 
\rightarrow - \psi_2^{abc}$.

\begin{figure}[h!]
                \centering
                \includegraphics[width=10cm]{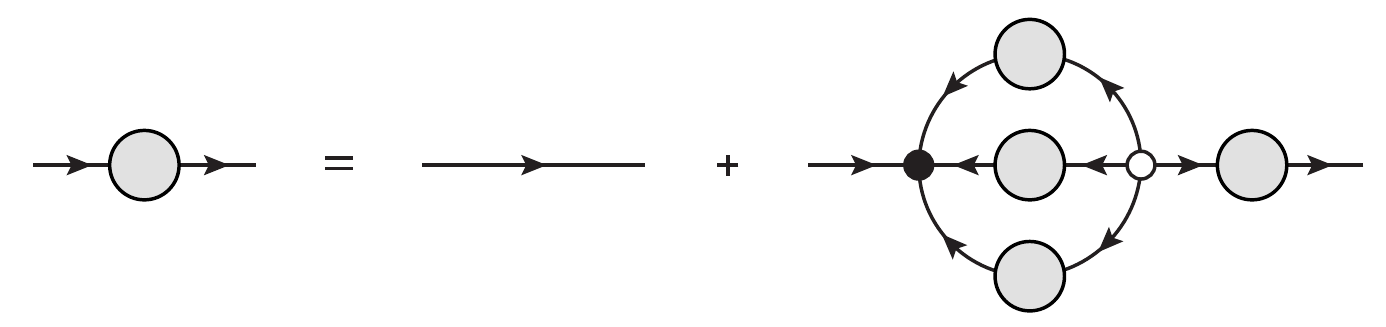}
                \caption{The graphical representation of the Schwinger-Dyson equation for the two-point function in the bipartite model.}
                \label{SD2p}
\end{figure}

In the large $N$ limit this model is dominated by the bipartite melonic diagrams. The two-point function 
\begin{align}
\langle T(\bar{\psi}^{abc}(t) \psi^{a'b'c'}(0))\rangle = \delta^{aa'}\delta^{bb'}\delta^{cc'}G(t)
\end{align}
satisfies the Schwinger-Dyson equation 
 \begin{align}
G(t_{1}-t_{2}) = G_{0}(t_{1}-t_{2})- g\bar{g}N^{3}\int dt dt' G_{0}(t_{1}-t)G(t'-t)^{3}G(t'-t_{2})\,,
\end{align}
which is graphically depicted in figure \ref{SD2p}. 

\noindent Since there is no chemical potential, 
we have $G(-t)=-G(t)$. Therefore, the Schwinger-Dyson equation may be written as
\begin{align}
G(t_{1}-t_{2}) = G_{0}(t_{1}-t_{2})+ g\bar{g}N^{3}\int dt dt' G_{0}(t_{1}-t)G(t-t')^{3}G(t'-t_{2})\,,
\end{align}
which is the same as in the $O(N)^3$ model with a real 3-tensor \cite{Klebanov:2016xxf}. The solution is given by the formula  
\begin{align}
G(t_{1}-t_{2})= -\Big(\frac{1}{4\pi \lambda^{2}}\Big)^{1/4}\ \frac{\sgn(t_{1}-t_{2})}{|t_{1}-t_{2}|^{1/2}}\,,
\label{twosolution}
\end{align}
where $\lambda^{2}=|g|^{2}N^{3}$ is kept fixed. This indicates that the fermion scaling dimension is $\Delta=1/4$.

Let us study the fermion bilinear operators 
\begin{align}
O_{n} = \bar{\psi}^{abc} \partial_{t}^{n} \psi^{abc}\ . 
\end{align}
For odd $n$ they are even under the $\mathbbm{Z}_{2}$ symmetry (\ref{Z2sym}), while for even $n$ they are odd 
under the $\mathbbm{Z}_{2}$. This is not hard to see using the real tensors $\psi_i$, $i=1,2$. As is well-known for the model with a real tensor \cite{Klebanov:2016xxf},
or equivalently for the SYK model \cite{Maldacena:2016hyu}, 
for even $n$ there are no primary operators of the
form  $\psi_i^{abc} \partial_{t}^{n} \psi_i^{abc}$. Therefore, 
the primary operators for even $n$ are the $\mathbbm{Z}_{2}$-odd operators $\psi_1^{abc} \partial_{t}^{n} \psi_2^{abc}$.

\begin{figure}[h!]
                \centering
                \includegraphics[width=10cm]{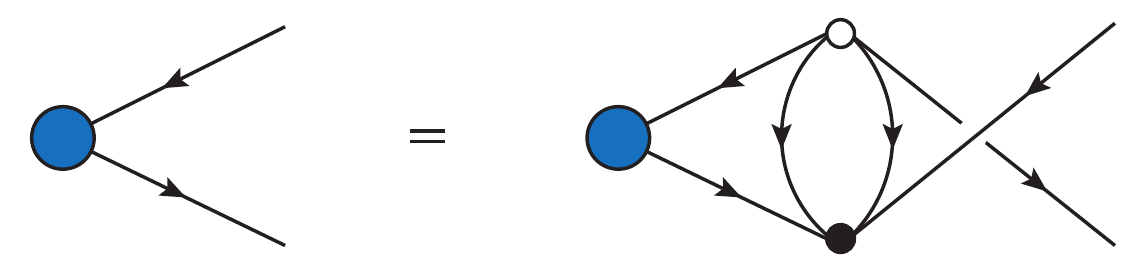}
                \caption{The graphical representation of the Schwinger-Dyson equation for the three-point function.}
                \label{SD3ptpic}
\end{figure}

To compute the scaling dimensions of $O_{n}$ we  consider a three point function 
\begin{align}
v_{n}(t_{0},t_{1},t_{2}) = \langle O_{n}(t_{0}) \psi^{abc}(t_{1})\bar{\psi}^{abc}(t_{2})\rangle\,.
\end{align}
Then one can derive the Schwinger-Dyson equation for the three point function in the IR region \cite{Gross:2016kjj}
\begin{align}
v_{n}(t_0,t_1,t_2) = \int dt_3 dt_4  K(t_{1},t_{2},t_{3},t_{4}) v_{n}(t_0, t_3, t_4)\,, \label{SD3pt}
\end{align}
where the kernel is 
\begin{align}
K(t_{1},t_{2},t_{3},t_{4})=  3\lambda^{2} G(t_{14})G(t_{23})G(t_{34})^2\,.
\end{align}
The  Schwinger-Dyson equation (\ref{SD3pt}) represented graphically in figure \ref{SD3ptpic}, where we have already dropped the bare term, which is irrelevant in the IR region.

\begin{figure}[h!]
                \centering
                \includegraphics[width=12cm]{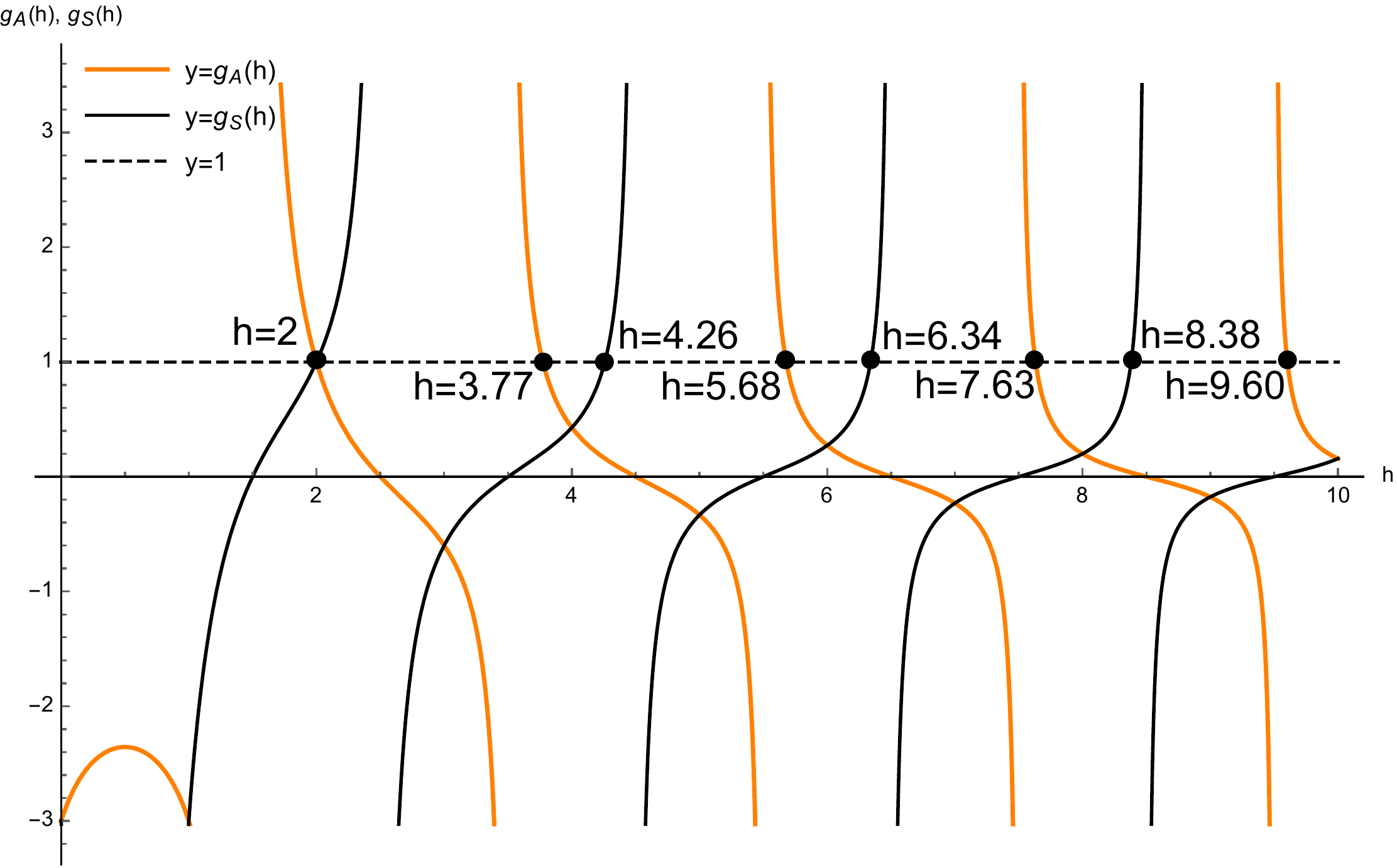}
                \caption{Graphical solution of the equations $g_{S}(h) = 1$ and  $g_{A}(h)=1$. 
}
                \label{AnomDim}
\end{figure}

\noindent For odd $n$ the three point function is antisymmetric $v(t_0,t_1,t_2)= - v(t_0, t_2, t_1)$ and is given by the general formula 
\begin{align}
v_{n}(t_0,t_1,t_2) = \frac{c_{n}\sgn(t_{1}-t_{2})}{|t_{01}|^{h}|t_{02}|^{h}|t_{12}|^{\frac{1}{2}-h}}\,, \label{ant3pt}
\end{align}
where $h$ is the anomalous dimension of the operator $O_{n}$ and $c_{n}$ is the structure constant\footnote{This method of finding anomalous dimension does not determine the structure constants $c_{n}$. One can find them by computing the whole four point function \cite{Maldacena:2016hyu}. }.  For arbitrary $h$ the three point function (\ref{ant3pt}) satisfies the equation 
\begin{align}
v_{n}(t_0,t_1,t_2) = g_{A}(h)\int dt_3 dt_4  K(t_{1},t_{2},t_{3},t_{4}) v_{n}(t_0, t_3, t_4)\,, \label{SD3pt2}
\end{align}
where
\begin{align}
g_{A}(h) = -\frac{3}{2}\frac{\tan(\frac{\pi}{2}(h-\frac{1}{2}))}{h-1/2}\,.
\end{align}
For even $n$ the three point function is symmetric $v(t_0,t_1,t_2)=  v(t_0, t_2, t_1)$ and is given by the general formula 
\begin{align}
v_{n}(t_0,t_1,t_2) = \frac{c_{n}\sgn(t_{0}-t_{1})\sgn(t_{0}-t_{2})}{|t_{01}|^{h}|t_{02}|^{h}|t_{12}|^{\frac{1}{2}-h}}\,, \label{sym3pt}
\end{align}
where $h$ is the anomalous dimension of the operator $O_{n}$ and $c_{n}$ is the structure constant.  For arbitrary $h$ the three point function (\ref{sym3pt}) satisfies the equation 
\begin{align}
v_{n}(t_0,t_1,t_2) = g_{S}(h)\int dt_3 dt_4  K(t_{1},t_{2},t_{3},t_{4}) v_{n}(t_0, t_3, t_4)\,, \label{SD3pt3}
\end{align}
where
\begin{align}
g_{S}(h) = \frac{3}{2}\frac{\tan(\frac{\pi}{2}(h+\frac{1}{2}))}{h-1/2}\,.
\end{align}
Now to find anomalous dimensions of the operators $O_{n}$ we have to find solutions to the equations 
\begin{align}
g_{S}(h) = 1, \quad g_{A}(h)=1\,.
\end{align}
We find a series of real solutions, which approach  $h_{n} \to n+\frac{1}{2}$ at large $n$. The first few values are $h= 2,\;2,\; 3.77,\; 4.26,\; 5.68,\;6.34\;$ corresponding to $n=1,2,3,4,5,6$, respectively. The plot is represented in figure \ref{AnomDim}. 
An interesting feature of this model is that both operators $O_{1}=\bar{\psi}^{abc}\partial_{t}\psi^{abc}$ and $O_{2}=\bar{\psi}^{abc}\partial_{t}^{2}\psi^{abc}$ have the same exact 
eigenvalue equal to $h=2$. For $n=1$ this eigenvalue corresponds to a normalizable state, but presumably this is not the case for $n=2$.

 The Schwinger -Dyson solution of the bipartite model has a new feature compared to the tensor models discussed in previous sections.
For $n=0$, which corresponds to the operator $O=\bar{\psi}^{abc}\psi^{abc}$, the solution of the equation $g_{S}(h)=1$ is complex:
\begin{align}
h\approx  \frac{1}{2}+1.5251 i\,. \label{comldim}
\end{align}
This signals a likely instability of this nearly-conformal model. From the point of view of the dual theory in $AdS_2$, the mass-squared of a a scalar field dual to an operator of dimension $h$ is
\begin{align}
m^2 = h (h-1) \,. 
\end{align}  
Therefore, a complex $h$ of the form $\frac{1}{2}+i \alpha$ corresponds to $m^2=m_{\rm BF}^2- \alpha^2$.
This is below the Breitenlohner-Freedman \cite{Breitenlohner:1982jf} stability bound $m_{\rm BF}^2=-\frac{1}{4}$, thus causing an instability. 

\section{Bosonic Tensor Models}
\label{Bonten}

It is of obvious interest to try extending the derivations above to quantum theories of bosonic tensors. It turns out that they 
frequently contain complex scaling dimensions, such as those noted in the previous section.
In \cite{Klebanov:2016xxf,Giombi:2017dtl} an $O(N)^3$ invariant theory of the scalar fields $\phi^{abc}$ was explored. When endowed with a quartic 
``tetrahedral" potential \cite{Carrozza:2015adg,Klebanov:2016xxf} 
\begin{align}
V= \frac{g}{4!}  \phi^{a_1 b_1 c_1} \phi^{a_1 b_2 c_2} \phi^{a_2 b_1 c_2} \phi^{a_2 b_2 c_1}\ ,
\end{align} 
this theory is super-renormalizable in $d<4$
and is formally solvable using the Schwinger-Dyson equations. However, this model has some instabilities. 
One problem is that the tetrahedral potential is not positive definite. Even if we ignore this and consider the large $N$ limit formally, we find that in $d<4$ the 
$O(N)^3$ invariant operator $\phi^{abc} \phi^{abc}$ has a complex dimension of the form $\frac{d}{2}+ i\alpha(d)$, which leads to instabilitiy of the resulting
large $N$ CFT.\footnote{Complex scaling dimensions of this form
appear in various other large $N$ theories; see, for example, \cite{Dymarsky:2005uh,Pomoni:2008de,Grabner:2017pgm,Prakash:2017hwq}.} 
From the dual AdS point of view, such a complex dimension corresponds to a scalar field whose $m^2$ is below the 
Breitenlohner-Freedman stability bound \cite{Breitenlohner:1982jf,Klebanov:1999tb}. 
The origin of the complex dimensions was elucidated using perturbation theory in $4-\epsilon$ dimensions: the fixed point was found to be at 
complex values of the couplings for the additional $O(N)^3$ invariant operators required by the renormalizability \cite{Klebanov:2016xxf,Giombi:2017dtl}. 
In \cite{Giombi:2017dtl} a $O(N)^5$ symmetric theory for tensor $\phi^{abcde}$ and sextic interactions was also considered. It was found that the
dimension of $\phi^{abcde}\phi^{abcde}$ is real in the narrow range $d_{\rm crit}< d< 3$, where $d_{\rm crit}\approx 2.97$. However, the scalar potential of this theory is
again unstable, so the theory may be defined only formally. 
 In spite of these problems, some interesting formal results on melonic scalar theories of this type were found recently \cite{Liu:2018jhs}. 

\begin{figure}[h!]
\begin{center}
\includegraphics[width=15.5cm]{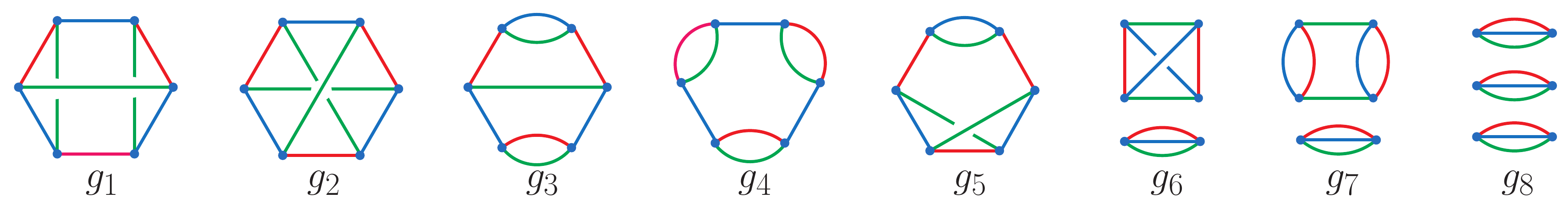}
\caption{Diagrammatic representation of the eight possible $O(N)^3$ invariant sextic interaction terms. }
\label{8inter}
\end{center}
\end{figure}

In \cite{Giombi:2018qgp}, the search was continued for stable non-supersymmetric large $N$ scalar theories with multiple $O(N)$ symmetry groups. 
The specific model studied was the $O(N)^3$ symmetric theory of scalar fields $\phi^{abc}$ with a sextic interaction, whose Euclidean action in $d$ dimensions is
\begin{equation}
S= \int d^d x \left ( \frac {1} {2} (\partial_\mu \phi^{abc})^2 + 
{g_1\over 6!} \phi^{a_1 b_1 c_1} \phi^{a_1 b_2 c_2} \phi^{a_2 b_1 c_2} \phi^{a_3 b_3 c_1} \phi^{a_3 b_2 c_3} \phi^{a_2 b_3 c_3}
\right )
\ .
\label{prism}
\end{equation}
This theory is super-renormalizable in $d<3$.
When fields $\phi^{abc}$ are represented by vertices and index contractions by edges, this interaction term looks like a prism \cite{Klebanov:2016xxf};
it is the leftmost diagram in figure \ref{8inter}.  In the large $N$ limit, $g_1 N^3$ is held fixed \cite{Giombi:2018qgp}.
We may call this theory ``prismatic" (note that the Feynman diagrams are not the same as the melonic diagrams which appear
in the sextic theory of \cite{Giombi:2017dtl} with the $O(N)^5$ symmetry).
Unlike with the quartic interaction of tetrahedral topology \cite{Klebanov:2016xxf,Giombi:2017dtl}, the action (\ref{prism}) is positive for $g_1>0$.
In fact, it is the bosonic part of the action of the supersymmetric theory with two supercharges and tetrahedral superpotential \cite{Klebanov:2016xxf},
$W\sim \Phi^{a_1 b_1 c_1} \Phi^{a_1 b_2 c_2} \Phi^{a_2 b_1 c_2} \Phi^{a_2 b_2 c_1}$.

The theory (\ref{prism}) may be viewed as a tensor counterpart of the bosonic theory with random couplings, which was introduced in section 6.2 of
\cite{Murugan:2017eto}. Since both theories are dominated by the same class of diagrams in the large $N$ limit, they have the same Schwinger-Dyson equations for the 2-point and
4-point functions. For $1.68 < d < 2.81$ there is a scalar bilinear operator with a complex dimension of the form $\frac{d}{2}+i \alpha(d)$, but outside of this range
there is no such complex scaling dimension \cite{Giombi:2018qgp}. This raises the possibility of a purely bosonic near-conformal quantum mechanical theory.\footnote{Another
quantum mechanical theory containing bosonic tensors, where the supersymmetry is broken spontaneously, was recently studied in \cite{Chang:2018sve}.} 

One can also use the renormalized perturbation theory to develop the $3-\epsilon$ expansion.
To carry out the beta function calculation at finite $N$ we need to include all the $O(N)^3$ invariant sextic terms in the action.
The 11 such single-sum terms are shown diagrammatically in figure 5 of \cite{Bulycheva:2017ilt}.
It is convenient to impose the additional constraint that the action is invariant under the permutation group $S_3$ which acts on the three $O(N)$ symmetry groups.
The most essential for achieving the large $N$ limit is the ``prism" term (\ref{prism});
it is positive definite and symmetric under the interchanges of the three $O(N)$ groups.
The 8 needed operators and associated couplings are 
exhibited in figure \ref{8inter}, with the prism operator shown on the left. 
The coupled two-loop beta functions, calculated in \cite{Giombi:2018qgp}, have a ``prismatic" fixed point for $N> 53$, where all the coupling constants are real.
As $N$ is increased, $g_1\sim N^{-3}$ at this fixed point, while the other
$7$ couplings scale to zero faster.
The resulting $3-\epsilon$ expansions are, in the large $N$ limit,
in agreement with the results from the Schwinger-Dyson equations used to solve the prismatic large $N$ theory. 
This constitutes an explicit perturbative check of the large $N$ solution of the prismatic theory \cite{Giombi:2018qgp}.

\section*{Acknowledgments}

These notes are an expanded version of the lectures presented by IRK at the TASI 2017 summer school in June 2017 in Boulder, Colorado, 
and at the Abdus Salam ICTP 2018 Spring School in March 2018 in Trieste, Italy.
IRK is grateful to the TASI 2017 co-organizers Mirjam Cvetic, Tom DeGrand and Oliver DeWolfe for creating a wonderful environment at the school.
He is also grateful to the organizers of the 2018 ICTP Spring School, especially Atish Dabholkar, for the invitation and hospitality.
Many thanks to the students at both school for the many good questions and useful discussions.
We thank our collaborators K. Bulycheva, S. Giombi, A. Milekhin, K. Pakrouski, and S. Prakash, with whom some of the results reviewed here were obtained.
We are  grateful to D. Gross, R. Gurau, J. Maldacena, V. Rosenhaus, D. Stanford, and E. Witten, for illuminating discussions. 
The work of IRK and FP was supported in part by the US NSF under Grant No.~PHY-1620059. The work of GT was supported by  the MURI grant W911NF-14-1-0003 from ARO and by DOE grant de-sc0007870.

\bibliographystyle{JHEP}
\bibliography{TASInotes}

\providecommand{\href}[2]{#2}\begingroup\raggedright\begin{thebibliography}{100}

\bibitem{Stanley:1968gx}
H.~E. Stanley, \emph{{Spherical model as the limit of infinite spin
  dimensionality}}, \href{https://doi.org/10.1103/PhysRev.176.718}{\emph{Phys.
  Rev.} {\bfseries 176} (1968) 718}.

\bibitem{Wilson:1973jj}
K.~G. Wilson and J.~B. Kogut, \emph{{The Renormalization group and the epsilon
  expansion}}, \href{https://doi.org/10.1016/0370-1573(74)90023-4}{\emph{Phys.
  Rept.} {\bfseries 12} (1974) 75}.

\bibitem{Moshe:2003xn}
M.~Moshe and J.~Zinn-Justin, \emph{{Quantum field theory in the large N limit:
  A Review}}, \href{https://doi.org/10.1016/S0370-1573(03)00263-1}{\emph{Phys.
  Rept.} {\bfseries 385} (2003) 69}
  [\href{https://arxiv.org/abs/hep-th/0306133}{{\ttfamily hep-th/0306133}}].

\bibitem{'tHooft:1973jz}
G.~'t~Hooft, \emph{{A Planar Diagram Theory for Strong Interactions}},
  \href{https://doi.org/10.1016/0550-3213(74)90154-0}{\emph{Nucl. Phys.}
  {\bfseries B72} (1974) 461}.

\bibitem{Veneziano:1976wm}
G.~Veneziano, \emph{{Some Aspects of a Unified Approach to Gauge, Dual and
  Gribov Theories}},
  \href{https://doi.org/10.1016/0550-3213(76)90412-0}{\emph{Nucl. Phys.}
  {\bfseries B117} (1976) 519}.

\bibitem{Maldacena:1997re}
J.~M. Maldacena, \emph{{The Large N limit of superconformal field theories and
  supergravity}}, \href{https://doi.org/10.1023/A:1026654312961,
  10.4310/ATMP.1998.v2.n2.a1}{\emph{Int. J. Theor. Phys.} {\bfseries 38} (1999)
  1113} [\href{https://arxiv.org/abs/hep-th/9711200}{{\ttfamily
  hep-th/9711200}}].

\bibitem{Gubser:1998bc}
S.~S. Gubser, I.~R. Klebanov and A.~M. Polyakov, \emph{{Gauge theory
  correlators from noncritical string theory}},
  \href{https://doi.org/10.1016/S0370-2693(98)00377-3}{\emph{Phys. Lett.}
  {\bfseries B428} (1998) 105}
  [\href{https://arxiv.org/abs/hep-th/9802109}{{\ttfamily hep-th/9802109}}].

\bibitem{Witten:1998qj}
E.~Witten, \emph{{Anti-de Sitter space and holography}},
  \href{https://doi.org/10.4310/ATMP.1998.v2.n2.a2}{\emph{Adv. Theor. Math.
  Phys.} {\bfseries 2} (1998) 253}
  [\href{https://arxiv.org/abs/hep-th/9802150}{{\ttfamily hep-th/9802150}}].

\bibitem{DeWolfe:2018dkl}
O.~DeWolfe, \emph{{TASI Lectures on Applications of Gauge/Gravity Duality}},
  in \emph{{Theoretical Advanced Study Institute in Elementary Particle
  Physics: Physics at the Fundamental Frontier (TASI 2017) Boulder, CO, USA,
  June 5-30, 2017}}, 2018, \href{https://arxiv.org/abs/1802.08267}{{\ttfamily
  1802.08267}}.

\bibitem{Erdmenger:2018xqz}
J.~Erdmenger, \emph{{Introduction to Gauge/Gravity Duality (TASI Lectures
  2017)}}, {\emph{PoS} {\bfseries TASI2017} (2017) 001}
  [\href{https://arxiv.org/abs/1807.09872}{{\ttfamily 1807.09872}}].

\bibitem{Harlow:2018fse}
D.~Harlow, \emph{{TASI Lectures on the Emergence of the Bulk in AdS/CFT}},
  \href{https://arxiv.org/abs/1802.01040}{{\ttfamily 1802.01040}}.

\bibitem{Klebanov:2000me}
I.~R. Klebanov, \emph{{TASI lectures: Introduction to the AdS / CFT
  correspondence}},  in \emph{{Strings, branes and gravity. Proceedings,
  Theoretical Advanced Study Institute, TASI'99, Boulder, USA, May 31-June 25,
  1999}}, pp.~615--650, 2000,
  \href{https://arxiv.org/abs/hep-th/0009139}{{\ttfamily hep-th/0009139}},
  \href{https://doi.org/10.1142/9789812799630_0007}{DOI}.

\bibitem{DHoker:2002nbb}
E.~D'Hoker and D.~Z. Freedman, \emph{{Supersymmetric gauge theories and the AdS
  / CFT correspondence}},  in \emph{{Strings, Branes and Extra Dimensions: TASI
  2001: Proceedings}}, pp.~3--158, 2002,
  \href{https://arxiv.org/abs/hep-th/0201253}{{\ttfamily hep-th/0201253}}.

\bibitem{Maldacena:2003nj}
J.~M. Maldacena, \emph{{TASI 2003 lectures on AdS / CFT}},  in \emph{{Progress
  in string theory. Proceedings, Summer School, TASI 2003, Boulder, USA, June
  2-27, 2003}}, pp.~155--203, 2003,
  \href{https://arxiv.org/abs/hep-th/0309246}{{\ttfamily hep-th/0309246}}.

\bibitem{Aharony:1999ti}
O.~Aharony, S.~S. Gubser, J.~M. Maldacena, H.~Ooguri and Y.~Oz, \emph{{Large N
  field theories, string theory and gravity}},
  \href{https://doi.org/10.1016/S0370-1573(99)00083-6}{\emph{Phys. Rept.}
  {\bfseries 323} (2000) 183}
  [\href{https://arxiv.org/abs/hep-th/9905111}{{\ttfamily hep-th/9905111}}].

\bibitem{Brezin:1977sv}
E.~Brezin, C.~Itzykson, G.~Parisi and J.~B. Zuber, \emph{{Planar Diagrams}},
  \href{https://doi.org/10.1007/BF01614153}{\emph{Commun. Math. Phys.}
  {\bfseries 59} (1978) 35}.

\bibitem{Gross:1989vs}
D.~J. Gross and A.~A. Migdal, \emph{{Nonperturbative Two-Dimensional Quantum
  Gravity}}, \href{https://doi.org/10.1103/PhysRevLett.64.127}{\emph{Phys. Rev.
  Lett.} {\bfseries 64} (1990) 127}.

\bibitem{Brezin:1990rb}
E.~Brezin and V.~A. Kazakov, \emph{{Exactly Solvable Field Theories of Closed
  Strings}}, \href{https://doi.org/10.1016/0370-2693(90)90818-Q}{\emph{Phys.
  Lett.} {\bfseries B236} (1990) 144}.

\bibitem{Douglas:1989ve}
M.~R. Douglas and S.~H. Shenker, \emph{{Strings in Less Than One-Dimension}},
  \href{https://doi.org/10.1016/0550-3213(90)90522-F}{\emph{Nucl. Phys.}
  {\bfseries B335} (1990) 635}.

\bibitem{Ginsparg:1993is}
P.~H. Ginsparg and G.~W. Moore, \emph{{Lectures on 2-D gravity and 2-D string
  theory}},  in \emph{{Proceedings, Theoretical Advanced Study Institute (TASI
  92): From Black Holes and Strings to Particles: Boulder, USA, June 1-26,
  1992}}.

\bibitem{Klebanov:1991qa}
I.~R. Klebanov, \emph{{String theory in two-dimensions}},  in \emph{{Spring
  School on String Theory and Quantum Gravity (to be followed by Workshop)
  Trieste, Italy, April 15-23, 1991}}, pp.~30--101, 1991,
  \href{https://arxiv.org/abs/hep-th/9108019}{{\ttfamily hep-th/9108019}}.

\bibitem{Polyakov:1981rd}
A.~M. Polyakov, \emph{{Quantum Geometry of Bosonic Strings}},
  \href{https://doi.org/10.1016/0370-2693(81)90743-7}{\emph{Phys. Lett.}
  {\bfseries B103} (1981) 207}.

\bibitem{Ambjorn:1990ge}
J.~Ambjorn, B.~Durhuus and T.~Jonsson, \emph{{Three-dimensional simplicial
  quantum gravity and generalized matrix models}},
  \href{https://doi.org/10.1142/S0217732391001184}{\emph{Mod. Phys. Lett.}
  {\bfseries A6} (1991) 1133}.

\bibitem{Sasakura:1990fs}
N.~Sasakura, \emph{{Tensor model for gravity and orientability of manifold}},
  \href{https://doi.org/10.1142/S0217732391003055}{\emph{Mod. Phys. Lett.}
  {\bfseries A6} (1991) 2613}.

\bibitem{Gross:1991hx}
M.~Gross, \emph{{Tensor models and simplicial quantum gravity in $> 2$-D}},
  \href{https://doi.org/10.1016/S0920-5632(05)80015-5}{\emph{Nucl. Phys. Proc.
  Suppl.} {\bfseries 25A} (1992) 144}.

\bibitem{Gurau:2009tw}
R.~Gurau, \emph{{Colored Group Field Theory}},
  \href{https://doi.org/10.1007/s00220-011-1226-9}{\emph{Commun. Math. Phys.}
  {\bfseries 304} (2011) 69} [\href{https://arxiv.org/abs/0907.2582}{{\ttfamily
  0907.2582}}].

\bibitem{Gurau:2011aq}
R.~Gurau and V.~Rivasseau, \emph{{The 1/N expansion of colored tensor models in
  arbitrary dimension}},
  \href{https://doi.org/10.1209/0295-5075/95/50004}{\emph{Europhys. Lett.}
  {\bfseries 95} (2011) 50004}
  [\href{https://arxiv.org/abs/1101.4182}{{\ttfamily 1101.4182}}].

\bibitem{Gurau:2011xq}
R.~Gurau, \emph{{The complete 1/N expansion of colored tensor models in
  arbitrary dimension}},
  \href{https://doi.org/10.1007/s00023-011-0118-z}{\emph{Annales Henri
  Poincare} {\bfseries 13} (2012) 399}
  [\href{https://arxiv.org/abs/1102.5759}{{\ttfamily 1102.5759}}].

\bibitem{Bonzom:2011zz}
V.~Bonzom, R.~Gurau, A.~Riello and V.~Rivasseau, \emph{{Critical behavior of
  colored tensor models in the large N limit}},
  \href{https://doi.org/10.1016/j.nuclphysb.2011.07.022}{\emph{Nucl. Phys.}
  {\bfseries B853} (2011) 174}
  [\href{https://arxiv.org/abs/1105.3122}{{\ttfamily 1105.3122}}].

\bibitem{Tanasa:2011ur}
A.~Tanasa, \emph{{Multi-orientable Group Field Theory}},
  \href{https://doi.org/10.1088/1751-8113/45/16/165401}{\emph{J. Phys.}
  {\bfseries A45} (2012) 165401}
  [\href{https://arxiv.org/abs/1109.0694}{{\ttfamily 1109.0694}}].

\bibitem{Bonzom:2012hw}
V.~Bonzom, R.~Gurau and V.~Rivasseau, \emph{{Random tensor models in the large
  N limit: Uncoloring the colored tensor models}},
  \href{https://doi.org/10.1103/PhysRevD.85.084037}{\emph{Phys. Rev.}
  {\bfseries D85} (2012) 084037}
  [\href{https://arxiv.org/abs/1202.3637}{{\ttfamily 1202.3637}}].

\bibitem{Carrozza:2015adg}
S.~Carrozza and A.~Tanasa, \emph{{$O(N)$ Random Tensor Models}},
  \href{https://doi.org/10.1007/s11005-016-0879-x}{\emph{Lett. Math. Phys.}
  {\bfseries 106} (2016) 1531}
  [\href{https://arxiv.org/abs/1512.06718}{{\ttfamily 1512.06718}}].

\bibitem{Witten:2016iux}
E.~Witten, \emph{{An SYK-Like Model Without Disorder}},
  \href{https://arxiv.org/abs/1610.09758}{{\ttfamily 1610.09758}}.

\bibitem{Klebanov:2016xxf}
I.~R. Klebanov and G.~Tarnopolsky, \emph{{Uncolored random tensors, melon
  diagrams, and the Sachdev-Ye-Kitaev models}},
  \href{https://doi.org/10.1103/PhysRevD.95.046004}{\emph{Phys. Rev.}
  {\bfseries D95} (2017) 046004}
  [\href{https://arxiv.org/abs/1611.08915}{{\ttfamily 1611.08915}}].

\bibitem{Giombi:2018qgp}
S.~Giombi, I.~R. Klebanov, F.~Popov, S.~Prakash and G.~Tarnopolsky,
  \emph{{Prismatic Large $N$ Models for Bosonic Tensors}},
  \href{https://arxiv.org/abs/1808.04344}{{\ttfamily 1808.04344}}.

\bibitem{Gurau:2011xp}
R.~Gurau and J.~P. Ryan, \emph{{Colored Tensor Models - a review}},
  \href{https://doi.org/10.3842/SIGMA.2012.020}{\emph{SIGMA} {\bfseries 8}
  (2012) 020} [\href{https://arxiv.org/abs/1109.4812}{{\ttfamily 1109.4812}}].

\bibitem{Tanasa:2015uhr}
A.~Tanasa, \emph{{The Multi-Orientable Random Tensor Model, a Review}},
  \href{https://doi.org/10.3842/SIGMA.2016.056}{\emph{SIGMA} {\bfseries 12}
  (2016) 056} [\href{https://arxiv.org/abs/1512.02087}{{\ttfamily
  1512.02087}}].

\bibitem{Delporte:2018iyf}
N.~Delporte and V.~Rivasseau, \emph{{The Tensor Track V: Holographic Tensors}},
   2018, \href{https://arxiv.org/abs/1804.11101}{{\ttfamily 1804.11101}}.

\bibitem{Sachdev:1992fk}
S.~Sachdev and J.~Ye, \emph{{Gapless spin fluid ground state in a random,
  quantum Heisenberg magnet}},
  \href{https://doi.org/10.1103/PhysRevLett.70.3339}{\emph{Phys. Rev. Lett.}
  {\bfseries 70} (1993) 3339}
  [\href{https://arxiv.org/abs/cond-mat/9212030}{{\ttfamily
  cond-mat/9212030}}].

\bibitem{1999PhRvB..59.5341P}
O.~{Parcollet} and A.~{Georges}, \emph{{Non-Fermi-liquid regime of a doped Mott
  insulator}}, \href{https://doi.org/10.1103/PhysRevB.59.5341}{\emph{Physical
  Review B} {\bfseries 59} (1999) 5341}
  [\href{https://arxiv.org/abs/cond-mat/9806119}{{\ttfamily
  cond-mat/9806119}}].

\bibitem{2000PhRvL..85..840G}
A.~{Georges}, O.~{Parcollet} and S.~{Sachdev}, \emph{{Mean Field Theory of a
  Quantum Heisenberg Spin Glass}},
  \href{https://doi.org/10.1103/PhysRevLett.85.840}{\emph{Physical Review
  Letters} {\bfseries 85} (2000) 840}
  [\href{https://arxiv.org/abs/cond-mat/9909239}{{\ttfamily
  cond-mat/9909239}}].

\bibitem{Kitaev:2015}
A.~Kitaev, \emph{{A simple model of quantum holography}}, .

\bibitem{Kitaev:2017awl}
A.~Kitaev and S.~J. Suh, \emph{{The soft mode in the Sachdev-Ye-Kitaev model
  and its gravity dual}},
  \href{https://doi.org/10.1007/JHEP05(2018)183}{\emph{JHEP} {\bfseries 05}
  (2018) 183} [\href{https://arxiv.org/abs/1711.08467}{{\ttfamily
  1711.08467}}].

\bibitem{Sarosi:2017ykf}
G.~Sarosi, \emph{{AdS$_{2}$ holography and the SYK model}},
  \href{https://doi.org/10.22323/1.323.0001}{\emph{PoS} {\bfseries Modave2017}
  (2018) 001} [\href{https://arxiv.org/abs/1711.08482}{{\ttfamily
  1711.08482}}].

\bibitem{Rosenhaus:2018dtp}
V.~Rosenhaus, \emph{{An introduction to the SYK model}},
  \href{https://arxiv.org/abs/1807.03334}{{\ttfamily 1807.03334}}.

\bibitem{Aharony:2008ug}
O.~Aharony, O.~Bergman, D.~L. Jafferis and J.~Maldacena, \emph{{N=6
  superconformal Chern-Simons-matter theories, M2-branes and their gravity
  duals}}, \href{https://doi.org/10.1088/1126-6708/2008/10/091}{\emph{JHEP}
  {\bfseries 10} (2008) 091} [\href{https://arxiv.org/abs/0806.1218}{{\ttfamily
  0806.1218}}].

\bibitem{Klebanov:1996un}
I.~R. Klebanov and A.~A. Tseytlin, \emph{{Entropy of near extremal black
  p-branes}}, \href{https://doi.org/10.1016/0550-3213(96)00295-7}{\emph{Nucl.
  Phys.} {\bfseries B475} (1996) 164}
  [\href{https://arxiv.org/abs/hep-th/9604089}{{\ttfamily hep-th/9604089}}].

\bibitem{Drukker:2010nc}
N.~Drukker, M.~Marino and P.~Putrov, \emph{{From weak to strong coupling in
  ABJM theory}}, \href{https://doi.org/10.1007/s00220-011-1253-6}{\emph{Commun.
  Math. Phys.} {\bfseries 306} (2011) 511}
  [\href{https://arxiv.org/abs/1007.3837}{{\ttfamily 1007.3837}}].

\bibitem{Herzog:2010hf}
C.~P. Herzog, I.~R. Klebanov, S.~S. Pufu and T.~Tesileanu, \emph{{Multi-Matrix
  Models and Tri-Sasaki Einstein Spaces}},
  \href{https://doi.org/10.1103/PhysRevD.83.046001}{\emph{Phys. Rev.}
  {\bfseries D83} (2011) 046001}
  [\href{https://arxiv.org/abs/1011.5487}{{\ttfamily 1011.5487}}].

\bibitem{Marino:2011eh}
M.~Marino and P.~Putrov, \emph{{ABJM theory as a Fermi gas}},
  \href{https://doi.org/10.1088/1742-5468/2012/03/P03001}{\emph{J. Stat. Mech.}
  {\bfseries 1203} (2012) P03001}
  [\href{https://arxiv.org/abs/1110.4066}{{\ttfamily 1110.4066}}].

\bibitem{Marino:2016new}
M.~Marino, \emph{{Localization at large N in Chern–Simons-matter theories}},
  \href{https://doi.org/10.1088/1751-8121/aa5f69}{\emph{J. Phys.} {\bfseries
  A50} (2017) 443007} [\href{https://arxiv.org/abs/1608.02959}{{\ttfamily
  1608.02959}}].

\bibitem{Pufu:2016zxm}
S.~S. Pufu, \emph{{The F-Theorem and F-Maximization}},
  \href{https://doi.org/10.1088/1751-8121/aa6765}{\emph{J. Phys.} {\bfseries
  A50} (2017) 443008} [\href{https://arxiv.org/abs/1608.02960}{{\ttfamily
  1608.02960}}].

\bibitem{Jafferis:2011zi}
D.~L. Jafferis, I.~R. Klebanov, S.~S. Pufu and B.~R. Safdi, \emph{{Towards the
  F-Theorem: N=2 Field Theories on the Three-Sphere}},
  \href{https://doi.org/10.1007/JHEP06(2011)102}{\emph{JHEP} {\bfseries 06}
  (2011) 102} [\href{https://arxiv.org/abs/1103.1181}{{\ttfamily 1103.1181}}].

\bibitem{Bulycheva:2017ilt}
K.~Bulycheva, I.~R. Klebanov, A.~Milekhin and G.~Tarnopolsky, \emph{{Spectra of
  Operators in Large $N$ Tensor Models}},
  \href{https://doi.org/10.1103/PhysRevD.97.026016}{\emph{Phys. Rev.}
  {\bfseries D97} (2018) 026016}
  [\href{https://arxiv.org/abs/1707.09347}{{\ttfamily 1707.09347}}].

\bibitem{Sachdev:2015efa}
S.~Sachdev, \emph{{Bekenstein-Hawking Entropy and Strange Metals}},
  \href{https://doi.org/10.1103/PhysRevX.5.041025}{\emph{Phys. Rev.} {\bfseries
  X5} (2015) 041025} [\href{https://arxiv.org/abs/1506.05111}{{\ttfamily
  1506.05111}}].

\bibitem{Davison:2016ngz}
R.~A. Davison, W.~Fu, A.~Georges, Y.~Gu, K.~Jensen and S.~Sachdev,
  \emph{{Thermoelectric transport in disordered metals without quasiparticles:
  The Sachdev-Ye-Kitaev models and holography}},
  \href{https://doi.org/10.1103/PhysRevB.95.155131}{\emph{Phys. Rev.}
  {\bfseries B95} (2017) 155131}
  [\href{https://arxiv.org/abs/1612.00849}{{\ttfamily 1612.00849}}].

\bibitem{Giombi:2016ejx}
S.~Giombi, \emph{{Higher Spin -- CFT Duality}},  in \emph{{Proceedings,
  Theoretical Advanced Study Institute in Elementary Particle Physics: New
  Frontiers in Fields and Strings (TASI 2015): Boulder, CO, USA, June 1-26,
  2015}}, pp.~137--214, 2017,
  \href{https://arxiv.org/abs/1607.02967}{{\ttfamily 1607.02967}},
  \href{https://doi.org/10.1142/9789813149441_0003}{DOI}.

\bibitem{Fei:2014yja}
L.~Fei, S.~Giombi and I.~R. Klebanov, \emph{{Critical $O(N)$ models in
  $6-\epsilon$ dimensions}},
  \href{https://doi.org/10.1103/PhysRevD.90.025018}{\emph{Phys. Rev.}
  {\bfseries D90} (2014) 025018}
  [\href{https://arxiv.org/abs/1404.1094}{{\ttfamily 1404.1094}}].

\bibitem{Kos:2013tga}
F.~Kos, D.~Poland and D.~Simmons-Duffin, \emph{{Bootstrapping the $O(N)$ vector
  models}}, \href{https://doi.org/10.1007/JHEP06(2014)091}{\emph{JHEP}
  {\bfseries 06} (2014) 091} [\href{https://arxiv.org/abs/1307.6856}{{\ttfamily
  1307.6856}}].

\bibitem{Rattazzi:2008pe}
R.~Rattazzi, V.~S. Rychkov, E.~Tonni and A.~Vichi, \emph{{Bounding scalar
  operator dimensions in 4D CFT}},
  \href{https://doi.org/10.1088/1126-6708/2008/12/031}{\emph{JHEP} {\bfseries
  12} (2008) 031} [\href{https://arxiv.org/abs/0807.0004}{{\ttfamily
  0807.0004}}].

\bibitem{Poland:2018epd}
D.~Poland, S.~Rychkov and A.~Vichi, \emph{{The Conformal Bootstrap: Theory,
  Numerical Techniques, and Applications}},
  \href{https://arxiv.org/abs/1805.04405}{{\ttfamily 1805.04405}}.

\bibitem{Klebanov:2002ja}
I.~R. Klebanov and A.~M. Polyakov, \emph{{AdS dual of the critical O(N) vector
  model}}, \href{https://doi.org/10.1016/S0370-2693(02)02980-5}{\emph{Phys.
  Lett.} {\bfseries B550} (2002) 213}
  [\href{https://arxiv.org/abs/hep-th/0210114}{{\ttfamily hep-th/0210114}}].

\bibitem{Vasiliev:1990en}
M.~A. Vasiliev, \emph{{Consistent equation for interacting gauge fields of all
  spins in (3+1)-dimensions}},
  \href{https://doi.org/10.1016/0370-2693(90)91400-6}{\emph{Phys. Lett.}
  {\bfseries B243} (1990) 378}.

\bibitem{Giombi:2012ms}
S.~Giombi and X.~Yin, \emph{{The Higher Spin/Vector Model Duality}},
  \href{https://doi.org/10.1088/1751-8113/46/21/214003}{\emph{J. Phys.}
  {\bfseries A46} (2013) 214003}
  [\href{https://arxiv.org/abs/1208.4036}{{\ttfamily 1208.4036}}].

\bibitem{Athenodorou:2016ebg}
A.~Athenodorou and M.~Teper, \emph{{SU(N) gauge theories in 2+1 dimensions:
  glueball spectra and k-string tensions}},
  \href{https://doi.org/10.1007/JHEP02(2017)015}{\emph{JHEP} {\bfseries 02}
  (2017) 015} [\href{https://arxiv.org/abs/1609.03873}{{\ttfamily
  1609.03873}}].

\bibitem{Klebanov:1997kc}
I.~R. Klebanov, \emph{{World volume approach to absorption by nondilatonic
  branes}}, \href{https://doi.org/10.1016/S0550-3213(97)00235-6}{\emph{Nucl.
  Phys.} {\bfseries B496} (1997) 231}
  [\href{https://arxiv.org/abs/hep-th/9702076}{{\ttfamily hep-th/9702076}}].

\bibitem{Polchinski:1996na}
J.~Polchinski, \emph{{Tasi lectures on D-branes}},  in \emph{{Fields, strings
  and duality. Proceedings, Summer School, Theoretical Advanced Study Institute
  in Elementary Particle Physics, TASI'96, Boulder, USA, June 2-28, 1996}}.

\bibitem{Polchinski:1995mt}
J.~Polchinski, \emph{{Dirichlet Branes and Ramond-Ramond charges}},
  \href{https://doi.org/10.1103/PhysRevLett.75.4724}{\emph{Phys. Rev. Lett.}
  {\bfseries 75} (1995) 4724}
  [\href{https://arxiv.org/abs/hep-th/9510017}{{\ttfamily hep-th/9510017}}].

\bibitem{Kim:1985ez}
H.~J. Kim, L.~J. Romans and P.~van Nieuwenhuizen, \emph{{The Mass Spectrum of
  Chiral N=2 D=10 Supergravity on S**5}},
  \href{https://doi.org/10.1103/PhysRevD.32.389}{\emph{Phys. Rev.} {\bfseries
  D32} (1985) 389}.

\bibitem{Minahan:2002ve}
J.~A. Minahan and K.~Zarembo, \emph{{The Bethe ansatz for N=4
  superYang-Mills}},
  \href{https://doi.org/10.1088/1126-6708/2003/03/013}{\emph{JHEP} {\bfseries
  03} (2003) 013} [\href{https://arxiv.org/abs/hep-th/0212208}{{\ttfamily
  hep-th/0212208}}].

\bibitem{Beisert:2010jr}
N.~Beisert et~al., \emph{{Review of AdS/CFT Integrability: An Overview}},
  \href{https://doi.org/10.1007/s11005-011-0529-2}{\emph{Lett. Math. Phys.}
  {\bfseries 99} (2012) 3} [\href{https://arxiv.org/abs/1012.3982}{{\ttfamily
  1012.3982}}].

\bibitem{Fiamberti:2007rj}
F.~Fiamberti, A.~Santambrogio, C.~Sieg and D.~Zanon, \emph{{Wrapping at four
  loops in N=4 SYM}},
  \href{https://doi.org/10.1016/j.physletb.2008.06.061}{\emph{Phys. Lett.}
  {\bfseries B666} (2008) 100}
  [\href{https://arxiv.org/abs/0712.3522}{{\ttfamily 0712.3522}}].

\bibitem{Bajnok:2008bm}
Z.~Bajnok and R.~A. Janik, \emph{{Four-loop perturbative Konishi from strings
  and finite size effects for multiparticle states}},
  \href{https://doi.org/10.1016/j.nuclphysb.2008.08.020}{\emph{Nucl. Phys.}
  {\bfseries B807} (2009) 625}
  [\href{https://arxiv.org/abs/0807.0399}{{\ttfamily 0807.0399}}].

\bibitem{Gromov:2009tv}
N.~Gromov, V.~Kazakov and P.~Vieira, \emph{{Exact Spectrum of Anomalous
  Dimensions of Planar N=4 Supersymmetric Yang-Mills Theory}},
  \href{https://doi.org/10.1103/PhysRevLett.103.131601}{\emph{Phys. Rev. Lett.}
  {\bfseries 103} (2009) 131601}
  [\href{https://arxiv.org/abs/0901.3753}{{\ttfamily 0901.3753}}].

\bibitem{Gromov:2013pga}
N.~Gromov, V.~Kazakov, S.~Leurent and D.~Volin, \emph{{Quantum Spectral Curve
  for Planar $\mathcal{N} = 4$ Super-Yang-Mills Theory}},
  \href{https://doi.org/10.1103/PhysRevLett.112.011602}{\emph{Phys. Rev. Lett.}
  {\bfseries 112} (2014) 011602}
  [\href{https://arxiv.org/abs/1305.1939}{{\ttfamily 1305.1939}}].

\bibitem{Gromov:2009zb}
N.~Gromov, V.~Kazakov and P.~Vieira, \emph{{Exact Spectrum of Planar ${\cal
  N}=4$ Supersymmetric Yang-Mills Theory: Konishi Dimension at Any Coupling}},
  \href{https://doi.org/10.1103/PhysRevLett.104.211601}{\emph{Phys. Rev. Lett.}
  {\bfseries 104} (2010) 211601}
  [\href{https://arxiv.org/abs/0906.4240}{{\ttfamily 0906.4240}}].

\bibitem{Gromov:2011bz}
N.~Gromov and S.~Valatka, \emph{{Deeper Look into Short Strings}},
  \href{https://doi.org/10.1007/JHEP03(2012)058}{\emph{JHEP} {\bfseries 03}
  (2012) 058} [\href{https://arxiv.org/abs/1109.6305}{{\ttfamily 1109.6305}}].

\bibitem{Gromov:2014bva}
N.~Gromov, F.~Levkovich-Maslyuk, G.~Sizov and S.~Valatka, \emph{{Quantum
  spectral curve at work: from small spin to strong coupling in $ \mathcal{N} $
  = 4 SYM}}, \href{https://doi.org/10.1007/JHEP07(2014)156}{\emph{JHEP}
  {\bfseries 07} (2014) 156} [\href{https://arxiv.org/abs/1402.0871}{{\ttfamily
  1402.0871}}].

\bibitem{Hegedus:2016eop}
A.~Hegedus and J.~Konczer, \emph{{Strong coupling results in the AdS$_{5}$ /CF
  T$_{4}$ correspondence from the numerical solution of the quantum spectral
  curve}}, \href{https://doi.org/10.1007/JHEP08(2016)061}{\emph{JHEP}
  {\bfseries 08} (2016) 061}
  [\href{https://arxiv.org/abs/1604.02346}{{\ttfamily 1604.02346}}].

\bibitem{Roiban:2011fe}
R.~Roiban and A.~A. Tseytlin, \emph{{Semiclassical string computation of
  strong-coupling corrections to dimensions of operators in Konishi
  multiplet}},
  \href{https://doi.org/10.1016/j.nuclphysb.2011.02.016}{\emph{Nucl. Phys.}
  {\bfseries B848} (2011) 251}
  [\href{https://arxiv.org/abs/1102.1209}{{\ttfamily 1102.1209}}].

\bibitem{Beccaria:2012xm}
M.~Beccaria, S.~Giombi, G.~Macorini, R.~Roiban and A.~A. Tseytlin,
  \emph{{'Short' spinning strings and structure of quantum $AdS_5 \times S^5$
  spectrum}}, \href{https://doi.org/10.1103/PhysRevD.86.066006}{\emph{Phys.
  Rev.} {\bfseries D86} (2012) 066006}
  [\href{https://arxiv.org/abs/1203.5710}{{\ttfamily 1203.5710}}].

\bibitem{Ambjorn:1985az}
J.~Ambjorn, B.~Durhuus and J.~Frohlich, \emph{{Diseases of Triangulated Random
  Surface Models, and Possible Cures}},
  \href{https://doi.org/10.1016/0550-3213(85)90356-6}{\emph{Nucl. Phys.}
  {\bfseries B257} (1985) 433}.

\bibitem{Das:1989fq}
S.~R. Das, A.~Dhar, A.~M. Sengupta and S.~R. Wadia, \emph{{New Critical
  Behavior in $d=0$ Large $N$ Matrix Models}},
  \href{https://doi.org/10.1142/S0217732390001165}{\emph{Mod. Phys. Lett.}
  {\bfseries A5} (1990) 1041}.

\bibitem{Klebanov:1994kv}
I.~R. Klebanov and A.~Hashimoto, \emph{{Nonperturbative solution of matrix
  models modified by trace squared terms}},
  \href{https://doi.org/10.1016/0550-3213(94)00518-J}{\emph{Nucl. Phys.}
  {\bfseries B434} (1995) 264}
  [\href{https://arxiv.org/abs/hep-th/9409064}{{\ttfamily hep-th/9409064}}].

\bibitem{Klebanov:2003wg}
I.~R. Klebanov, J.~M. Maldacena and N.~Seiberg, \emph{{Unitary and complex
  matrix models as 1-d type 0 strings}},
  \href{https://doi.org/10.1007/s00220-004-1183-7}{\emph{Commun. Math. Phys.}
  {\bfseries 252} (2004) 275}
  [\href{https://arxiv.org/abs/hep-th/0309168}{{\ttfamily hep-th/0309168}}].

\bibitem{Dalley:1991qg}
S.~Dalley, C.~V. Johnson and T.~R. Morris, \emph{{Multicritical complex matrix
  models and nonperturbative 2-D quantum gravity}},
  \href{https://doi.org/10.1016/0550-3213(92)90217-Y}{\emph{Nucl. Phys.}
  {\bfseries B368} (1992) 625}.

\bibitem{Ferrari:2017ryl}
F.~Ferrari, \emph{{The Large D Limit of Planar Diagrams}},
  \href{https://arxiv.org/abs/1701.01171}{{\ttfamily 1701.01171}}.

\bibitem{Ferrari:2017jgw}
F.~Ferrari, V.~Rivasseau and G.~Valette, \emph{{A New Large N Expansion for
  General Matrix-Tensor Models}},
  \href{https://arxiv.org/abs/1709.07366}{{\ttfamily 1709.07366}}.

\bibitem{Giombi:2017dtl}
S.~Giombi, I.~R. Klebanov and G.~Tarnopolsky, \emph{{Bosonic tensor models at
  large $N$ and small $\epsilon$}},
  \href{https://doi.org/10.1103/PhysRevD.96.106014}{\emph{Phys. Rev.}
  {\bfseries D96} (2017) 106014}
  [\href{https://arxiv.org/abs/1707.03866}{{\ttfamily 1707.03866}}].

\bibitem{Klebanov:2017nlk}
I.~R. Klebanov and G.~Tarnopolsky, \emph{{On Large $N$ Limit of Symmetric
  Traceless Tensor Models}},
  \href{https://arxiv.org/abs/1706.00839}{{\ttfamily 1706.00839}}.

\bibitem{Benedetti:2017qxl}
D.~Benedetti, S.~Carrozza, R.~Gurau and M.~Kolanowski, \emph{{The $1/N$
  expansion of the symmetric traceless and the antisymmetric tensor models in
  rank three}},  \href{https://arxiv.org/abs/1712.00249}{{\ttfamily
  1712.00249}}.

\bibitem{Carrozza:2018ewt}
S.~Carrozza, \emph{{Large $N$ limit of irreducible tensor models: $O(N)$
  rank-$3$ tensors with mixed permutation symmetry}},
  \href{https://doi.org/10.1007/JHEP06(2018)039}{\emph{JHEP} {\bfseries 06}
  (2018) 039} [\href{https://arxiv.org/abs/1803.02496}{{\ttfamily
  1803.02496}}].

\bibitem{Narayan:2017qtw}
P.~Narayan and J.~Yoon, \emph{{SYK-like Tensor Models on the Lattice}},
  \href{https://doi.org/10.1007/JHEP08(2017)083}{\emph{JHEP} {\bfseries 08}
  (2017) 083} [\href{https://arxiv.org/abs/1705.01554}{{\ttfamily
  1705.01554}}].

\bibitem{Gubser:2018yec}
S.~S. Gubser, C.~Jepsen, Z.~Ji and B.~Trundy, \emph{{Higher melonic theories}},
   \href{https://arxiv.org/abs/1806.04800}{{\ttfamily 1806.04800}}.

\bibitem{Pakrouski:2018jcc}
K.~Pakrouski, I.~R. Klebanov, F.~Popov and G.~Tarnopolsky, \emph{{Spectrum of
  Majorana Quantum Mechanics with $O(4)^3$ Symmetry}},
  \href{https://arxiv.org/abs/1808.07455}{{\ttfamily 1808.07455}}.

\bibitem{Klebanov:2018nfp}
I.~R. Klebanov, A.~Milekhin, F.~Popov and G.~Tarnopolsky, \emph{{Spectra of
  eigenstates in fermionic tensor quantum mechanics}},
  \href{https://doi.org/10.1103/PhysRevD.97.106023}{\emph{Phys. Rev.}
  {\bfseries D97} (2018) 106023}
  [\href{https://arxiv.org/abs/1802.10263}{{\ttfamily 1802.10263}}].

\bibitem{Polchinski:2016xgd}
J.~Polchinski and V.~Rosenhaus, \emph{{The Spectrum in the Sachdev-Ye-Kitaev
  Model}}, \href{https://doi.org/10.1007/JHEP04(2016)001}{\emph{JHEP}
  {\bfseries 04} (2016) 001}
  [\href{https://arxiv.org/abs/1601.06768}{{\ttfamily 1601.06768}}].

\bibitem{Maldacena:2016hyu}
J.~Maldacena and D.~Stanford, \emph{{Comments on the Sachdev-Ye-Kitaev model}},
  \href{https://doi.org/10.1103/PhysRevD.94.106002}{\emph{Phys. Rev.}
  {\bfseries D94} (2016) 106002}
  [\href{https://arxiv.org/abs/1604.07818}{{\ttfamily 1604.07818}}].

\bibitem{Gross:2016kjj}
D.~J. Gross and V.~Rosenhaus, \emph{{A Generalization of Sachdev-Ye-Kitaev}},
  \href{https://doi.org/10.1007/JHEP02(2017)093}{\emph{JHEP} {\bfseries 02}
  (2017) 093} [\href{https://arxiv.org/abs/1610.01569}{{\ttfamily
  1610.01569}}].

\bibitem{Garcia-Garcia:2016mno}
A.~M. Garcia-Garcia and J.~J.~M. Verbaarschot, \emph{{Spectral and
  thermodynamic properties of the Sachdev-Ye-Kitaev model}},
  \href{https://doi.org/10.1103/PhysRevD.94.126010}{\emph{Phys. Rev.}
  {\bfseries D94} (2016) 126010}
  [\href{https://arxiv.org/abs/1610.03816}{{\ttfamily 1610.03816}}].

\bibitem{Cotler:2016fpe}
J.~S. Cotler, G.~Gur-Ari, M.~Hanada, J.~Polchinski, P.~Saad, S.~H. Shenker
  et~al., \emph{{Black Holes and Random Matrices}},
  \href{https://doi.org/10.1007/JHEP05(2017)118}{\emph{JHEP} {\bfseries 05}
  (2017) 118} [\href{https://arxiv.org/abs/1611.04650}{{\ttfamily
  1611.04650}}].

\bibitem{Gur-Ari:2018okm}
G.~Gur-Ari, R.~Mahajan and A.~Vaezi, \emph{{Does the SYK model have a spin
  glass phase?}},  \href{https://arxiv.org/abs/1806.10145}{{\ttfamily
  1806.10145}}.

\bibitem{Benedetti:2018goh}
D.~Benedetti and R.~Gurau, \emph{{2PI effective action for the SYK model and
  tensor field theories}},
  \href{https://doi.org/10.1007/JHEP05(2018)156}{\emph{JHEP} {\bfseries 05}
  (2018) 156} [\href{https://arxiv.org/abs/1802.05500}{{\ttfamily
  1802.05500}}].

\bibitem{Choudhury:2017tax}
S.~Choudhury, A.~Dey, I.~Halder, L.~Janagal, S.~Minwalla and R.~Poojary,
  \emph{{Notes on melonic $O(N)^{q-1}$ tensor models}},
  \href{https://doi.org/10.1007/JHEP06(2018)094}{\emph{JHEP} {\bfseries 06}
  (2018) 094} [\href{https://arxiv.org/abs/1707.09352}{{\ttfamily
  1707.09352}}].

\bibitem{Krishnan:2016bvg}
C.~Krishnan, S.~Sanyal and P.~N. Bala~Subramanian, \emph{{Quantum Chaos and
  Holographic Tensor Models}},
  \href{https://doi.org/10.1007/JHEP03(2017)056}{\emph{JHEP} {\bfseries 03}
  (2017) 056} [\href{https://arxiv.org/abs/1612.06330}{{\ttfamily
  1612.06330}}].

\bibitem{Krishnan:2017txw}
C.~Krishnan and K.~V.~P. Kumar, \emph{{Towards a Finite-$N$ Hologram}},
  \href{https://doi.org/10.1007/JHEP10(2017)099}{\emph{JHEP} {\bfseries 10}
  (2017) 099} [\href{https://arxiv.org/abs/1706.05364}{{\ttfamily
  1706.05364}}].

\bibitem{Krishnan:2018hhu}
C.~Krishnan and K.~V. Pavan~Kumar, \emph{{Exact Solution of a Strongly Coupled
  Gauge Theory in 0+1 Dimensions}},
  \href{https://doi.org/10.1103/PhysRevLett.120.201603}{\emph{Phys. Rev. Lett.}
  {\bfseries 120} (2018) 201603}
  [\href{https://arxiv.org/abs/1802.02502}{{\ttfamily 1802.02502}}].

\bibitem{Symanzik:1972wj}
K.~Symanzik, \emph{{On Calculations in conformal invariant field theories}},
  \href{https://doi.org/10.1007/BF02824349}{\emph{Lett. Nuovo Cim.} {\bfseries
  3} (1972) 734}.

\bibitem{Almheiri:2014cka}
A.~Almheiri and J.~Polchinski, \emph{{Models of AdS$_{2}$ backreaction and
  holography}}, \href{https://doi.org/10.1007/JHEP11(2015)014}{\emph{JHEP}
  {\bfseries 11} (2015) 014} [\href{https://arxiv.org/abs/1402.6334}{{\ttfamily
  1402.6334}}].

\bibitem{Jevicki:2016bwu}
A.~Jevicki, K.~Suzuki and J.~Yoon, \emph{{Bi-Local Holography in the SYK
  Model}}, \href{https://doi.org/10.1007/JHEP07(2016)007}{\emph{JHEP}
  {\bfseries 07} (2016) 007}
  [\href{https://arxiv.org/abs/1603.06246}{{\ttfamily 1603.06246}}].

\bibitem{Maldacena:2016upp}
J.~Maldacena, D.~Stanford and Z.~Yang, \emph{{Conformal symmetry and its
  breaking in two dimensional Nearly Anti-de-Sitter space}},
  \href{https://doi.org/10.1093/ptep/ptw124}{\emph{PTEP} {\bfseries 2016}
  (2016) 12C104} [\href{https://arxiv.org/abs/1606.01857}{{\ttfamily
  1606.01857}}].

\bibitem{Engelsoy:2016xyb}
J.~Engelsoy, T.~G. Mertens and H.~Verlinde, \emph{{An investigation of
  AdS$_{2}$ backreaction and holography}},
  \href{https://doi.org/10.1007/JHEP07(2016)139}{\emph{JHEP} {\bfseries 07}
  (2016) 139} [\href{https://arxiv.org/abs/1606.03438}{{\ttfamily
  1606.03438}}].

\bibitem{Jensen:2016pah}
K.~Jensen, \emph{{Chaos in AdS$_2$ Holography}},
  \href{https://doi.org/10.1103/PhysRevLett.117.111601}{\emph{Phys. Rev. Lett.}
  {\bfseries 117} (2016) 111601}
  [\href{https://arxiv.org/abs/1605.06098}{{\ttfamily 1605.06098}}].

\bibitem{Geloun:2013kta}
J.~Ben~Geloun and S.~Ramgoolam, \emph{{Counting Tensor Model Observables and
  Branched Covers of the 2-Sphere}},
  \href{https://arxiv.org/abs/1307.6490}{{\ttfamily 1307.6490}}.

\bibitem{Beccaria:2017aqc}
M.~Beccaria and A.~A. Tseytlin, \emph{{Partition function of free conformal
  fields in 3-plet representation}},
  \href{https://doi.org/10.1007/JHEP05(2017)053}{\emph{JHEP} {\bfseries 05}
  (2017) 053} [\href{https://arxiv.org/abs/1703.04460}{{\ttfamily
  1703.04460}}].

\bibitem{Itoyama:2017xid}
H.~Itoyama, A.~Mironov and A.~Morozov, \emph{{Ward identities and combinatorics
  of rainbow tensor models}},
  \href{https://doi.org/10.1007/JHEP06(2017)115}{\emph{JHEP} {\bfseries 06}
  (2017) 115} [\href{https://arxiv.org/abs/1704.08648}{{\ttfamily
  1704.08648}}].

\bibitem{Mironov:2017aqv}
A.~Mironov and A.~Morozov, \emph{{Correlators in tensor models from character
  calculus}},  \href{https://arxiv.org/abs/1706.03667}{{\ttfamily 1706.03667}}.

\bibitem{Diaz:2017kub}
P.~Diaz and S.-J. Rey, \emph{{Orthogonal Bases of Invariants in Tensor
  Models}},  \href{https://arxiv.org/abs/1706.02667}{{\ttfamily 1706.02667}}.

\bibitem{deMelloKoch:2017bvv}
R.~de~Mello~Koch, D.~Gossman and L.~Tribelhorn, \emph{{Gauge Invariants,
  Correlators and Holography in Bosonic and Fermionic Tensor Models}},
  \href{https://arxiv.org/abs/1707.01455}{{\ttfamily 1707.01455}}.

\bibitem{Dartois:2013he}
S.~Dartois, V.~Rivasseau and A.~Tanasa, \emph{{The $1/N$ expansion of
  multi-orientable random tensor models}},
  \href{https://doi.org/10.1007/s00023-013-0262-8}{\emph{Annales Henri
  Poincare} {\bfseries 15} (2014) 965}
  [\href{https://arxiv.org/abs/1301.1535}{{\ttfamily 1301.1535}}].

\bibitem{Fu:2016vas}
W.~Fu, D.~Gaiotto, J.~Maldacena and S.~Sachdev, \emph{{Supersymmetric
  Sachdev-Ye-Kitaev models}}, \href{https://doi.org/10.1103/PhysRevD.95.069904,
  10.1103/PhysRevD.95.026009}{\emph{Phys. Rev.} {\bfseries D95} (2017) 026009}
  [\href{https://arxiv.org/abs/1610.08917}{{\ttfamily 1610.08917}}].

\bibitem{Bulycheva:2017uqj}
K.~Bulycheva, \emph{{A note on the SYK model with complex fermions}},
  \href{https://doi.org/10.1007/JHEP12(2017)069}{\emph{JHEP} {\bfseries 12}
  (2017) 069} [\href{https://arxiv.org/abs/1706.07411}{{\ttfamily
  1706.07411}}].

\bibitem{Yoon:2017nig}
J.~Yoon, \emph{{SYK Models and SYK-like Tensor Models with Global Symmetry}},
  \href{https://doi.org/10.1007/JHEP10(2017)183}{\emph{JHEP} {\bfseries 10}
  (2017) 183} [\href{https://arxiv.org/abs/1707.01740}{{\ttfamily
  1707.01740}}].

\bibitem{Peng:2017spg}
C.~Peng, M.~Spradlin and A.~Volovich, \emph{{Correlators in the $\mathcal{N}=2$
  Supersymmetric SYK Model}},
  \href{https://doi.org/10.1007/JHEP10(2017)202}{\emph{JHEP} {\bfseries 10}
  (2017) 202} [\href{https://arxiv.org/abs/1706.06078}{{\ttfamily
  1706.06078}}].

\bibitem{Chaturvedi:2018uov}
P.~Chaturvedi, Y.~Gu, W.~Song and B.~Yu, \emph{{A note on the complex SYK model
  and warped CFTs}},  \href{https://arxiv.org/abs/1808.08062}{{\ttfamily
  1808.08062}}.

\bibitem{Jaewon:2018}
J.~Kim, \emph{{Large $N$ Tensor and SYK Models, Princeton University Senior
  Thesis, 2018}}, .

\bibitem{Gurau:2016lzk}
R.~Gurau, \emph{{The complete $1/N$ expansion of a SYK--like tensor model}},
  \href{https://arxiv.org/abs/1611.04032}{{\ttfamily 1611.04032}}.

\bibitem{Breitenlohner:1982jf}
P.~Breitenlohner and D.~Z. Freedman, \emph{{Stability in Gauged Extended
  Supergravity}},
  \href{https://doi.org/10.1016/0003-4916(82)90116-6}{\emph{Annals Phys.}
  {\bfseries 144} (1982) 249}.

\bibitem{Dymarsky:2005uh}
A.~Dymarsky, I.~R. Klebanov and R.~Roiban, \emph{{Perturbative search for fixed
  lines in large N gauge theories}},
  \href{https://doi.org/10.1088/1126-6708/2005/08/011}{\emph{JHEP} {\bfseries
  08} (2005) 011} [\href{https://arxiv.org/abs/hep-th/0505099}{{\ttfamily
  hep-th/0505099}}].

\bibitem{Pomoni:2008de}
E.~Pomoni and L.~Rastelli, \emph{{Large N Field Theory and AdS Tachyons}},
  \href{https://doi.org/10.1088/1126-6708/2009/04/020}{\emph{JHEP} {\bfseries
  04} (2009) 020} [\href{https://arxiv.org/abs/0805.2261}{{\ttfamily
  0805.2261}}].

\bibitem{Grabner:2017pgm}
D.~Grabner, N.~Gromov, V.~Kazakov and G.~Korchemsky, \emph{{Strongly
  $\gamma$-Deformed $\mathcal{N}=4$ Supersymmetric Yang-Mills Theory as an
  Integrable Conformal Field Theory}},
  \href{https://doi.org/10.1103/PhysRevLett.120.111601}{\emph{Phys. Rev. Lett.}
  {\bfseries 120} (2018) 111601}
  [\href{https://arxiv.org/abs/1711.04786}{{\ttfamily 1711.04786}}].

\bibitem{Prakash:2017hwq}
S.~Prakash and R.~Sinha, \emph{{A Complex Fermionic Tensor Model in $d$
  Dimensions}}, \href{https://doi.org/10.1007/JHEP02(2018)086}{\emph{JHEP}
  {\bfseries 02} (2018) 086}
  [\href{https://arxiv.org/abs/1710.09357}{{\ttfamily 1710.09357}}].

\bibitem{Klebanov:1999tb}
I.~R. Klebanov and E.~Witten, \emph{{AdS / CFT correspondence and symmetry
  breaking}}, \href{https://doi.org/10.1016/S0550-3213(99)00387-9}{\emph{Nucl.
  Phys.} {\bfseries B556} (1999) 89}
  [\href{https://arxiv.org/abs/hep-th/9905104}{{\ttfamily hep-th/9905104}}].

\bibitem{Liu:2018jhs}
J.~Liu, E.~Perlmutter, V.~Rosenhaus and D.~Simmons-Duffin,
  \emph{{$d$-dimensional SYK, AdS Loops, and $6j$ Symbols}},
  \href{https://arxiv.org/abs/1808.00612}{{\ttfamily 1808.00612}}.

\bibitem{Murugan:2017eto}
J.~Murugan, D.~Stanford and E.~Witten, \emph{{More on Supersymmetric and 2d
  Analogs of the SYK Model}},
  \href{https://doi.org/10.1007/JHEP08(2017)146}{\emph{JHEP} {\bfseries 08}
  (2017) 146} [\href{https://arxiv.org/abs/1706.05362}{{\ttfamily
  1706.05362}}].

\bibitem{Chang:2018sve}
C.-M. Chang, S.~Colin-Ellerin and M.~Rangamani, \emph{{On Melonic Supertensor
  Models}},  \href{https://arxiv.org/abs/1806.09903}{{\ttfamily 1806.09903}}.

\end{thebibliography}\endgroup

\end{document}